\documentclass[aps,prb,showpacs,twocolumn,floatfix,letterpaper,%
superscriptaddress,10pt]{revtex4-1}

\usepackage[intlimits,sumlimits]{amsmath}
\usepackage{amsfonts,amssymb}
\usepackage{bm}
\usepackage{graphicx}
\usepackage{hyperref}
\usepackage{tabularx}
\usepackage{dcolumn}

\usepackage{hyperref}
\hypersetup{%
pdftitle={Generalized spectral method for near-field optical microscopy},%
pdfauthor={B.-Y. Jiang et al.},%
pdfpagemode={UseNone},%
pdfstartview={FitH},%
breaklinks=true,%
citecolor=blue,%
colorlinks=true,%
linkcolor=blue,%
urlcolor=blue}

\providecommand\tsub[1]{\ensuremath{_{\text{#1}}}} %
\providecommand\tsup[1]{\ensuremath{^{\text{#1}}}} %

\providecommand\zfint[1][\infty]{\int_{0}^{#1}} %
\providecommand\dd{\ensuremath{\textrm{d}}} %
\providecommand\zfsum[2][0]{\ensuremath{\sum_{#2=#1}^{\infty}}} %
\providecommand\ofsum[1]{\zfsum[1]{#1}} %
\providecommand\sysum[2][\infty]{\ensuremath{\sum_{#2=-#1}^{#1}}}

\providecommand\eqc{\,,\quad} %

\providecommand\bhat[1]{\ensuremath{\hat{\boldsymbol{#1}}}} %

\DeclareMathOperator{\re}{Re}
\DeclareMathOperator{\im}{Im}

\DeclareMathOperator{\arccosh}{arccosh}

\providecommand\bhat[1]{\bvec{\hat{#1}}} %

\providecommand\fhalf[1]{\ensuremath{\frac{#1}{2}}} 

\providecommand\prm{\ensuremath{^{\prime}{}}} 

\def\XXint#1#2#3{{\setbox0=\hbox{$#1{#2#3}{\int}$}
     \vcenter{\hbox{$#2#3$}}\kern-.5\wd0}}

\providecommand\tExt{\tsub{ext}} %

\providecommand\eps[1]{\ensuremath{\epsilon_{#1}}} %
\providecommand\kz[1]{\ensuremath{k_{#1}^{z}{}}} %

\providecommand\zp{\ensuremath{z\tsub{p}}} %
\providecommand\ztip{\ensuremath{z\tsub{tip}}} %
\providecommand\rr{\ensuremath{r_{\alpha}}} %
\def\imagewidth{2.31in}

\begin{document}
\newcommand\UCSD{Department of Physics, University of California San Diego, 9500 Gilman Drive, La Jolla, California 92093, USA}

\title{Generalized spectral method for near-field optical microscopy}

\author{B.-Y.~Jiang}
\affiliation{\UCSD}

\author{L.~M.~Zhang}
\affiliation{\UCSD}

\author{A.~H.~Castro~Neto}
\affiliation{Department of Physics, Boston University, 590 Commonwealth Avenue, Boston, Massachusetts 02215}
\affiliation{Centre for Advanced 2D Materials and Graphene Research Centre, National University of Singapore, Singapore 117542, Singapore}

\author{D.~N.~Basov}
\affiliation{\UCSD}

\author{M.~M.~Fogler}
\affiliation{\UCSD}

\date{\today}

\begin{abstract}

Electromagnetic interaction between a sub-wavelength particle (the `probe')
and a material surface (the `sample') is studied theoretically.
The interaction is shown to be governed by a series of resonances 
corresponding to surface polariton modes localized near the probe. 
The resonance parameters depend on the dielectric function and geometry of the probe,
as well as the surface reflectivity of the material.
Calculation of such resonances is carried out
for several types of axisymmetric probes: spherical, spheroidal, and pear-shaped. 
For spheroids an efficient numerical method is developed,
capable of handling cases of large or strongly momentum-dependent surface reflectivity.
Application of the method to highly resonant materials such as aluminum oxide (by itself or covered with graphene)
reveals a rich structure of multi-peak spectra and nonmonotonic  approach curves,
i.e., the probe-sample distance dependence.
These features also strongly depend on the probe shape and
optical constants of the model.
For less resonant materials such as silicon oxide,
the dependence is weak, so that the spheroidal model is reliable.
The calculations are done within the quasistatic approximation with radiative damping included perturbatively.

\end{abstract}


\maketitle

\section{Introduction}
\label{sec:Introduction}
The problem of electromagnetic interaction between a material surface and a small external particle is fundamental to numerous physical phenomena and spectroscopic techniques,
including surface-enhanced Raman scattering, surface fluorescence, adsorbed molecules spectroscopy, and near-field microscopy.
From the point of view of electromagnetic theory, it is a special kind of scattering problem where the scatterer resides in a uniform half-space, e.g., vacuum, while the effect of the other half-space --- the sample --- is represented by the surface reflectivity $\rr(q, \omega)$.
The reflectivity may depend on the in-plane momentum $q$, frequency $\omega$, and polarization $\alpha = \mathrm{P}$ or $\mathrm{S}$.
Far-field optics describes the regime $q < \omega / c$.
Momenta $q \gg \omega / c$,
which correspond to in-plane distances $\Delta \rho$ much smaller than the diameter $c / \omega$ of Wheeler's radian sphere,~\cite{Wheeler1959ras} are the domain of near-field optics.

This work is motivated by recent advancements of the scattering-type near-field optical microscopy~\cite{Keilmann2004nfm, Atkin2012n} (s-SNOM),
which has become one of the leading tools for measuring optical response of diverse materials on spatial scales as short as $5$--$20\,\mathrm{nm}$.
Thanks to technical improvements and the development of tunable and broad-band infrared sources,~\cite{Amarie2009min, Fei2011ino, Amarie2011bia, Huth2011isn}
the s-SNOM has provided insights into properties of complex oxides,~\cite{Ma2006gan, Qazilbash2007mti, Zhan2007tmi, Qazilbash2009Isa, Frenzel2009ies, Jones2010noi, Lai2010mpr} organic monolayers,~\cite{Nikiforov2009ppa} graphene, and other two-dimensional crystals.~\cite{Fei2011ino, Fei2012gto, Chen2012oni, Dai2014tsp}

\begin{figure*}[tbh]
	\begin{center}
		\includegraphics[height=1.6in]{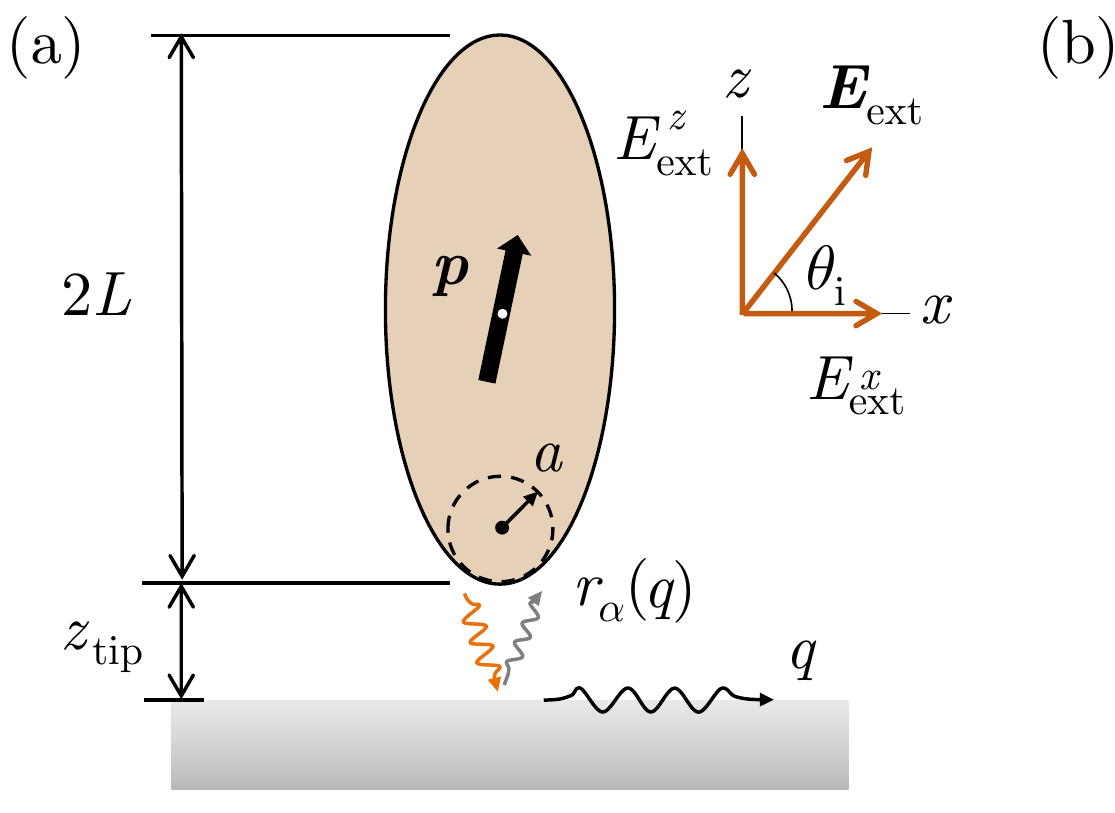}
		\includegraphics[height=1.55in]{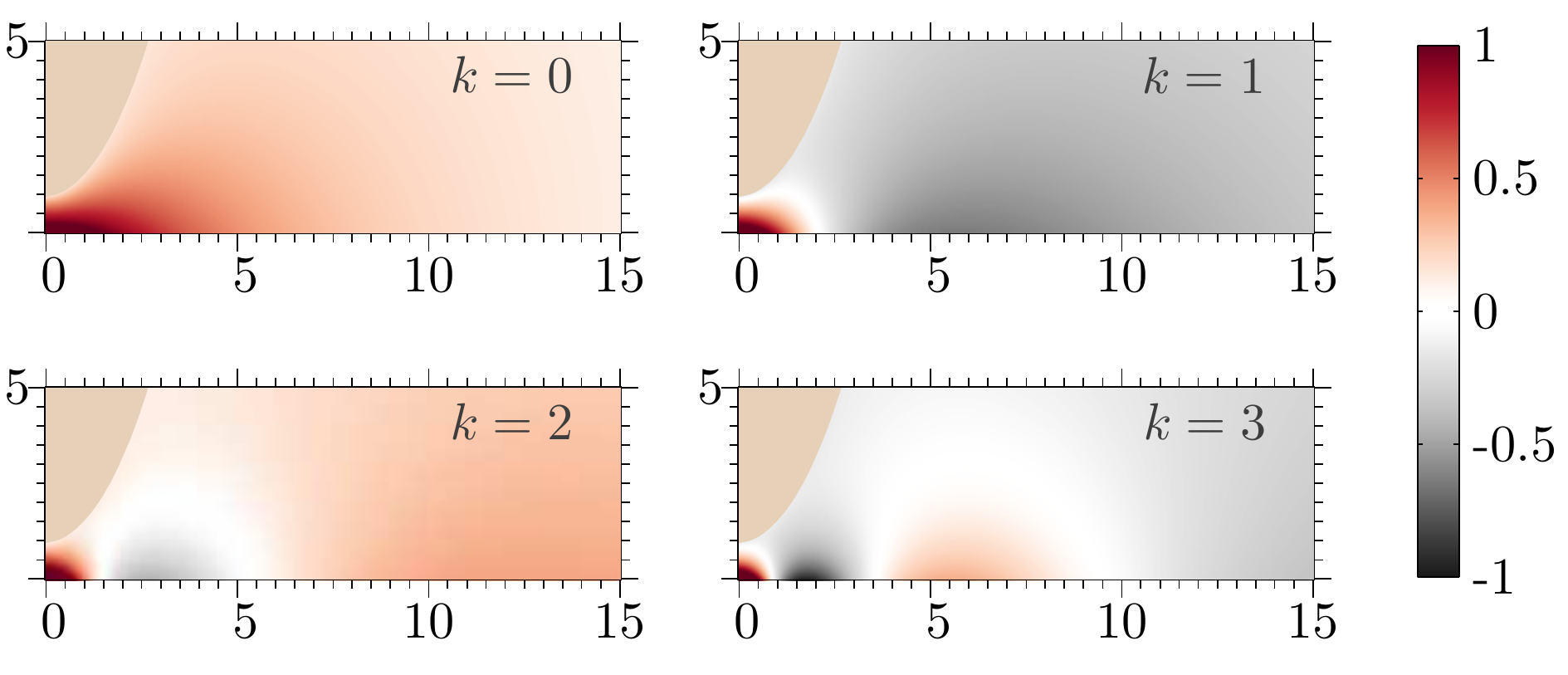}
	\end{center}
	\caption{(Color online) (a) Schematics of an s-SNOM experiment in which a polarizable probe is used to examine a sample characterized by the surface reflectivity $r_\alpha(q)$.
		The external electric field $\bm{E}\tsub{ext}$ incident on the system creates evanescent waves inside the probe-sample gap.
	  This modifies the dipole moment $\bm{p}$ of the probe, which is detectable by its far-field radiation.
	  (b) The real-space potential distribution for the first four eigenmodes of the probe polarizability $\chi^\bot$ computed numerically for a spheroidal probe of half-length $L = 25a$.
	  The axes are the $x$- and $z$-coordinates in units of $a$, the curvature radius of the apex of the probe.
	  The probe's location is represented by the uniformly shaded beige area in the upper left corner of each panel.
	  }
	\label{fig:modes}
\end{figure*}

The schematics of an s-SNOM experiment is shown in Fig.~\ref{fig:modes}(a).
A sharp elongated probe is brought into close proximity of a sample and is illuminated by an external electromagnetic wave with electric field $\bm{E}_{\text{ext}} e^{-i \omega t}$.
Its interaction with the probe creates scattered waves $e^{i \bm{q} \bm{\rho} + i k^z z - i \omega t}$, $\bm{\rho}=(x, y)$, with arbitrary in-plane momentum $\bm{q}$, including large-$q$ evanescent waves, $k^z = \sqrt{(\omega / c)^2 - q^2} \simeq i q$.
Multiple reflections of these waves inside the probe-sample nanogap cause small but important changes in the total radiating dipole moment $\bm{p}  e^{-i \omega t}$ of the probe.
These changes are detected by measuring the far-field scattering signal as a function of the probe coordinates.
This signal is proportional to the probe polarizabilities,
\begin{equation}
\chi^{\bot} \equiv p^{z} / E_{\text{ext}}^{z}\,,
\quad
\chi^{\parallel} \equiv p^{x} / E_{\text{ext}}^{x}\,,
\label{eqn:chi}
\end{equation}
which have the dimension of volume.

The goal of this paper is to study the properties of functions $\chi^\bot$ and $\chi^\parallel$.
For simplicity, we consider only axisymmetric probes.
We are especially interested in probes of large aspect ratio.
In the experiment, strongly elongated probes are used
because of high longitudinal polarizability $\chi^\bot$,
which promotes an efficient coupling between evanescent and far-field radiation modes --- the ``antenna'' effect --- making the detection of the near-field component possible.

We assume that the length of the probe is much smaller than the diameter of the radian sphere $c / \omega$,
so that the scattering problem can be treated within the quasistatic approximation.
The probe shape we examine the most is a prolate spheroid.
At first glance both of these assumptions are unrealistic because actual probes are not spheroidal and their length (typically, tens of $\mu\mathrm{m}$) can often exceed $c / \omega$ for $\omega$ in infrared or optical frequency domain.
Yet this model was previously found to yield quantitative agreement with
the s-SNOM experimental data for many materials.
This apparent agreement can be expected in cases where
the surface reflectivity $r_\alpha(q, \omega)$ of the sample is not too large,
and the aspect ratio of the probe does not vary greatly from one experiment to the next.
Under such conditions
the gross features of the s-SNOM scattering amplitude should indeed have only a modest dependence on the exact shape of the probe and other experimental parameters.
However, fine details of the scattering amplitude are shape-dependent even in this case~\cite{McLeod2014lrm} and they may be discerned
as the instrumental resolution improves.
Furthermore, for samples with high reflectivity,
even the gross features become sensitive to the shape and size of the probe.
To demonstrate these trends in this paper we study the longitudinal and the transverse polarizabilities in great detail.
We will ignore the S-polarization reflectivity $r_{\text{S}}(q, \omega)$ because
for most materials it becomes very small at $q \gg \omega / c$.
Hence, $\chi^\nu$ are  functionals of the remaining reflectivity function $r_{\mathrm{P}}(q, \omega)$ and the probe-sample distance $\ztip$.
We show that such functionals
can be quite complicated, especially for strongly
momentum-dependent reflectivity typical of layered and/or ultrathin materials.
Therefore, it is good to start with a simpler case
of a \emph{bulk} medium with a $q$-independent reflectivity
\begin{equation}
\beta(\omega) \equiv r_{\mathrm{P}}(q, \omega)\,,
\label{eqn:beta_definition}
\end{equation}
so that for a fixed $\ztip$ and $\omega$,
the probe polarizabilities are functions of a single parameter $\beta$.

\begin{figure*}[htb]
\begin{center}
\includegraphics[width=\imagewidth]{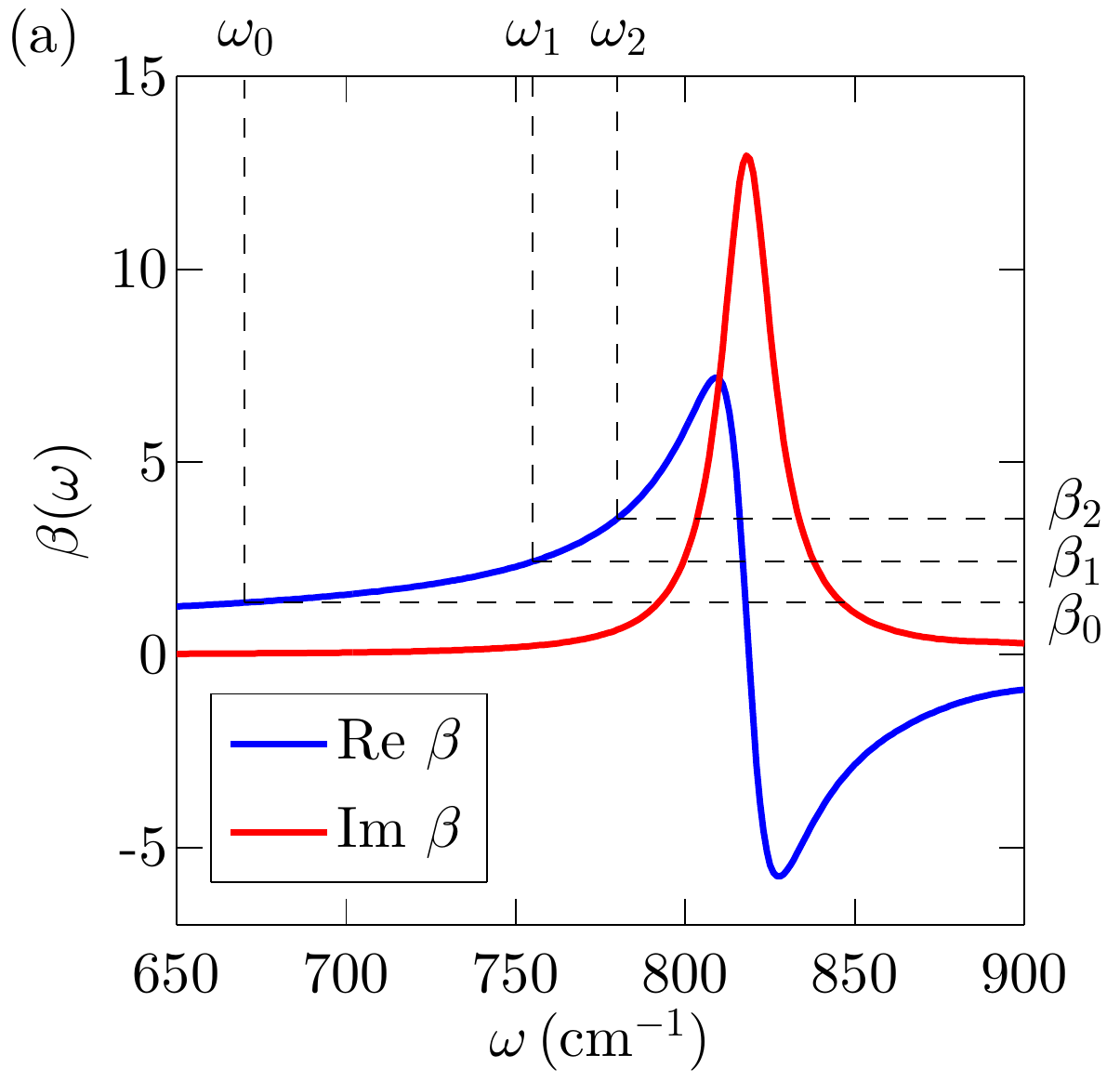}
	\hspace{0.3 in}
    \includegraphics[width=1.9 in]{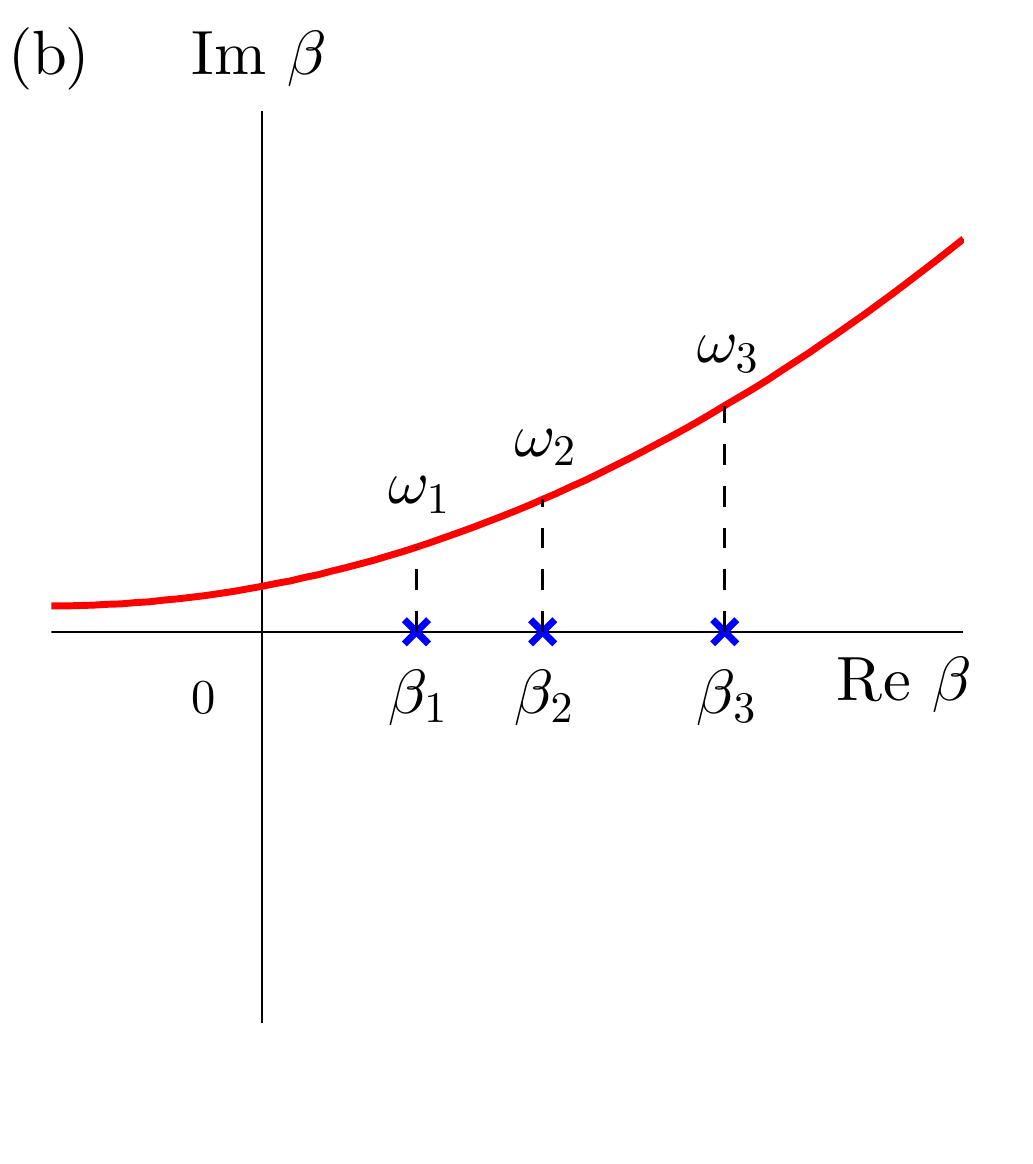}    
	\hspace{0.2 in}
    \includegraphics[width=1.9 in]{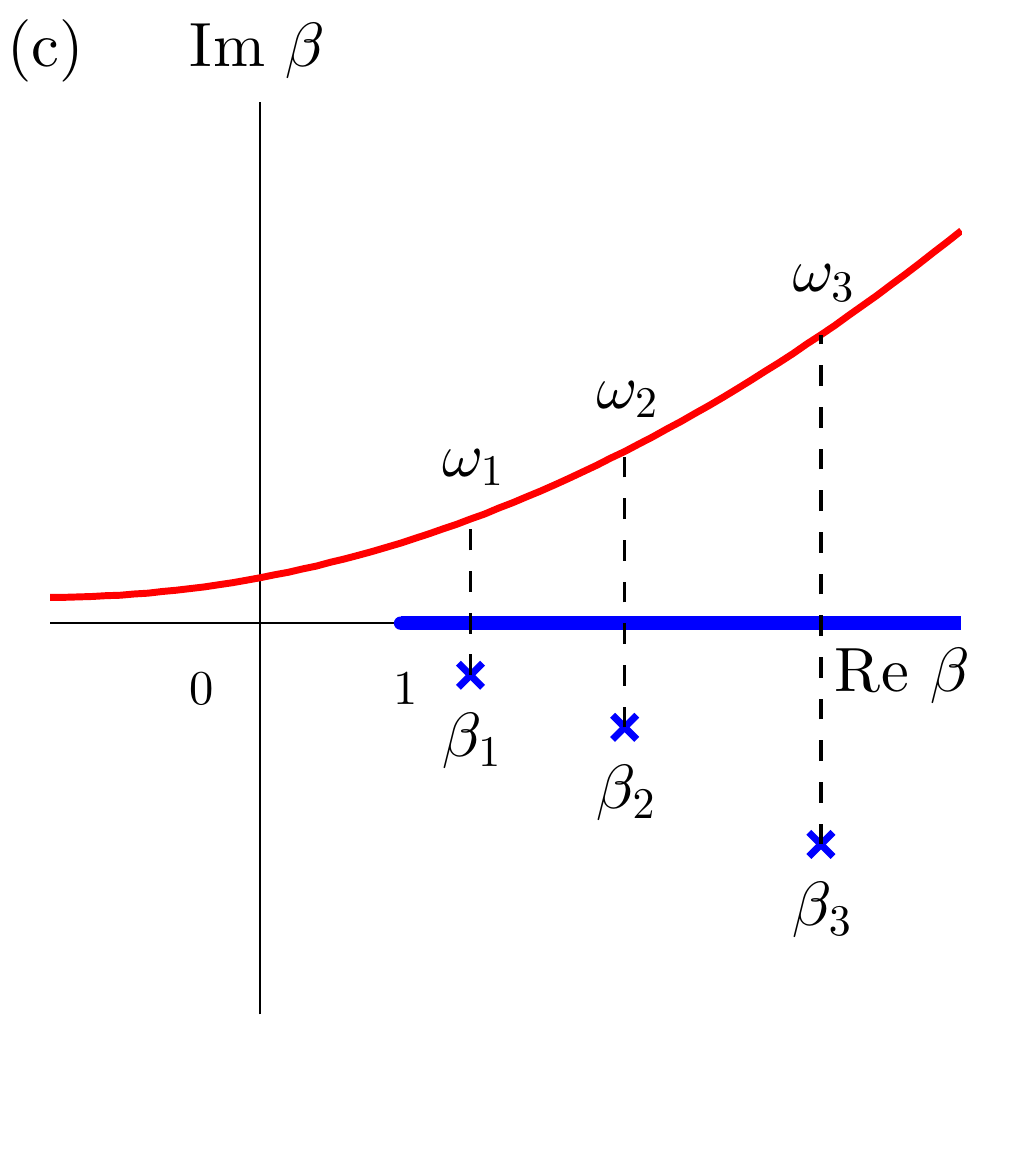}
\end{center}
\caption{(Color online) (a) Near-field reflectivity $\beta(\omega)$ of bulk Al$_2$O$_3$ discussed in Sec.~\ref{sec:result}.
Whenever the condition $\re \beta(\omega)=\beta_k^\nu$ is met,
a local maximum appears in $\im \chi^\nu$.
The frequencies of three such resonances are indicated by the dashed lines.
(b) In the complex plane of $\beta$, the poles $\beta_k^\nu$ lie on the positive real axis, while real materials trace curves in the upper half plane, shown in red.
(c) A full electrodynamic treatment predicts that the poles shift into the lower half-plane and an additional nonanalyticity in the form of a branch cut $[1,\infty)$ appears.
}
\label{fig:Al2O3_beta}
\end{figure*}

It should be clarified that while the absolute reflectivity may not exceed unity for the radiative modes $q < \omega / c$
because of energy conservation,
for the evanescent ones $q > \omega / c$ it is allowed do so.
Large $\beta$'s are indicative of weakly damped surface modes in a material, e.g.,
surface phonons in dielectrics or surface plasmons in metals.
We use the umbrella term ``surface polaritons'' for all such modes. 
The energy loss due to evanescent modes is governed not by $|\beta|$ but by $\textrm{Im}\, \beta$
which must be nonnegative at $q > \omega / c$.
(To compute the losses $\textrm{Im}\, \beta$ needs to be integrated over $q$ with a weight that depends on the probe-sample distance.~\cite{Ford1984eio})
In the limit of vanishingly small dissipation,
$\textrm{Im}\, \beta(\omega)$ tends to a $\delta$-function peak at the mode frequency.
In practice,
$\textrm{Im}\, \beta$ (and consequently $|\beta|$) as high as $10$--$20$ is possible for well-ordered crystalline solids, e.g., aluminum oxide Al$_2$O$_3$ possessing sharp phonon modes [Fig.~\ref{fig:Al2O3_beta}(a)].
Therefore, a robust theoretical formalism must be capable of computing functions $\chi^{\nu} (\beta)$ in the entire upper complex half-plane.
To meet this requirement such a formalism must correctly reproduce the analytic properties of functions
$\chi^{\nu} (\beta)$.
We adopt a version of the generalized spectral method (GSM) in which the total field outside the probe and sample is decomposed into eigenfunctions of an auxiliary homogeneous problem, and the role of eigenvalues are played by the reflectivity $\beta$, the so-called
$\beta$-method in the terminology of Ref.~\onlinecite{Agranovich1999gme}.
(Similar formalism is also known in the theory of conductivity of heterogeneous media.~\cite{Bergman1978dcc, Bergman1979dct})
We show that for any probe-sample distance $\ztip > 0$ functions $\chi^{\nu} (\beta)$ are meromorphic.
In other words, they admit the series representations
\begin{equation}
\chi^{\nu} (\beta)
  = \zfsum{k} \frac{R_{k}^{\nu}}{\beta_{k}^{\nu} - \beta} \,,
\quad
\nu = \,\bot \text{ or} \parallel\,,
\label{eqn:chi_eff_general_form}
\end{equation}
where the sequence of poles $\beta_{k}^{\nu}$ has no accumulation points, and so, no upper limit.
Additionally, we will show that if the probe is made of an ideal conductor and no other sources of dissipation are present,
then the poles $\beta_{k}^{\nu} > 1$ and the residues $R_{k}^{\nu} > 0$ are real.
If the dielectric constant $\epsilon_\mathrm{tip}$ of the probe is considered fixed,
$R_{k}^{\nu}$ and $\beta_{k}^{\nu}$ depend only on the geometric factors: the probe shape, size, and its distance $\ztip$ to the surface.
All these results comply with the general theory of the $\beta$-method developed in Ref.~\onlinecite{Agranovich1999gme}.

Both $\beta_{k}^{\nu}$ and $R_{k}^{\nu}$ grow exponentially with $k$ but
their ratios remain bounded and satisfy the sum rule
\begin{equation}
  \zfsum{k} \frac{R_{k}^{\nu}}{\beta_{k}^{\nu}} = \chi_0^{\nu}\,.
\label{eqn:chi_sum_rule}
\end{equation}
Here $\chi_0^{\nu} \equiv \chi^{\nu}(\beta = 0)$ is the polarizability of an isolated probe, which does not depend on $\ztip$.
These properties ensure convergence of the series~\eqref{eqn:chi_eff_general_form} at any $\beta \neq \beta_{k}^{\nu}$.
On the other hand, if a material-specific $\beta(\omega)$ approaches any of $\beta_{k}^{\nu}$, a resonant peak in $\chi^{\nu}$ and ultimately, in the near-field signal, would be observed. 

The divergence of $\chi^\nu$ at a given pole implies that a nonzero dipole, i.e., free oscillations
may exist in the absence of any external field.
Physical intuition about this regime is aided by the method of images,
according to which real charges $Q_i$ on the probe interact with their virtual images $-\beta Q_i$ inside the sample and
for $\beta > 1$ achieve a runaway positive feedback.
However, one must keep in mind that these eigenmodes arise only in the auxiliary problem where the sample is substituted by a fictitious material of reflectivity $\beta_k^\nu$.
The divergence never actually happens in real materials due to their inherent dissipation, which enters in the form of a positive imaginary part in $\beta$ as shown in
Fig.~\ref{fig:Al2O3_beta}(a) and~\ref{fig:Al2O3_beta}(b).
The resonances are further damped due to shifting of the poles $\beta_k^\nu$ to the lower complex half-plane when radiative corrections are considered  [Fig.~\ref{fig:Al2O3_beta}(c)], as discussed in more detail in Sec.~\ref{sec:shape}.
For a generic probe that ends in a rounded tip,
the amplitude of the eigenmodes is the greatest near the tip,
as illustrated in Fig.~\ref{fig:modes}(b) for a spheroidal probe.
Overall, this physical picture of tip-localized eigenmodes is an elegant and economical approach to understanding the mechanism of probe-sample coupling.

The main objective of the present work is to elucidate the 
analytical properties of the coefficients $\beta_{k}^{\nu}$ and $R_{k}^{\nu}$.
We focus on the practically interesting case where the probe length $L$ is much larger than 
the curvature radius $a$ of the probe tip.
We show that for such strongly elongated probes three regimes can be distinguished.
The first is the short-distance limit $\ztip \ll a$
where the behavior of $\beta_{k}^{\nu}$ is universal.
We show that it can be derived from the known exact solutions for spherical particles (Sec.~\ref{sec:limits}).
The second is the long-distance limit, $\ztip \gg L$,
where the probe acts as a point-dipole
and the functional form of the resonance parameters is again universal.
The remaining third regime $a < \ztip < L$ is the most nontrivial one where $\beta_{k}^{\nu}$ and $R_{k}^{\nu}$ depend on the probe shape.

For all the probe geometries we study the poles $\beta_{k}^{\nu}$ grows exponentially with $k$, and so for moderate values of $\beta$
it is permissible to truncate the series in Eq.~\eqref{eqn:chi_eff_general_form} after one or a few leading terms.
This truncation is effectively
done in simplified models~\cite{Hillenbrand2000coc, Taubner2004nrt, Aizpurua2008sei,Ocelic2007qnf, Amarie2011bia} of the probe-sample coupling, see Sec.~\ref{sec:Conclusions}.
However, this simplification may lead to qualitatively and quantitatively wrong results at small $\ztip$ and/or for large $\beta$.
The latter characterize highly polar materials such as SiO$_2$ \cite{Zhang2012nfs}
(a commonly used substrate)
and the already mentioned Al$_2$O$_3$ (an important reference material of infrared optics).

Besides addressing analytical properties of the probe polarizabilities,
we also discuss methods for their numerical computation.
For the simplest case of a momentum-independent reflectivity,
the calculation can be made virtually instantaneous with the help of Eq.~\eqref{eqn:chi_eff_general_form} once the first few $\beta_k^\nu$ and $R_k^\nu$ are computed and stored.
For specific case of a spheroidal probe, this calculation can be further accelerated using the spheroidal harmonics basis instead of the standard boundary element method (BEM).
Since the number of relevant poles and residues is relatively small,
for further convenience, they can be fitted to analytical forms,
see an example for $L = 25a$ spheroidal probe in Table~\ref{tab:table}.
The speed becomes a crucial consideration if the calculations have to be done repeatedly.
An important example is extracting optical constants of the sample from near-field spectroscopy data by curve-fitting algorithms.~\cite{McLeod2014lrm}
One may anticipate to find a considerable speed-up if this inverse problem
were treated using the GSM.
The acceleration occurs because the unknown physical parameter $\beta = \beta(\omega)$ of the sample and the geometric parameters $\beta_{k}^{\nu}$ and $R_{k}^{\nu}$ of the probe stand clearly separated.
The GSM also applies for momentum-dependent $r_\mathrm{P}(q, \omega)$,
e.g., for layered samples; however, in the current implementation
the speed-up compared to the BEM is less significant. 

\newcolumntype{j}{D{.}{.}{4}}
\newcolumntype{d}{D{.}{.}{5}}
\newcolumntype{e}{D{.}{.}{7}}
\newcolumntype{f}{D{.}{.}{1}}
\newcolumntype{g}{D{.}{.}{9}}
\newcolumntype{h}{D{.}{.}{3}}
\newcolumntype{i}{D{.}{.}{0}}
\begin{table*}
\caption{Coefficients of the nine-pole rational fits $\log \beta_k = \sum\limits_{i=0}^5 a_{i} \alpha^{i}/\sum\limits_{i=0}^4 b_{i} \alpha^{i}$ and $R_k a^{-3} \mathcal{Z}^{-1} = \sum\limits_{i=0}^5 c_i \mathcal{Z}^{i} / \sum\limits_{i=0}^3 d_i \mathcal{Z}^{i} $ for $L = 25a$ and $0.003 < \mathcal{Z} < 10$, where $\mathcal{Z}\equiv \ztip/a$.
The fits for the residues apply only to the first eight poles, $k = 0$ through $7$. 
The remaining residue $R_8$ is constrained to obey the sum rule~\eqref{eqn:chi_sum_rule}.
}
\begin{tabular*}{7in}{@{\extracolsep{\fill} }r f d d d h g c d d d d} 
\hline \hline\\[-0.1in] 
$k$ & 
\multicolumn{1}{c}{$a_5$} &
\multicolumn{1}{c}{$a_4$} &
\multicolumn{1}{c}{$a_3$} &
\multicolumn{1}{c}{$a_2$} &
\multicolumn{1}{c}{$a_1$} &
\multicolumn{1}{c}{$a_0$} &
\multicolumn{1}{c}{$b_4$} &
\multicolumn{1}{c}{$b_3$} &
\multicolumn{1}{c}{$b_2$} &
\multicolumn{1}{c}{$b_1$} &
\multicolumn{1}{c}{$b_0$} 
\\ [0.5ex] 

\hline 
\\[-0.1in]
0&            3 &     -36.399  &     234.56  &    -762.76   &    1783.1&    -0.015667        &    1   &   -10.345   &    83.048    &  -417.42   &    1522.2\\
1&            5 &     -25.733  &     111.01  &    -93.002   &    290.11& 		0     &      1 &     -3.4964 &      27.841  &     11.949 &      87.231\\
2&            7 &     -33.029  &     157.55  &    -118.29   &    1391.3& 0     &      1 &     -0.9961 &     -2.6274  &     149.14 &      253.32\\
3&            9 &     -36.251  &      173.3   &   -40.018    &   1879.9 & 0       &     1  &   -0.12625 &     -8.2396  &     185.03 &      246.31\\
4&           11&      -47.517 &      221.75 &     -85.286  &     2292.3&  -0.00017314     &       1&      -0.3866&      -8.4205 &       180.3 &      237.17\\
5&           13&      -45.678 &      233.45 &      42.551  &     3435.5&  -0.000094847     &       1&       1.0728 &     -20.329 &      253.41&       291.59\\
6&           15&      -46.254 &      254.23 &      223.28  &       2547 &  0        &    1   &  -0.75722  &      7.941    &   175.21   &    185.07\\
7&           17&      -27.808 &      235.93 &      770.95   &    1961.3&  0      &      1 &     -1.8957 &      34.251  &     163.71 &      122.51\\
8&           19&      -65.583&       251.72 &     -308.28 &      402.67&  0.000032595&  1  &     1.6624   &     4.104     & -1.1364      &  23.24\\
 [1ex] 
\hline \hline\\
\end{tabular*}
\begin{tabular*}{7in}{@{\extracolsep{\fill} }c h h h h h h c d e g} 
\hline\hline\\[-0.1 in]
$k$ & 
\multicolumn{1}{c}{$c_5$} &
\multicolumn{1}{c}{$c_4$} &
\multicolumn{1}{c}{$c_3$} &
\multicolumn{1}{c}{$c_2$} &
\multicolumn{1}{c}{$c_1$} &
\multicolumn{1}{c}{$c_0$} &
\multicolumn{1}{c}{$d_3$} &
\multicolumn{1}{c}{$d_2$} &
\multicolumn{1}{c}{$d_1$} &
\multicolumn{1}{c}{$d_0$}  \\ [0.5ex] 

\hline 
\\[-0.1in]
0&       3.9999&       303.23&       5141.1&       4811.1&       282.17&       1.4941&            1&      0.77084&     0.023552&  0.000027594\\
1&       12.001&       916.75&        17089&        33226&       2371.9&       12.255&            1&       1.4472&      0.11961&    0.0003052\\
2&       24.001&       1844.7&        35207&        90005&       36.881&       11.908&            1&       1.9006&     0.012624&   0.00025669\\
3&           40&       3166.2&        65393&       288144&       224067&       4417.7&            1&       4.1622&       2.6905&      0.09134\\
4&       60.028&       4584.9&        87200&       196974&       5632.8&        28.67&            1&       1.5802&     0.066244&   0.00048749\\
5&       84.685&       6304.9&       122561&       226585&       4364.5&        4.237&            1&       1.2714&     0.044389&  0.000077993\\
6&       116.26&       8216.9&       166927&       316568&        98948&        843.4&            1&       1.4088&      0.44188&    0.0087351\\
7&       146.83&        10354&       214160&       350367&        77249&       334.56&            1&       1.1606&      0.28331&    0.0034808\\[1ex]
\hline \hline
\label{tab:table}
\end{tabular*}
\end{table*}

The remainder of the article is organized as follows.
In Sec.~\ref{sec:limits} we analyze the universal aspects of the short- and the long-distance regimes.
In Sec.~\ref{sec:spheroidal} the spheroidal probe model is considered.
The equations for the poles and residues are presented and the results of their numerical solution for the case of a $q$-independent $r_\mathrm{P}$ are discussed.
In Sec.~\ref{sec:weak-local} we explore the effects due to a weakly $q$-dependent surface reflectivity.
In Sec.~\ref{sec:result} we apply our numerical method to computing the near-field response of bulk Al$_2$O$_3$,
a strongly polar material.
In Sec.~\ref{sec:q-depend} we perform the calculation for the same Al$_2$O$_3$ substrate but covered with graphene,
which is a system with a strongly $q$-dependent reflectivity.
In Sec.~\ref{sec:shape} we discuss the effects of the probe shape and retardation on these calculations.
We also do a similar comparison for SiO$_2$, a less polar material.
In Sec.~\ref{sec:Conclusions} we discuss
prior theoretical work and close with concluding remarks.
Technical details of the derivations are summarized in Appendix.
The source code of our computer program is available as the online Supplemental Material for this article.

\section{Probe-sample interaction in short- and long-distance limits}
\label{sec:limits}

We start with a qualitative analysis of the short-distance regime defined by the condition $\ztip \ll a$.
In this limit the structure of the localized polariton modes
can be understood intuitively by analogy~\cite{Rendell1981spc} to electromagnetic modes in an open cavity.
The probe-sample gap can be approximated by a cavity with height $z(\rho) \simeq \ztip + (\rho^2 / 2 a)$ gradually increasing as a function of the radial position $\rho$.
For simplicity, let us assume the surface reflectivity of the probe is equal to unity, as for an ideal conductor.
To have free oscillations exist in such a cavity the surface reflectivity $\beta$ of the sample must exceed unity,
compensating for the exponential decay of the evanescent waves.
The condition of the self-sustained oscillations is $\beta \exp\bigl(2 i k^z(\rho) z(r)\bigr) = 1$.
Accordingly, the local radial momentum $q(\rho) \simeq -i k^z(\rho) = \log \beta\, /\, 2 z(\rho)$.
Imposing the quasiclassical quantization condition $\int_0^\infty d \rho q(\rho) = \pi [k + \mathcal{O}(1)]$ for mode number $k$,
we obtain
\begin{equation}
\log \beta_k \simeq \left[k + \mathcal{O}(1)\right] \sqrt{\frac{8 \ztip}{a}}\,,
\quad \ztip \ll a\,.
\label{eqn:beta_k_short}
\end{equation}
The mode is localized at distances $\rho \lesssim \sqrt{\ztip a}\,$.
The validity of this qualitative analysis is supported by the exact results for spherical particles.
For the $\nu =\,\parallel$ part,
the following compact formulas for the poles and residues are available \cite{Rendell1981spc, Aravind1982uop, Aravind1983teo}:
\begin{align}
    \beta_{k}^\parallel (\alpha) &= e^{(2k+3)\alpha} \,,
\label{eqn:chi_b_sphere_S}\\
		R_{k}^\parallel (\alpha) &= 4 (k+1)(k+2) a^3 \, \sinh^3 \alpha \,,	\label{eqn:chi_R_sphere_S}
\end{align}
where
\begin{equation}
  \alpha = \arccosh \left(\frac{\ztip}{a} + 1\right)\,.
  \label{eqn:alpha}
\end{equation}
It is easy to check that Eqs.~\eqref{eqn:beta_k_short} and \eqref{eqn:chi_b_sphere_S} agree in the limit of small $\alpha$.
(Dependence of $\beta_{k}^\bot$ on $\alpha$ is qualitatively similar;
however, the residues scale as  $R_{k}^\bot \sim k a^3 \alpha^2$ at small $\alpha$,
see Appendix~\ref{sec:spherical_limit}.)
It is reasonable to think that the behavior of $\beta_k^\nu(\alpha)$ at $\alpha \ll 1$ should be common for any shape ending in a rounded tip.
As long as the modes are localized at $\rho \ll a$, they should be affected weakly by the rest of the probe.
This hypothesis is supported by numerical calculations presented later in this article.

Consider next the long-distance limit $\ztip \gg L$.
In this case the probe-sample interaction can be analyzed using the multipole expansion.
For the lowest resonance $k = 0$
it is sufficient to include only the dipole term.
The dipole moment of the probe is given by $p^\nu = \chi_0^\nu E^\nu_{\mathrm{tot}}$,
where $E^\nu_{\mathrm{tot}} = E^\nu_{\mathrm{ext}} + E^\nu_{\mathrm{ind}}$ is the total field at the probe position
and $E^\nu_{\mathrm{ind}}$ is the field induced by the image dipole.
In particular, $E^\bot_{\mathrm{ind}} = \beta p^z / 4 \ztip^3$ and 
$E^\parallel_{\mathrm{ind}} = \beta p^\parallel / 8 \ztip^3$.
Solving these equation for $p^\nu$ and casting the result for $\chi^\nu = p^\nu / E_{\mathrm{ext}}^\nu$ in the form~\eqref{eqn:chi_eff_general_form},
we get
\begin{subequations}
\begin{align}
\beta_0^\bot &\simeq 4 \ztip^3 / \chi_0^\bot\,,
&
R_0^\bot &\simeq 4 \ztip^3\,,
\label{eqn:beta_0_dip_perp}\\
\beta_0^\parallel &\simeq 8 \ztip^3 / \chi_0^\parallel\,,
&
R_0^\parallel &\simeq 8 \ztip^3\,.
&\label{eqn:beta_0_dip_parallel}
\end{align}
\end{subequations}
For the sphere $\chi_0^\nu = a^3$, so that the last pair of equations agrees with the exact result~\eqref{eqn:chi_b_sphere_S} and \eqref{eqn:chi_R_sphere_S}.
The $k > 0$ resonances are dominated by higher-order multipoles.
The principal dependence of the poles and residues on $\alpha$ is expected to be the same as for the sphere, i.e.,
\begin{equation}
  \beta_k^\nu \sim \ztip^{2k + 3}\,,
  \ R_k^\nu \sim4 g^\nu (k+1)(k+2)\ztip^3
  \text{ if } \ztip \gg L\,,
  \label{eqn:far}
\end{equation}
where $g^{\bot} = 1/2$ and $g^{\parallel} = 1 $. The forms for $R^\nu_k$ are verified numerically in a later section.
Equations~\eqref{eqn:beta_0_dip_perp}--\eqref{eqn:far} imply that in the large $\ztip$ limit the sum rule~\eqref{eqn:chi_sum_rule} is saturated by the $k = 0$ mode alone.

The case of a $q$-dependent reflectivity can be treated similarly.
Thus, for $k = 0$ one finds~\cite{Aizpurua2008sei}
\begin{align}
\chi^{\nu} (\omega, \ztip) &= \frac{\chi_0^{\nu}}{1 - \chi_0^{\nu} g^{\nu}(\omega, \ztip)}\,,
\label{eqn:chi_eff_pdp_q}\\
g^{\nu}(\omega, \ztip) &= c^{\nu} \zfint r_{\text{P}}(q, \omega) e^{-2q\ztip} q^{2} d q \,,
\label{eqn:g}
\end{align}
where $c^{\bot} = 1$ and $c^{\parallel} = 1 / 2$.
Note that the integral in Eq.~\eqref{eqn:g} is dominated by the in-plane momenta $q \sim 1 / \ztip$,
which we assume to be well outside the light cone, $q \gg \omega / c$.
At $\ztip > c / \omega$ this condition no longer holds and one has to include retardation effects, see Sec.~\ref{sec:shape}.

In summary, in this section we presented arguments that the limiting case formulas~\eqref{eqn:beta_k_short} and \eqref{eqn:beta_0_dip_perp}--\eqref{eqn:g} apply to perfectly conducting probes of arbitrary shapes.
For the sphere $L = a$ and for probes of modest aspect ratio $L \gtrsim a$,
these formulas match by the order of magnitude at $\ztip \sim a$.
However, for strongly elongated probes $L \gg a$ an additional intermediate regime $a \ll \ztip \ll L$ exists which requires further study.
The simplest example of such a shape is the prolate spheroid and
we discuss it in the next section.

\section{Intermediate distances. Spheroidal probe}
\label{sec:spheroidal}

\begin{figure}[b]
	\begin{center}
		\includegraphics[width=\imagewidth]{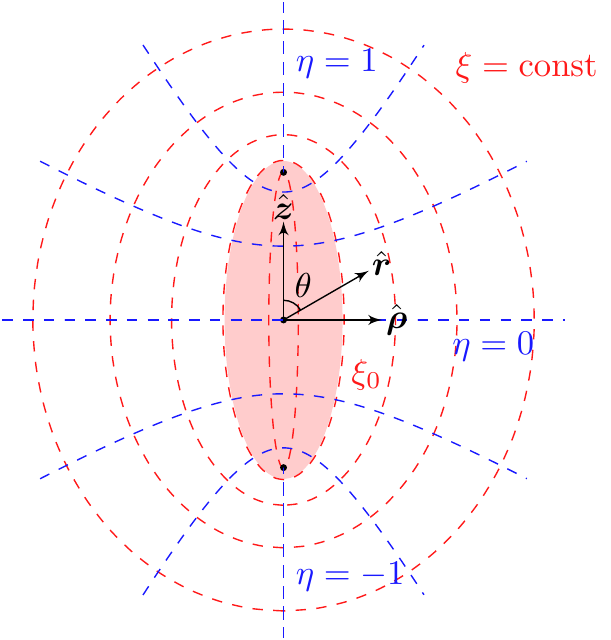}
	\end{center}
	\caption{The prolate spheroidal coordinate system.
	Contours of constant $\xi$ ($\eta$) are confocal spheroids (hyperboloids).
	The unit vector $\bhat{r}$ and polar angle $\theta$ of spherical polar coordinates and unit vectors $\bhat{\rho}$ and $\bhat{z}$ of cylindrical polar coordinates are also shown for reference.
	}
	\label{fig:pro_sph_coord}
\end{figure}

Unlike the problem of a sphere, that of a spheroidal probe
cannot be solved analytically.
However, we can take advantage of the separation of variables in prolate spheroidal coordinates (Fig.~\ref{fig:pro_sph_coord}),
which enables a more efficient numerical solution.~\cite{Bobbert1987psp}
In this coordinate system the spheroid is a surface of constant $\xi = L / F \equiv \xi_{0}$.
The focal length $F$, the major semi-axis $L$, the minor semi-axis $W$,
and the curvature radius $a$ of the spheroid apex are related by
\begin{equation}
	F = \sqrt{L^{2} - W^{2}}
	\,,\quad
	a = {W^2} /\, {L} \,.
  \label{eqn:spheroid_lengths}
\end{equation}
This implies $\xi_0 = [1 - (a / L)]^{-1/2}$.
We assume that the major axis of the spheroid is along the $z$-axis.
If the distance between the spheroid and the sample is $\ztip$,
the sample surface is at $z = -L - \ztip$.

\begin{figure*}[t]
\begin{center}
	\includegraphics[width=\imagewidth]{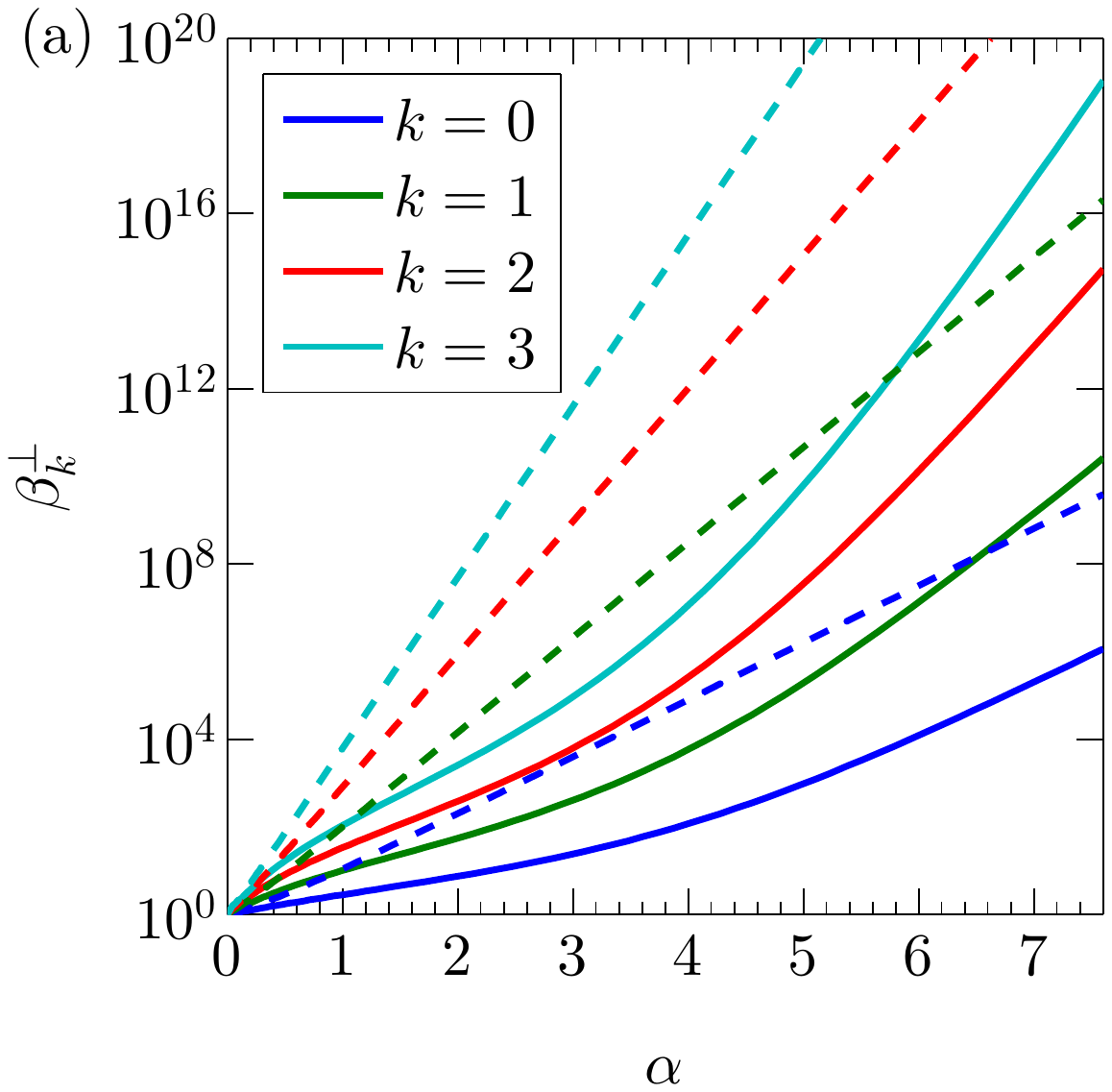}
	\includegraphics[width=\imagewidth]{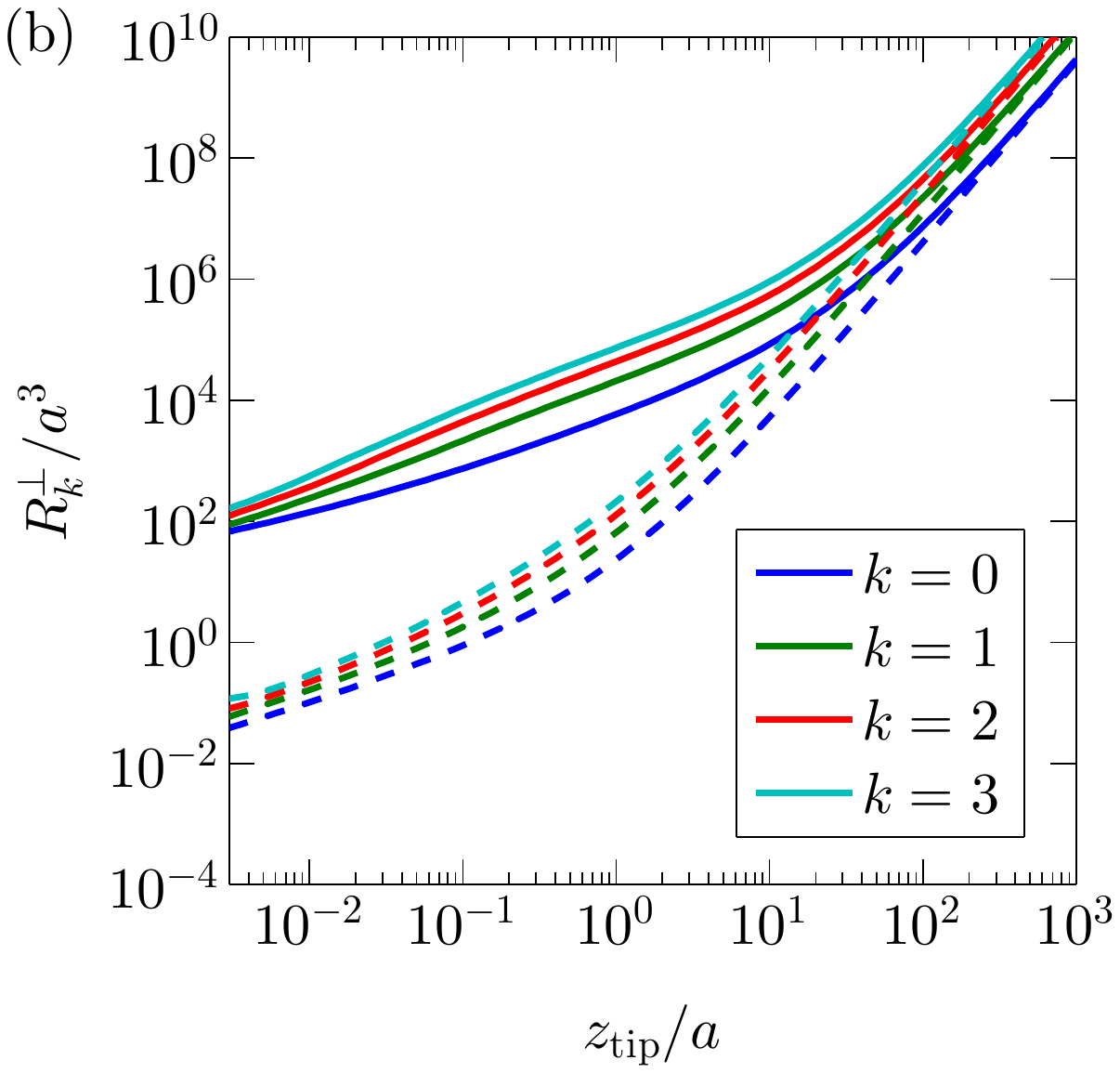}
\\
   \includegraphics[width=\imagewidth]{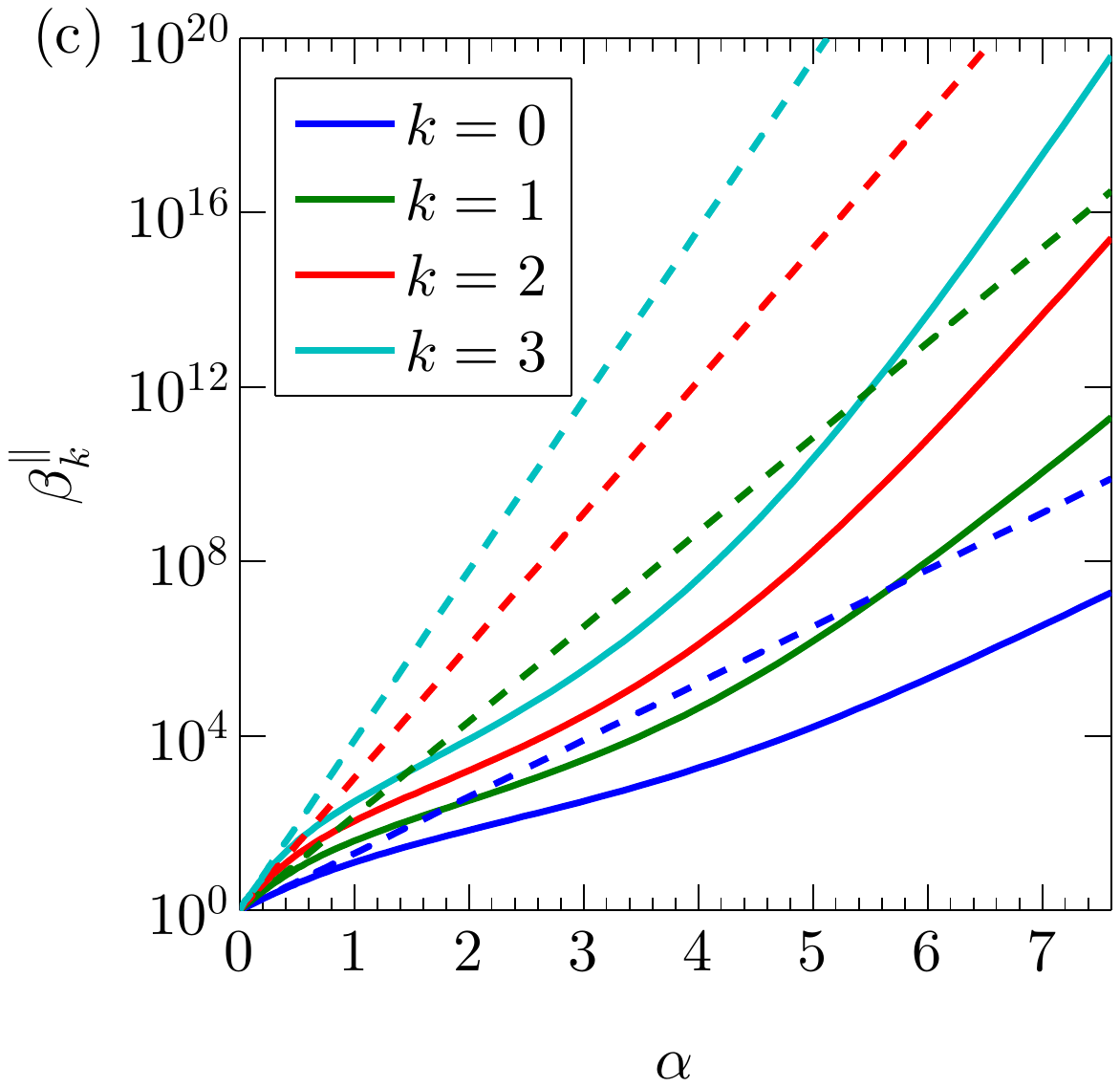}
    \includegraphics[width=\imagewidth]{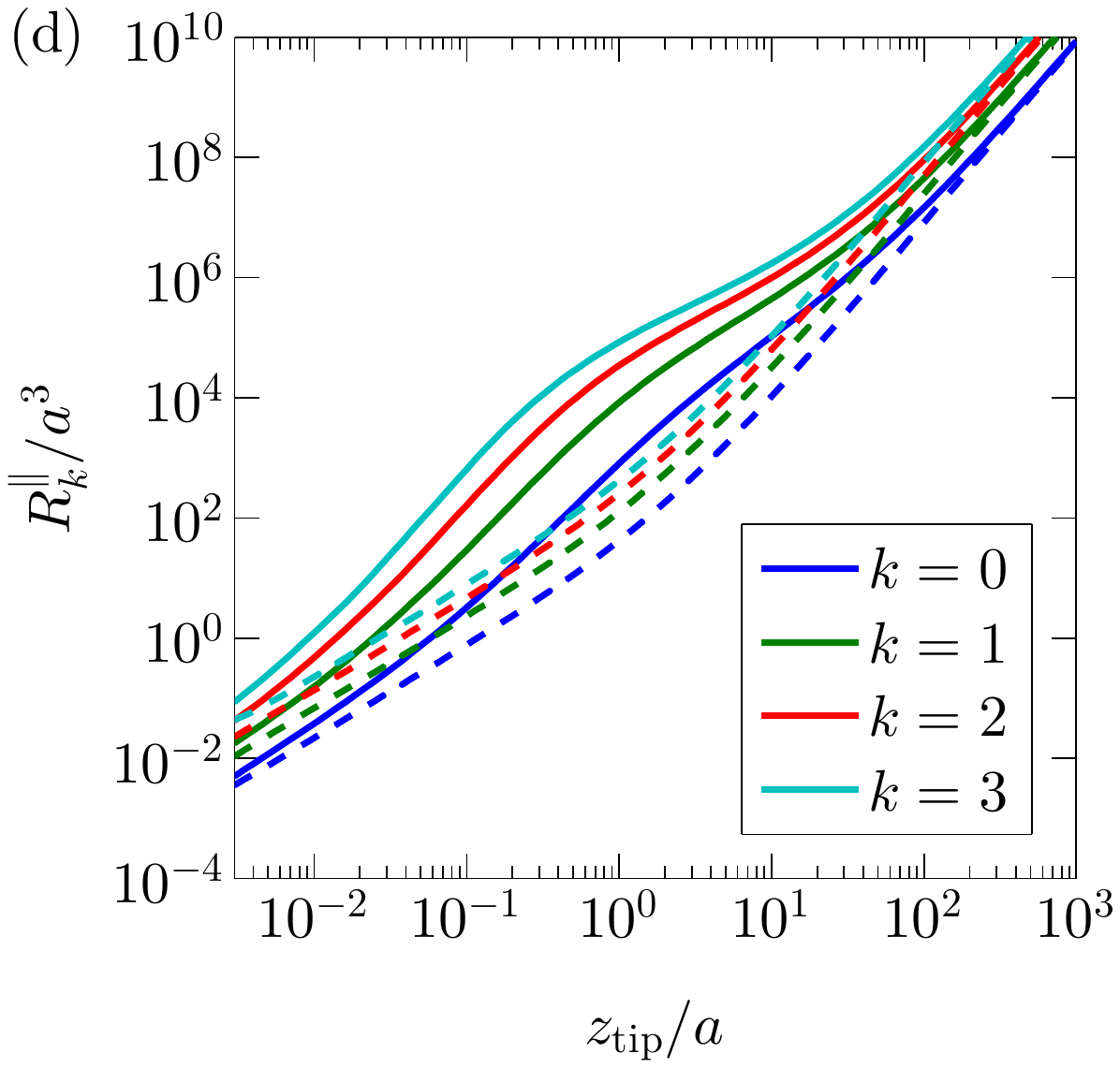}
\end{center}
\caption{(Color online) (a) The first four poles $\beta_{k}^\nu$ of the polarizability $\chi^\nu$ for perfectly conducting spheroids.
The probe-sample distance is parameterized by $\alpha = \arccosh(1 + {\ztip}/{a})$ [Eq.~\eqref{eqn:alpha}].
The solid lines are for a strongly elongated spheroid $L = 25a$, the dashed lines are for a nearly spherical one $L = 1.01a$.
The external field is in the $z$-direction, $\nu = \,\bot$.
(b) The corresponding residues $R_k^\bot$ divided by $a^3$. 
Poles for different shapes converge at small $\ztip$, while residues converge at large $\ztip$.
(c), (d) Similar plots for the external field in the $x$-$y$ plane, $\nu = \,\parallel$.
\label{fig:pz_beta_resdu}
}
\end{figure*}

We consider the quasistatic limit where the scalar potential has the harmonic time dependence $\propto e^{-i\omega t}$.
Its spatial part must obey the Laplace equation
in the domain outside both the spheroid and the sample.
Therefore, it can be expanded into spheroidal harmonics,
which are products of the generalized Legendre functions of the first and second kind $P_{l}^{m}(x)$ and $Q_{l}^{m}(x)$.
Here $m = 0, \pm 1, \pm 2, \ldots$ is the $z$-axis angular momentum and $l$ must be greater or equal to $|m|$.
As shown in Appendix~\ref{sec:spheroidal_model},
the expansion coefficients $A^{m}{}_{l}$ can be related to the charge distribution on the spheroid.
For example, $A^{m}{}_{0}$ is proportional to the total oscillating charge of the spheroid $\propto e^{-i\omega t}$.
For a passive probe, $A^{m}{}_{0} = 0$.
The $l = 1$ terms determine the components of the dipole moment induced on the probe:
\begin{equation}
	p_{z} = -\frac{1}{3} F^3 A^{0}{}_{1}
	\,,\quad
	p_{x} - i p_{y} = \frac{2}{3} F^3 A^{1}{}_{1} \,.
\label{eqn:p_from_A}
\end{equation}
For each $m$ the set of coefficients $A^{m}{}_{l}$
satisfies the infinite-order system of linear equations
\begin{equation}
	\ofsum{l\prm} \left( \Lambda^{m}_{ll\prm} - H_{ll\prm}
	 \right)
	A^{m}{}_{l\prm} = b^{m}{}_{l} \,,
	\label{eqn:sphrd_char_eqns}
\end{equation}
where $\Lambda^{m}_{ll\prm}$ and $H_{ll\prm}$ are defined by Eqs.~\eqref{eqn:Lambda_def} and \eqref{eqn:H_def} below.
According to Eq.~\eqref{eqn:p_from_A}, to find $\bm{p}$ we need to consider only $m=0$ and $m = 1$.
The requisite coefficients $b^{m}{}_{l}$ on the right-hand side of Eq.~\eqref{eqn:sphrd_char_eqns} are given by
\begin{align}
b^{0}{}_{1} &= -\frac43 E^z\,,
\label{eqn:b_0}\\
b^{1}{}_{1} &= \frac43 (E_x - i E_y)\,,
\quad
b^{-1}{}_{1} = \frac13 (E_x + i E_y)\,.
\label{eqn:b_1}
\end{align}
If the external field
$\bm{E}\tExt = E_{x} \bhat{x} + E_{y} \bhat{y} + E_{z} \bhat{z}$ is uniform,
all other $b^{m}{}_{l}$ vanish.
Once we solve the system~\eqref{eqn:sphrd_char_eqns} for $m = 0$,
we can find the transverse polarizability from
\begin{equation}
 \chi^\bot = \frac{p^z}{E^z} = \frac49 F^3\, \frac{A^{0}{}_{1}}{b^{0}{}_{1}}\,.
\label{eqn:chi_perp}
\end{equation}
In turn, the solution for $m = 1$ would give us $A^{1}{}_{1}$ and
\begin{equation}
 \chi^\parallel = \frac{p_x - i p_y}{E_x - i E_y} = \frac89 F^3\, \frac{A^{1}{}_{1}}{b^{1}{}_{1}}\,.
\label{eqn:chi_parallel}
\end{equation}
Equation~\eqref{eqn:sphrd_char_eqns} we wish to solve can be cast in a matrix form
\begin{equation}
\left(\bm{\Lambda}^m - \mathbf{H} \right) \mathbf{A}^m = \mathbf{b}^m\,.
\label{eqn:sphrd_char_eqns_matrix}
\end{equation}
Matrix $\bm{\Lambda}^m$ is diagonal, $\Lambda^{m}{}_{ll\prm} = \Lambda^{m}_{l} \delta_{l l\prm}$, where
\begin{equation}
	\Lambda^{m}_{l}
	= \frac{(-1)^{m}}{2 l + 1}
	\frac{4}{\epsilon\tsub{tip} - 1}
	\left[ {\epsilon\tsub{tip}
		\frac{Q_{l}^{m}(\xi_0)}{P_{l}^{m}(\xi_0)} - 
		\frac{\frac{d}{d\xi_0} Q_{l}^{m}(\xi_0)}
		{\frac{d}{d\xi_0} P_{l}^{m}(\xi_0)}}
	\right]
	\label{eqn:Lambda_def}
\end{equation}
and $\eps{\text{tip}}$ is again the dielectric constant of the spheroid.
If the probe is made of an ideal conductor,
$\eps{\text{tip}} \to \infty$,
then Eq.~\eqref{eqn:Lambda_def} simplifies to
\begin{equation}
	\Lambda^{m}_{l}
	= (-1)^{m}\frac{4}{2 l + 1}
		\frac{Q_{l}^{m}(\xi_0)}{P_{l}^{m}(\xi_0)}\,.
\label{eqn:Lambda_def_metal}
\end{equation}
All these $\Lambda^{m}_{l}$ are actually positive numbers because the factor $(-1)^m$ is compensated by the same factor in the definition of $Q_{l}^{m}(\xi_0)$.
The behavior of $\Lambda^{m}_{l}$ at large $l$ is approximately exponential,
as can be deduced from the asymptotic formula
\begin{equation}
(-1)^m \frac{Q_{l}^{m}(\xi_0)}{P_{l}^{m}(\xi_0)}
\simeq \pi e^{-(2l + 1) \alpha_0}\,,
\quad \alpha_0 \equiv \arccosh \xi_0\,.
\label{eqn:PQ_asym}
\end{equation}
In Sec.~\ref{sec:result} we also consider the case where $\eps{\text{tip}}$ is a finite positive number.
In this case the decay of $\Lambda^{m}_{l}$ at large $l$ is also exponential but with a different factor in front.

If the dielectric function of the probe is real and negative, then $\Lambda_l^m$ can be negative, too.
It can also be zero,
which corresponds to a plasmon (or phonon) polariton resonance of the probe.
The resonances occur at discrete $\eps{\text{tip}}$ values
\begin{equation}
\eps{\text{tip}, l}^m = \frac{\frac{d}{d \xi_0} \ln Q_l^m(\xi_0)}
{\frac{d}{d \xi_0} \ln P_l^m(\xi_0)}\,,
\quad l = 1, 2,\ldots,
\end{equation}
see, e.g., Refs.~\onlinecite{Caldwell2014scv, Sun2015hoh} (where, in fact, a more general case of anisotropic $\eps{\text{tip}}$ is treated).
For each $m$, the sequence $\eps{\text{tip}, l}^m$ asymptotically approaches $-1$ as $l \to \infty$.
The smallest, i.e., the most negative value in each sequence is the starting one.
It can be alternatively written as 
\begin{equation}
\eps{\text{tip}, 1}^m = 1 -\frac{1}{L^\nu}\,,
\end{equation}
where $\nu =\,\bot$ for $m = 0$, $\nu =\,\parallel$ for $m = 1$,
and $L^\nu$ are the depolarization factors of the spheroid~\cite{Bohren2004asl}
\begin{align}
L^\bot &= (\xi_0^2 - 1)\left[\frac{1}{2}\,\xi_0
\ln \left(\frac{\xi_0 + 1}{\xi_0 - 1}\right) - 1\right]\,,
\label{eqn:L_bot}\\
L^\parallel &= \frac{1 - L^\perp}{2}\,.
\label{eqn:L_parallel}
\end{align}
For prolate spheroids, these depolarization factors obey the inequalities $0 < L^\bot < L^\parallel < \frac12$,
and so
$\eps{\text{tip}, 1}^0 < \eps{\text{tip}, 1}^1$.
For example, if $L = 25 a$, which we use in our calculations below, then $\xi_0 = \sqrt{25 / 24}$,
$\eps{\text{tip}, 1}^0 = -16.9$, and
$\eps{\text{tip}, 1}^1 = -1.11$.
If the probe is made out of platinum or iridium,
which are common materials for AFM tips,
its dielectric function can indeed be negative.
It is in principle possible to achieve plasmonic resonances in such probes somewhere
in the near-infrared or visible spectral range.
On the other hand, at mid-infrared frequencies,
for which we do calculations in this paper,
the dielectric functions of such metals
are in the range of hundreds or thousands.
Such probes are very far from any of the plasmonic resonances and
the approximation of the ideal conductor,
Eq.~\eqref{eqn:Lambda_def_metal}, can be safely used.
We do so in the remainder of this Section.

The elements of matrix $\mathbf{H}$ in Eq.~\eqref{eqn:sphrd_char_eqns_matrix} are given by
\begin{equation}
	{H}_{ll\prm} \equiv 2 \pi \zfint
	r_\mathrm{P}(q, \omega) I_{l + \frac12} (q F)
	I_{l\prm + \frac12} (q F)
	e^{-2 q \zp} \frac{d q}{q}
	\label{eqn:H_def}
\end{equation}
where $I_{\nu}(z)$ are the modified Bessel functions of the first kind and
\begin{equation}
\zp \equiv \ztip + L\,.
\label{eqn:zp}
\end{equation}
As mentioned in Sec.~\ref{sec:Introduction}, the reflectivity $r_\mathrm{P}(q, \omega)$ may have strong peaks at the dispersion curves $\omega(q)$ of the surface polaritons of the sample.
In practice, $r_\mathrm{P}(q, \omega)$ is always finite,
so that the integrand in Eq.~\eqref{eqn:H_def} is well-behaved and exponentially decreasing.
A fast method of computing ${H}_{ll\prm}$ numerically is explained
in Supplemental Material.
In the remainder of this section we will assume that $r_\mathrm{P}(q, \omega)$ is $q$-independent.
We will show that the polarizabilies of the spheroidal probe are meromorphic functions as stated in Sec.~\ref{sec:Introduction}.
We will also present our analytical and numerical results concerning the behavior of their poles and residues.

If $r_\mathrm{P}(q, \omega) = \beta = \mathrm{const}$,
then matrix $\mathbf{H}$ factorizes
$\mathbf{H} = \beta\, \bar{\mathbf{H}}$ and
Eq.~\eqref{eqn:sphrd_char_eqns_matrix} becomes
\begin{equation}
\left(\bm{\Lambda}^m -  \beta\, \bar{\mathbf{H}} \right) \mathbf{A}^m
 = \mathbf{b}^m\,.
	\label{eqn:sphrd_char_eqns_beta}
\end{equation}
A particular case of this equation for $\ztip = 0$ was previously derived in Ref.~\onlinecite{Bobbert1987psp}.
In general, Eq.~\eqref{eqn:sphrd_char_eqns_beta} implies that $\mathbf{A}^m$ as a function of $\beta$ has poles $\beta_k^\nu$
that are the solutions of the eigenvalue problem
\begin{equation}
\left(\bm{\Lambda}^m - \beta_k^\nu\, \bar{\mathbf{H}}\right) \mathbf{u}_k
 = 0\,.
\label{eqn:u_k}
\end{equation}
The substitution $\mathbf{u}_k = (\bm{\Lambda}^m)^{-1/2} \mathbf{v}_k$ transforms it to
\begin{equation}
\mathbf{v}_k = \beta_k^\nu\, \mathbf{M} \mathbf{v}_k\,,
\quad \mathbf{M} = (\bm{\Lambda}^m)^{-1/2} \bar{\mathbf{H}}\, (\bm{\Lambda}^m)^{-1/2}\,.
\label{eqn:v_k}
\end{equation}
Since all $\Lambda_l^m$ are assumed to be positive,
matrix $\mathbf{M}$ is real and symmetric, and so
its eigenvalues $(\beta_k^\nu)^{-1}$ are real
and its eigenvectors $\mathbf{v}_k$ can be chosen to be orthonormal.
Assuming $\mathbf{v}_k$ form a complete basis,
the solution $\mathbf{A}^m$ of Eq.~\eqref{eqn:sphrd_char_eqns_beta} can be sought as a linear combination of the corresponding $\mathbf{u}_k$.
Taking into account Eqs.~\eqref{eqn:chi_perp} and\eqref{eqn:chi_parallel},
we arrive at Eq.~\eqref{eqn:chi_eff_general_form} with
\begin{equation}
\frac{R_k^\nu}{\beta_k^\nu} = \chi_0^\nu \left|\left(\mathbf{v}_k\right)_0\right|^2\,,
\quad
\chi_0^\nu = \frac49 \frac{m + 1}{\Lambda^{m}_{1}} F^3\,,
\label{eqn:R_k}
\end{equation}
where, once again, $m = 0$ for $\nu =\,\bot$, $m = 1$ for $\nu = \,\parallel$,
and $\left(\mathbf{v}_k\right)_0$ is the first component of vector $\mathbf{v}_k$.
The completeness of the basis entails $\sum_{k} \left|\left(\mathbf{v}_k\right)_0\right|^2 = 1$,
leading to the sum rule~\eqref{eqn:chi_sum_rule}.
The explicit formulas for $\chi_0^\nu$ that follow from Eqs.~\eqref{eqn:Lambda_def_metal}, \eqref{eqn:R_k},
\eqref{eqn:L_bot}, and \eqref{eqn:L_parallel} are
\begin{subequations}
\begin{align}
\chi_0^\bot &= \frac{L^3}{3 \xi_0^3} \left[
\frac12 \ln \left(\frac{\xi_0 + 1}{\xi_0 - 1}\right) - \frac{1}{\xi_0}
\right]^{-1}
= \frac{V}{4\pi L^\bot}\,,
\label{eqn:chi_0_perp}\\
\chi_0^\parallel &= \frac{2 L^3}{3 \xi_0^3} \left[
\frac{\xi_0}{\xi_0^2 - 1} - \frac12 \ln \left(\frac{\xi_0 + 1}{\xi_0 - 1}\right) 
\right]^{-1}
= \frac{V}{4\pi L^\parallel}\,,
\label{eqn:chi_0_parallel}
\end{align}
\end{subequations}
where $V = (4\pi / 3) L^2 a$ is the volume of the spheroid.
These formulas should be familiar from classical electrostatics or from the theory of light scattering by small particles.~\cite{Bohren2004asl}
For strongly elongated spheroid $L \gg a$, $\xi_0 \simeq 1$, they yield
\begin{subequations}
\begin{align}
\chi_0^\bot &\simeq \frac23 \frac{L^3}{\ln (4 L / a)}\,,
\label{eqn:chi_0_perp_asym}\\
\chi_0^\parallel &\simeq \frac23 L^2 a\,.
\label{eqn:chi_0_parallel_asym}
\end{align}
\end{subequations}

In Sec.~\ref{sec:Introduction} we stated that the sequence $\beta_k^\nu$ may not have accumulation points.
For the present case of a spheroidal probe this can be proven directly from the properties of matrix 
$\mathbf{M}$.
The first step is to
show that the matrix elements of $\bar{\mathbf{H}}$ obey the asymptotic bound
\begin{equation}
\ln \bar{H}_{ll\prm} < -(l + l\prm + 1) \arccosh \left(\frac{\zp}{F}\right)
\label{eqn:bar_H_asym}
\end{equation}
at large $l$ and $l\prm$.
This can be established using the saddle-point integration in Eq.~\eqref{eqn:H_def}.
Together with Eqs.~\eqref{eqn:Lambda_def_metal} and \eqref{eqn:PQ_asym},
this bound ensures that at $\ztip > 0$ the high-order matrix elements of $\mathbf{M}$ decay exponentially,
\[
\ln M_{ll\prm} < -(l + l\prm + 1)  \left[
\arccosh \left(\cosh \alpha + \frac{\ztip}{F}\right) - \alpha
\right].
\]
Here $\alpha$ [Eq.~\eqref{eqn:alpha}] parametrizes the probe-sample distance $\ztip$.
Hence, the double series $\sum_{ll\prm} M_{ll\prm}^2 = \text{tr}\, \mathbf{M}^2$ is convergent.
Considering the identity
\begin{equation}
\sum_{k = 0}^\infty (\beta_k^\nu)^{-2}
 = \text{tr}\, \mathbf{M}^2 < \infty
\label{eqn:beta-2_sum}
\end{equation}
we see that the accumulation points are ruled out.
On the contrary, $\text{tr}\, \mathbf{M}^2$ diverges at $\ztip = 0$ and one accumulation point does exist: $\beta = 1$.
For the sphere this can be found directly from Eq.~\eqref{eqn:chi_R_sphere_S} by setting $\alpha = 0$.

In the spherical limit $\xi_0 \to \infty$
an analytical solution of our equations exists
although it is not obvious.
We deduced the form of this solution from the method of images, see Appendix~\ref{sec:spherical_limit}.
At finite $\xi_0$ we resorted to solving the problem numerically.
As already mentioned,
due to an exponential growth of $\beta_{k}^\nu$ with $k$,
only a first few of such poles are usually needed for evaluating the polarizabilities in question $\chi^\nu$.
To compute such $\beta_{k}^\nu$ and the corresponding $R_{k}^\nu$
we used the following procedure.
Given $L / a$ and $\alpha$, we would generate an $N \times N$ matrix
made of the first $N$ rows and columns of the full infinite matrix $\mathbf{M}$.
We would diagonalize this finite-size matrix by standard library routines.~\cite{MATLAB}
The obtained eigenvalues approximate the first $N$ poles $\beta_{k}^\nu$.
We would gradually increase the matrix size until the poles we are interested in would show no variation as a function of $N$ within the desired accuracy.
The larger $L / a$ and the smaller $\alpha$, the higher $N$ was needed.
We found this procedure workable as long as $N$ did not exceed about $500$.
As a rule, the higher eigenvalues of larger matrices would either fail to reach the accuracy or would show an $\alpha$-dependence inconsistent with physical principles.
This behavior stems most likely from roundoff errors.
In principle, one can combat them by utilizing higher-precision arithmetic but we did not pursue this route.
For $L = 25a$ the computation of the first nine poles with at least two-digit accuracy was possible for $\alpha > 0.08$, i.e., $\ztip > 0.003 a$.
The residues $R_k^\nu$ were obtained from the eigenvectors of the truncated matrix  $\mathbf{M}$ using Eqs.~\eqref{eqn:R_k}--\eqref{eqn:chi_0_parallel}.
In the interval $0 < \alpha < 0.08$ we used the linear interpolation between $\beta_k^\nu(\alpha = 0.08)$ and $\beta_k^\nu(\alpha = 0) = 1$.

The results of these calculations are presented in Fig.~\ref{fig:pz_beta_resdu}
for the first four modes, $k=0$ to $3$.
The solid lines in panels (a) and (c)
show $\beta_{k}^\bot$ and $\beta_{k}^\parallel$, respectively,
as a function of $\alpha$.
The corresponding quantities for a sphere are shown by the dashed lines.
The residues $R_{k}^\nu /a ^3$ are plotted in panels (b) and (d).
The first nine pole-residue pairs of the spheroid for $\nu = \,\bot$ have also been fitted with an error of 5\% or smaller to a combination of elementary functions in the range $0.003a < \ztip < 10 a$.
The fitting formulas and their coefficients are cataloged in Table~\ref{tab:table}.
The residue $R_8^\bot$ behaves differently from the others because it was constrained to satisfy the sum rule~\eqref{eqn:chi_sum_rule}.
Using these formulas one can find the response $\chi^\bot$ with negligible computational cost for any $\beta(\omega)$ as long as its value is not extremely large.
Note that although these results are for perfectly conducting spheroids $\epsilon\tsub{tip} = \infty$,
calculations for arbitrary finite $\epsilon\tsub{tip}$ can be done in the same way
except one has to use Eq.~\eqref{eqn:Lambda_def} instead of Eq.~\eqref{eqn:Lambda_def_metal}.

Let us now compare the obtained dependence of $\beta_k^\nu$ on $\ztip$ with the limiting asymptotic behavior predicted in Sec.~\ref{sec:limits}.
First, at $\ztip \ll a$,
the poles of the spheroid approach that of a sphere,
as expected,
see Fig.~\ref{fig:pz_beta_resdu}(a) and (c).
The other limit is $\ztip \gg L$,
where the point-dipole formulas~\eqref{eqn:beta_0_dip_perp}--\eqref{eqn:beta_0_dip_parallel} should apply.
In Fig.~\ref{fig:pz_beta_resdu} it is seen that the lowest eigenvalue of both shapes indeed have the correct behavior.
The intermediate regime $a \ll \ztip \ll L$ is the most nontrivial one.
We argue that in this regime function $\beta_0^\bot(\ztip)$ behaves as
\begin{equation}
\beta_0^\bot(\ztip) = c \ln (\ztip / a)\,,
\quad
a \ll \ztip \ll L\,,
\label{eqn:beta_0_middle}
\end{equation}
with some coefficient $c \sim 1$ independent of $L$.
To arrive at this formula we
first find bounds on $\beta_0^\bot$ using the following theorem.
Consider two perfectly conducting probes of different sizes.
If the surface of one probe can be inscribed into the other,
then the first probe must have a larger $\beta_0^\nu$.
This statement is physically natural because self-sustained oscillations around the smaller body require a larger compensation from the surface reflectivity, cf.~Sec.~\ref{sec:limits}.
It can also be proven mathematically from the variational principle.~\cite{Agranovich1999gme, Morse_book}
To place bounds on $\beta_0^\bot$ of the spheroid we can consider two other probes, a larger one and a smaller one.
We get
\begin{equation}
\beta_0^{\mathrm{cone},\nu}<\beta_0^{\nu}<\beta_0^{\mathrm{ss},\nu}\,,
\label{eqn:beta_0_middle_bounds}
\end{equation}
where $\beta_0^{\mathrm{cone},\nu}$ is the lowest pole of a cone with a vertex touching the sample and enveloping the spheroid;
$\beta_0^{\mathrm{ss},\nu}$ is the lowest pole of a spheroid of shorter length $L = \ztip$.
It can be shown~\cite{Fogler2015} that $\beta_0^{\mathrm{cone},\bot} \simeq (1/\pi) \ln (\ztip/a)$.
As for the smaller spheroid, the point-dipole formula should apply by order of magnitude, $\beta_0^{\mathrm{ss},\bot} \sim 6\ln (\ztip / a)$, cf.~Eqs.~\eqref{eqn:beta_0_dip_perp} and \eqref{eqn:chi_0_perp_asym}. 
Since the functional form of these bounds coincides with Eq.~\eqref{eqn:beta_0_middle} up to numerical coefficients,
we argue that $\beta_0^\bot(\ztip)$ should obey the same equation as well.
The graph shown in Fig.~\ref{fig:pz_beta_resdu}(a) is consistent with this prediction.
However, due to numerical limitations $L/\ztip$ and $\ztip/a$ could not be very large in our simulation and we could obtain only a crude estimate $1 < c < 3$ of the coefficient $c$.
The poles $\beta_k^\parallel$ of the in-plane polarizability,
which are plotted in Fig.~\ref{fig:pz_beta_resdu}(c) as a function of $\alpha$,
also show crossovers among three regimes (short, long, and intermediate distances) and can be understood in a similar way.

The behavior of the residues $R_{k}^\nu$
is more difficult to analyze.
At large distances $\ztip\gg L$, the residues of the spheroid approach those of the sphere [Eq.~\eqref{eqn:far}]. 
At small distances, where the poles behave as $\ln \beta_k^\nu \sim (2 k + 3) \alpha$, the polarizability is determined by a large number $\sim 1 / \alpha$ of terms in the pole-residue series.
The sum rule~\eqref{eqn:chi_sum_rule} implies that the sum of these dominant residues must be of the order of $\chi_0^\nu$ for each shape.
Indeed, the residues of the sphere, which have the form $R_k^\bot\propto k a^3 \alpha^2$ and $R_k^\parallel\propto (k+1)(k+2) a^3 \alpha^3$ (Appendix~\ref{sec:spherical_limit}), obey this requirement.
The residues of the spheroid are always larger than those of the sphere,
consistent with the higher $\chi_0^\nu$.
The intermediate-distance behavior of $R_{k}$ defies an obvious characterization.
It is intriguing that at small distances only the residues are affected by the aspect ratio of the probe, while at large distances only the poles are altered.

Information about the probe-sample coupling complementary to the properties of the poles and residues can be obtained by examining the potential distribution of the polariton modes in real space.
The examples for the $\bot$ modes are depicted in Fig.~\ref{fig:modes}(b).
The potential is strongly peaked near the tip of the spheroid,
demonstrating the localized nature of near-field coupling.
Note that the number of times the potential changes sign along $x$ is equal to $k$.

\section{Momentum-dependence of the probe-sample coupling}
\label{sec:weak-local}

\begin{figure*}[htb]
	\begin{center}
		\includegraphics[width=\imagewidth]{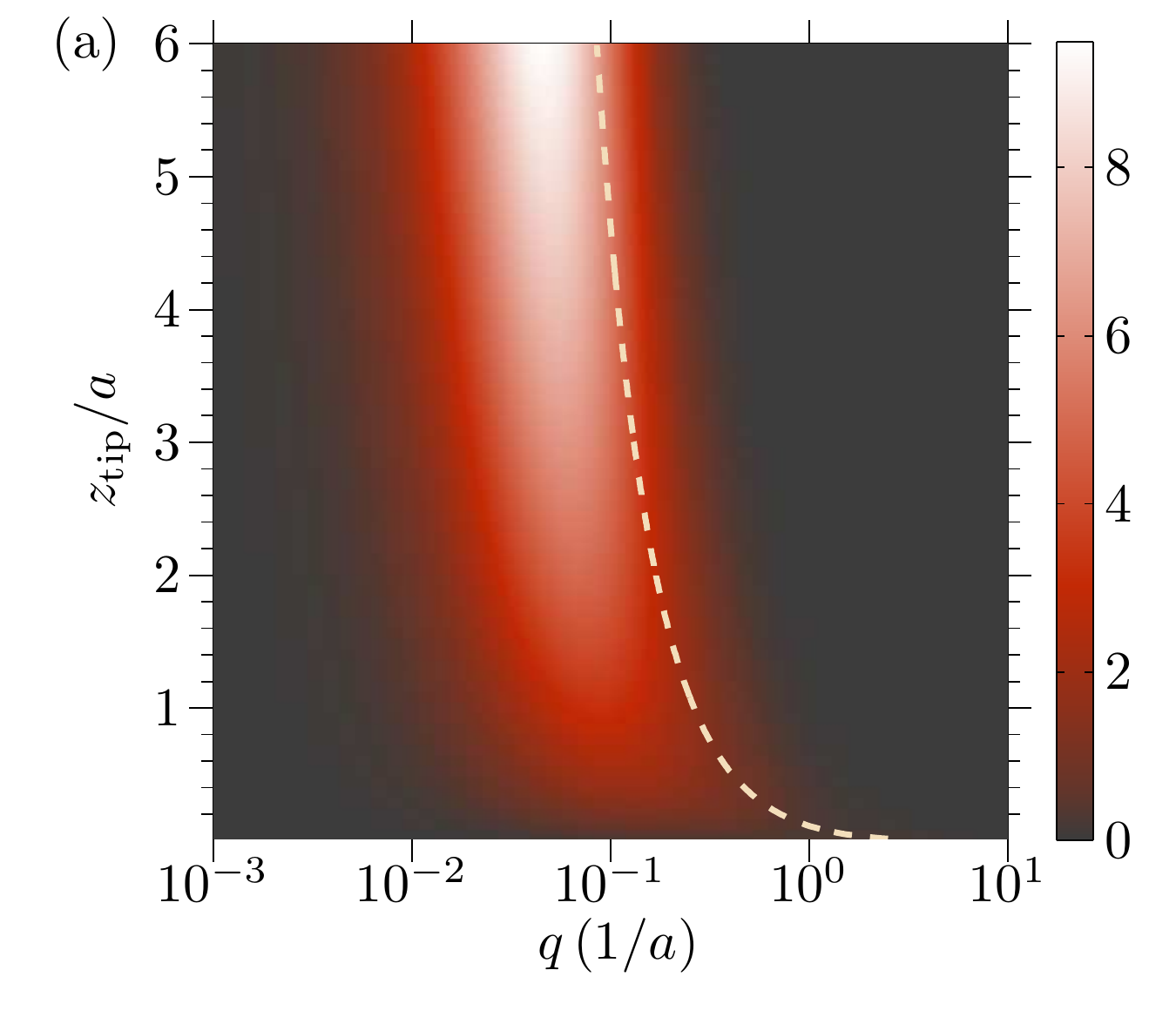}	
		\includegraphics[width=\imagewidth]{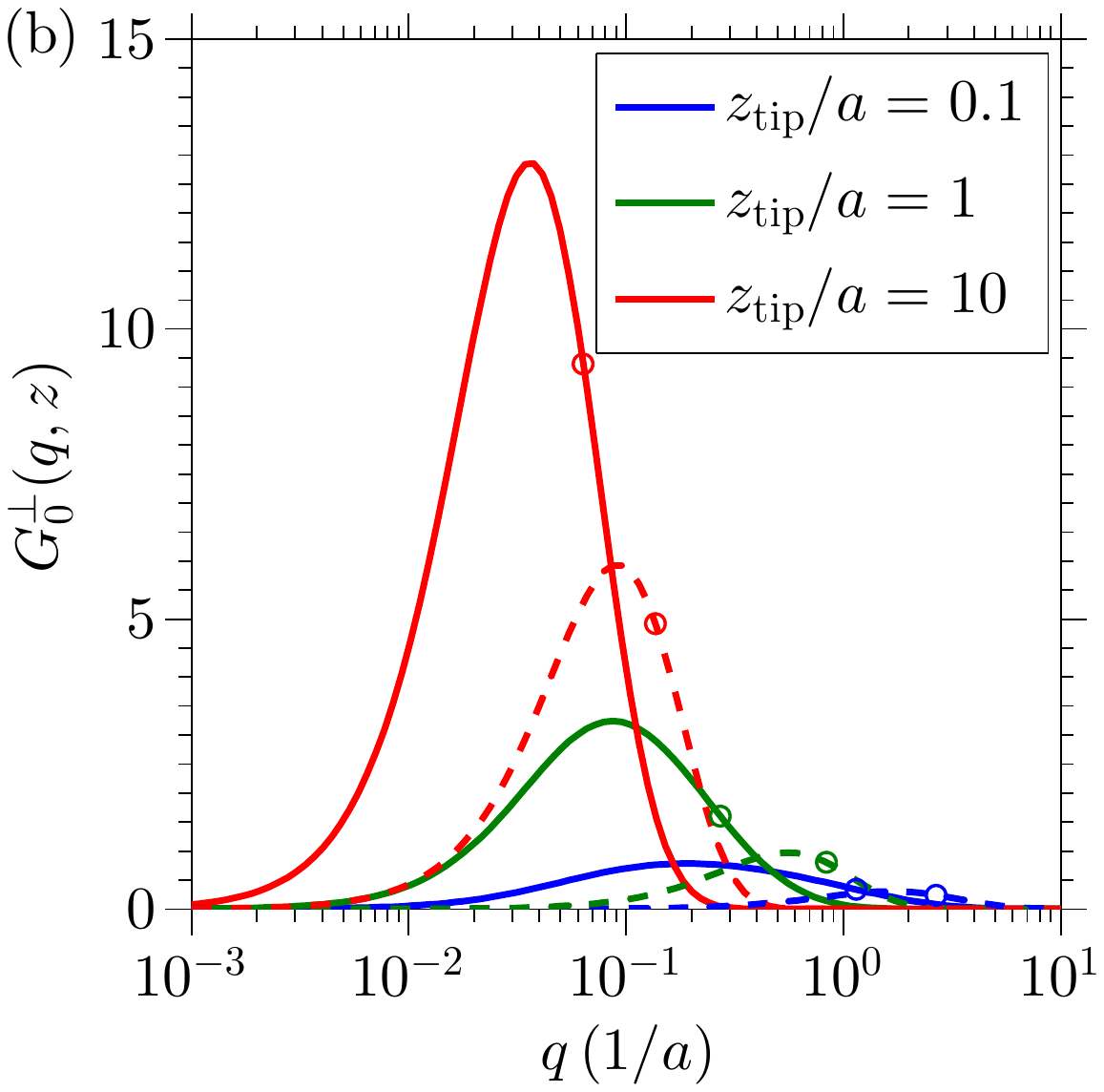}	
\hspace{-0.2in}
		\includegraphics[width=\imagewidth]{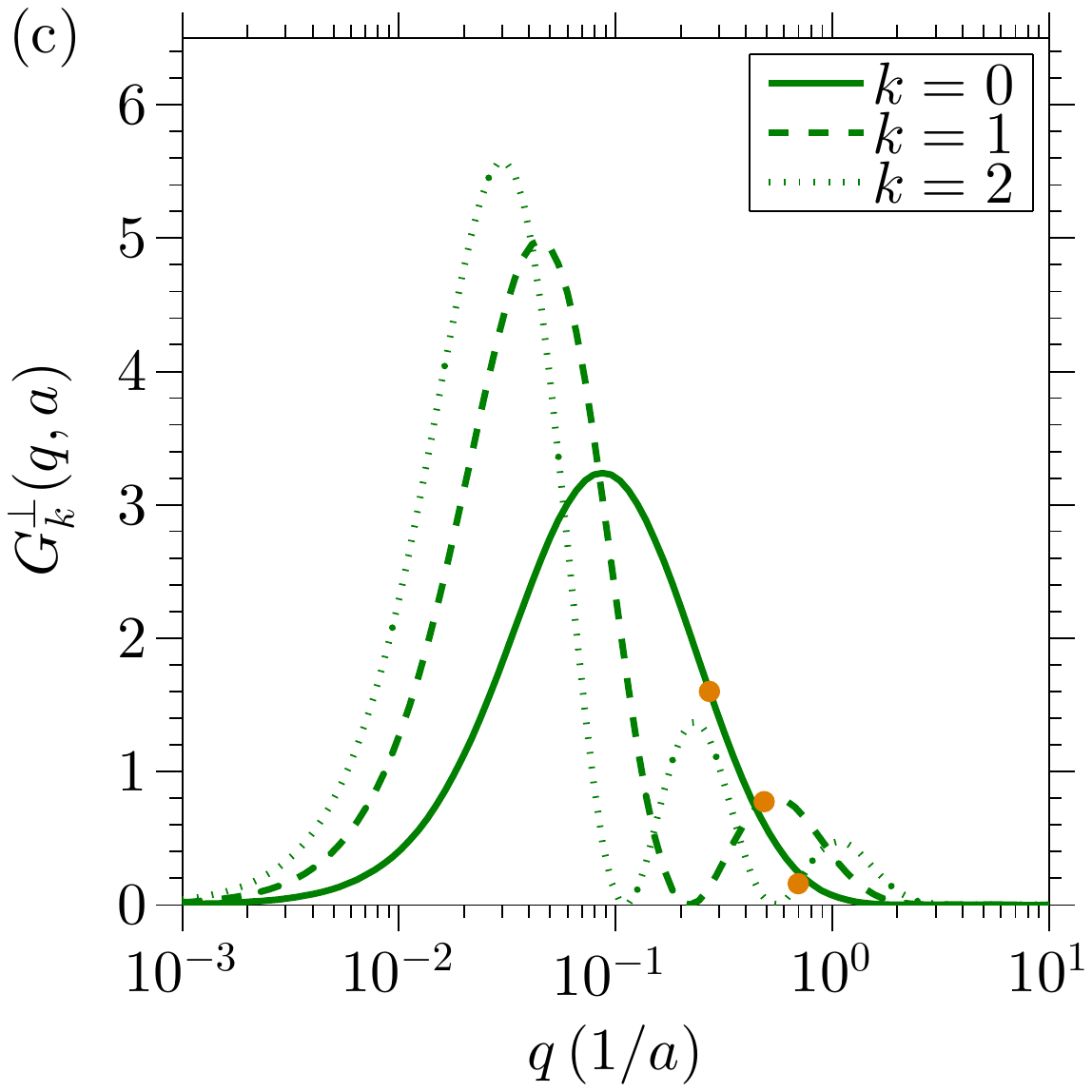}
	\end{center}
  \caption{(Color online) (a) The weight function $G_0^\bot(q, \ztip)$ for the spheroid with $L = 25a$ and $\epsilon_{\mathrm{tip}} = \infty$.
  The dashed line shows $\bar{q}_0^\perp(\ztip)$.
  (b) $G_0^\bot(q, \ztip)$ for several $\ztip$, with circles indicating $\bar{q}_0^\perp$. The solid lines are for the spheroid, the dashed lines are for the sphere. 
The spheroid is more sensitive to small $q$ compared to the sphere,
while both shapes are more sensitive to large $q$ as $\ztip$ decreases.
(c) The first three $G_k^\bot(q)$ for $\ztip = a$, with solid circles indicating $\bar{q}_k^\perp$ . The number of nodes in  $G_k^\bot(q)$ is equal to $k$, while $\bar{q}_k^\perp$ increases with $k$.
  The logarithmic scaling of the horizontal axes is used to show the small-$q$ structure more clearly.}
	\label{fig:G_q_z}
\end{figure*}

A simple physical picture of the s-SNOM that served as an important insight in the early days of the field and still remains popular today is the notion that the probe
couples predominantly to momenta $q \sim 1 / a$.
Accordingly, the s-SNOM signal is collected from a very small region of size $\sim a$ directly underneath the tip.
Modern applications of s-SNOM to two-dimensional and layered systems require going beyond this oversimplified picture because
the $q$-dependence of the reflectivity $r_\mathrm{P}(\omega, q)$ of such systems can be very sharp due to presence of dispersive collective modes (Sec.~\ref{sec:q-depend}).
Recall that for a momentum-independent reflectivity $\beta$ [Eq.~\eqref{eqn:beta_definition}], the poles and residues of the polariton eigenmodes are determined solely by the permittivity and geometry of the probe.
Unfortunately, for a $q$-dependent reflectivity such a clean separation of the probe and sample properties in the eigenproblem is not
possible.
While one can still define the eigenmodes
by suitably modifying Eq.~\eqref{eqn:chi_eff_general_form},
the corresponding poles and residues will be, in general, complicated functionals of $r_\mathrm{P}(\omega, q)$.
However, if the $q$-dependence of the reflectivity is weak,
it can be treated as a perturbation, and the sample-independent resonant modes are retained.
As we show in this Section, in this perturbative case one can precisely define the probe-sample coupling as a function of $q$ and the `dominant' momentum as a function of $\ztip$. 

Consider a small $q$-dependent correction to the reflectivity:
\begin{equation}
	r_\mathrm{P}(q) = \beta + \delta r_\mathrm{P}(q)\,.
	\label{eqn:rp_pert}
\end{equation}
The $k$th pole $\beta_k^\nu$ of the probe-sample eigenmodes is a functional of $r_\mathrm{P}$.
The key question is how
this pole is affected by the nonlocal correction to $r_\mathrm{P}$.
The answer can be written in terms of $-G_k^\nu(q, \ztip)$,
the first variational derivative of $\beta_k^\nu[r_\mathrm{P}(q)]$ with respect to $r_\mathrm{P}$:
\begin{equation}
	\delta\beta_k^\nu(\ztip) = - \zfint[\infty]
	G_k^\nu(q,\ztip) \delta r_\mathrm{P}(q) d q\,.
	\label{eqn:G_def}
\end{equation}
This is the desired relation to the leading order in $\delta r_\mathrm{P}$.
A few general properties of function $G_k^\nu$ at $q < 1 / \ztip$ can be established.
First, this function decays exponentially at large $q$:
\begin{equation}
G_k^\nu(q,\ztip) \sim e^{-2 q \ztip}.
	\label{eqn:G_large_q}
\end{equation}
This is so because the probe-sample interaction is mediated by multiple reflections of evanescent waves (Sec.~\ref{sec:Introduction})
and the shortest distance such waves have to travel is $2\ztip$.
Next, it is easy to see that $G_k^\nu$ is normalized:
\begin{equation}
\zfint[\infty] G_k^\nu(q, \ztip)d q = 1\,.
\label{eqn:G_constraint}
\end{equation}
Using a variation principle 
one can also show that for a perfectly conducting probe $G_k^\nu(q, \ztip)$ is nonnegative.
Therefore, functions $G_k^\nu(q, \ztip)$ can be considered \textit{weight} functions for the perturbation $\delta r_\mathrm{P}(q)$.
To put it another way, this set of functions quantifies the momentum dependence of the probe-sample coupling.
Below we show that the properties of these functions paint a much more nuanced physical picture than the naive idea that the coupling is maximized at a single momentum $q \sim 1 / a$.
However, if one insists on characterizing the entire distribution of relevant momenta by a single number,
the logical candidates are the average momenta
\begin{equation}
\bar{q}_k^\nu = \zfint[\infty] G_k^\nu(q,\ztip) q d q\,.
\label{eqn:q_avg}
\end{equation}
The idea is that unless $G_k^\nu(q, \ztip)$ has a complicated structure or a slow decay,
$\bar{q}_k^\nu$ should play the role of a characteristic momentum that determines $k$th polariton pole $\beta_k^\nu$.
Accordingly, we may expect that $1 / \bar{q}_k^\nu$ should give an improved estimate of the spatial resolution of the probe in the context of near-field imaging by s-SNOM.
Interestingly, $\bar{q}_k^\nu$ can be found by differentiating $\beta_k^\nu(\ztip)$:
\begin{equation}
\bar{q}_k^\nu(\ztip) = \frac12\, \frac{\partial}{\partial \ztip} \log \beta_k^\nu\,.
\label{eqn:q_avg_z}
\end{equation}
To obtain this formula consider first a sample with a $q$-independent reflectivity $\beta$ and let the probe-sample separation be $\ztip = z + d z$.
This system is equivalent to another one: the probe separated by $\ztip = z$ from a fictitious \emph{two-component} medium composed of a vacuum layer of thickness $d z$ plus the original sample.
The surface reflectivity of such a two-component medium is $q$-dependent,
$r_\mathrm{P}(q) = \beta e^{-2 q d z}$,
so that it has the form~\eqref{eqn:rp_pert} with $\delta r_\mathrm{P}(q) = -2 q \beta d z$.
Evidently, such a $\delta r_\mathrm{P}(q)$ shifts the resonant pole from $\beta = \beta_k^\nu(z)$ to $\beta = \beta_k^\nu(z + \delta z)$,
i.e., causes a differential change $\delta \beta_k^\nu = (\partial \beta_k^\nu / \partial z) d z$.
Substituting these relations into Eq.~\ref{eqn:G_def}, we get Eq.~\eqref{eqn:q_avg_z}.
Note that as $\beta_k^\nu$ rises more steeply with $\ztip$ for larger $k$, $\bar{q}_k^\nu$ increases with $k$.
 
An equivalent description of the effect of a $q$-dependent perturbation is that it induces a correction to the surface reflectivity.
The effective reflectivity $\beta^{\mathrm{eff}}$ is different for each $k$ and $\nu$,
\begin{equation}
\beta_k^{\nu,\mathrm{eff}} \equiv
\beta - \delta\beta_k^\nu
= \zfint[\infty]
G_k^\nu(q,\ztip)  r_\mathrm{P}(q) d q\,.
\label{eqn:beta_eff_k}
\end{equation}
The corresponding polarizability $\chi^\nu$ is given by
\begin{equation}
\chi^{\nu}
  = \zfsum{k} \frac{R_{k}^{\nu}}{\beta_{k}^{\nu} - \beta_k^{\nu,\mathrm{eff}}} \,.
\label{eqn:chi_eff_q_dep}
\end{equation}
%
In the following we focus on function $G_0^\nu(q, \ztip)$ because $k = 0$ is the dominant resonance at all but very small $\ztip$.
Actually, the large-distance limit of this function has the universal form
\begin{equation}
G_0^\nu(q,\ztip) \simeq 4 \ztip^3 q^2 e^{-2 q \ztip}\,,
\quad \ztip \gg L\,,
\label{eqn:G_0_far}
\end{equation}
same for both $\nu$.
Equation~\eqref{eqn:G_0_far} follows from Eqs.~\eqref{eqn:chi_eff_pdp_q} and \eqref{eqn:G_def}
and is consistent with the surmised large-$q$ behavior~\eqref{eqn:G_large_q}.
As one can see, Eq.~\eqref{eqn:G_0_far} gives $G_0^\nu(q,\ztip)$ that is
normalized, nonnegative, and has a single maximum at $q = 1 / \ztip$.
The average momentum is $\bar{q}_0^\nu \simeq 3 / (2 \ztip)$.

In the intermediate-distance regime functions $G_k^\nu(q, \ztip)$ are not expected to be universal.
The specific example we treat in detail is again the conducting spheroidal probe.
Combining Eq.~\eqref{eqn:q_avg_z} with
the results of Secs.~\ref{sec:limits} and \ref{sec:spheroidal},
for the strongly elongated spheroid we obtain the following:
\begin{equation}
\frac{1}{\bar{q}_0^\bot(\ztip)} \sim
\left\{
\begin{array}{lr}
(a \ztip)^{1 / 2}\,, & \ztip \ll a\,,\\
2 \ztip \log\left(\dfrac{\ztip}{a}\right)\,,\  & a \ll \ztip \ll \tilde{L}\,,\\
\dfrac{2 \ztip}{3}\,, & \ztip \gg L\,.
\end{array}
\right.
\label{eqn:q_avg_0}
\end{equation}
Since the left-hand side has the physical meaning of the spatial resolution of the probe,
we expect it to monotonically decrease as $\ztip$ decreases.
Therefore, the length scale $\tilde{L}$ appearing on the second line of Eq.~\eqref{eqn:q_avg_0} should be of the order of ${L}\,/\,{3 \log (L / a)}$.
The presence of a large logarithmic factor $\log(\ztip / a)$ in the intermediate-distance regime $a \ll \ztip \ll \tilde{L}$ indicates that function 
$G_0^\bot(q,\ztip)$ has a considerable weight at $q$ parametrically smaller than $1 / \ztip$.
In other words, a strongly elongated spheroidal probe senses electric fields beyond its immediate vicinity $\rho < \ztip$. (A similar point was made previously in Ref.~\onlinecite{Zhang2012nfs}.)
As $L / a$ decreases, $\tilde{L}$ comes close to $a$,
and this intermediate regime disappears.
For example, the sphere acts essentially as a local probe.

The calculation of $G_k^\nu(q, \ztip)$ for the spheroid can be done as follows.
Applying the first-order perturbation theory to the linear system~\eqref{eqn:sphrd_char_eqns_beta},
one finds
\begin{equation}
G_k^\nu(q, \ztip) = \frac{\mathbf{u}_k^\dagger \mathbf{H}^\prime \mathbf{u}_k}
                    {\mathbf{u}_k^\dagger \bar{\mathbf{H}} \mathbf{u}_k} \,,
\label{eqn:G_explicit}
\end{equation}
where $\mathbf{H}^\prime$ is the matrix with elements
\begin{equation}
H^{\prm}_{ll\prm} = \frac{2 \pi}{q} I_{l+\frac12} (q F) I_{l\prm+\frac12} (q F) e^{-2 q \zp}\,. 
\label{eqn:H_prime}
\end{equation}
Once the eigenvectors $\mathbf{u}_k$ are found, e.g., as described in Sec.~\ref{sec:spheroidal},
function $G_k^\nu(q, \ztip)$ can be readily computed.

Our numerical investigation of $G_k^\nu(q, \ztip)$ was limited mainly to $k = 0$ and $\nu = \,\bot$.
We observed that the eigenvector components approximately followed the geometric series $(\mathbf{u}_0)_j \sim t^j$.
The quotient $t$ is somewhat larger than unity for small $\ztip$.
As $\ztip$ increases, $t$ becomes less than unity, so that the first component $(\mathbf{u}_0)_0$ dominates.
Neglecting all other components and expressing the modified Bessel function $I_{3 / 2}(z)$ in terms of elementary functions,
we obtain the analytical approximation from Eqs.~\eqref{eqn:G_explicit} and \eqref{eqn:H_prime}:
\begin{equation}
G_0^\nu(q, \ztip) = \frac{c_0}{q^4}
 \left(q F \cosh q F - \sinh q F\right)^2 e^{-2 q \zp}\,,
\label{eqn:G_explicit_far}
\end{equation}
where $c_0$ is a normalization constant.
At $\ztip \gg L$ we can focus on the range of momenta less than $1 / L$
because at larger $q$ this function is already exponentially small.
For such $q$ the bracketed expression on the right-hand side can be replaced by $(F q)^6 / 9$ and $\zp = \ztip + L$ by $\ztip$,
which yields the asymptotic form~\eqref{eqn:G_0_far}.

To examine small and intermediate distances we used the direct numerical evaluation of 
$\mathbf{u}_0$ and $G_0^\nu(q, \ztip)$.
As in Sec.~\ref{sec:spheroidal},
we considered two aspect ratios: $L/a = 25$ and $L/a = 1$.
Only $\nu = \,\bot$ part was studied.
The results for $L/a = 25$ are shown using the false color scale in Fig.~\ref{fig:G_q_z}(a).
It can be seen that as $\ztip$ decreases,
both $\bar{q}_0^\bot(\ztip)$ and the position of the maximum of $G_0^\bot(q, \ztip)$ as a function of $q$ shift toward larger values.
This implies that the probe becomes more sensitive to finer spatial features of the sample,
as discussed above.
The line plot of $G_0^\bot(q, \ztip)$ for several $\ztip$ presented in Fig.~\ref{fig:G_q_z}(b) depicts the same trend.
The average momentum $\bar{q}_0^\bot$ and the position of the $G_0^\bot(q)$ maximum are of the same order of magnitude except at very short distances where $\bar{q}_0^\bot$ increases more rapidly as $\ztip$ decreases.
Note that Eq.~\eqref{eqn:q_avg_0} predicts that $\bar{q}_0^\bot$ diverges at $\ztip = 0$.
From Fig.~\ref{fig:G_q_z}(b) we also see
that for the same $\ztip$ the maximum of $G_0^\bot(q, \ztip)$ is found at $q$ smaller by a factor of $3$--$10$ for the spheroid compared to the sphere.
This confirms that the spheroid is much more sensitive to small in-plane momenta than the sphere, i.e.,
the response of a strongly elongated spheroid is affected by a relatively wide range of lengthscales.

For $k > 0$, $G_k^\bot(q, \ztip)$ has nodes as a function $q$ at fixed $\ztip$.
The number of nodes is equal to $k$, see Fig.~\ref{fig:G_q_z}(c). 
Apparently, at such $q$ near-field coupling between oscillatory charge distributions on the probe and the sample exactly vanishes.
Therefore, small perturbations at such discrete $q$ do not affect the $k$th resonant mode.
Finally, although $\bar{q}_k^\bot$ increase with $k$ for the reasons explained above, the maxima of $G_k^\bot$ show the opposite trend,
which is presently not understood.

\section{From near-field polarizabilities to far-field observables}
\label{sec:demodulation}

In order to apply our theory to simulation of s-SNOM experiments,
we need to include a few more ingredients in our calculation.
The first one is the so-called far-field factor
(FFF) $F^\nu(\omega)$.
This factor accounts for the fact that the probe is illuminated not only by the incident wave but also by its reflection from the sample.
In experiment $\mathrm{P}$-polarized incident field is usually used,
to take advantage of the high transverse polarizability of the probe.
Assuming the sample surface is flat, uniform, and its linear dimensions are much longer than the radian sphere diameter $c / \omega$,
the reflection of the incident wave is described by the coefficient $r_\mathrm{P}(q_s, \omega)$,
where
\begin{equation}
q_s = \frac{\omega}{c} \sin \theta
\label{eqn:q_s}
\end{equation}
is the in-plane photon momentum and $\theta$ is the angle of incidence.
Hence, the ratio of $\nu$-component of the electric field at the surface to that of the incident wave is $1 \pm r_\mathrm{P}(q_s, \omega)$ for $\nu =\,\bot$ and $\parallel$, respectively.
The FFF also takes into account that the field scattered by the probe reaches the detector in two waves: directly and after reflection from the sample surface.
Usually, the backscattered field is measured.
It has the in-plane momentum $-q_s$ and therefore the same reflection coefficient $r_\mathrm{P}(-q_s, \omega) = r_\mathrm{P}(q_s, \omega)$ as the incident wave.
The total FFFs for this setup are given by
\begin{subequations}
\begin{align}
 F^\bot(\omega) &= [1 + r_\mathrm{P}(q_s, \omega)]^2 \sin^2\theta\,,
	\label{eqn:FFF_perp}\\
 F^\parallel(\omega) &= [1 - r_\mathrm{P}(q_s, \omega)]^2 \cos^2\theta\,.
	\label{eqn:FFF_parallel}
\end{align}
\label{eqn:FFF}
\end{subequations}
The trigonometric factors on the right-hand side take care of 
conversion between the total electric field $E_{\mathrm{ext}}$ of the waves and their $\bot$, $\parallel$ components.
Note that our assumption of the plane-wave illumination is not entirely realistic.
In experiment, a focused Gaussian beam is typically used, in which case the FFFs are effectively averaged out over a range of angles $\theta$. Numerical apertures $\sim 0.4$ are common.
We must also stress that Eq.~\eqref{eqn:FFF} should be modified if the system studied by s-SNOM is nonuniform on scales shorter than $c / \omega$.
Typical examples include a small sample residing on some substrate~\cite{Zhang2012nfs} or
measurements done close to a boundary of two different materials.

Another point we have to discuss is signal demodulation.
In the experiment the probe is made to oscillate mechanically,
which causes periodic variation of the probe-sample distance:
\begin{equation}
	\ztip(\varphi) = z_{0} + \Delta{z} \left( 1 - \cos \varphi \right)\,,
	\varphi \equiv \Omega t\,.
	\label{eqn:z_tip_def}
\end{equation}
The oscillation amplitude is typically $\Delta{z} = 20$--$90\,\mathrm{nm}$, comparable to the radius of curvature $a \sim 30\,\mathrm{nm}$ of the probe.
The minimal approach distance $z_0 \geq 0$ can be equal to zero if the probe
taps the sample.
The tapping frequency $\Omega$ is many orders of magnitude smaller than the laser frequency $\omega$,
and so the motion of the tip does not affect the electromagnetic response.
Effectively, the experiment consists of measuring the scattered signal for many static configurations with different $\ztip$.
The $n$th Fourier harmonic of the backscattered field is
referred to as the demodulated signal $s_{n}$.
(Here we define $s_n$ as a complex number but in experimental literature
it is common to discuss the amplitude and the phase of $s_n$ separately.) The primary purpose of demodulation is to suppress the far-field background signal created by reflections from the body of the tip, the cantilever, \textit{etc}.
This background is large but depends on $\ztip$ very weakly (linearly) and thus contributes predominantly to the $n = 1$ harmonic.
Unfortunately, demodulation strongly diminishes the signal amplitude, making it more susceptible to experimental noise.
In practice, $n = 2$ or $3$ usually gives the best approximation of the true near-field signal.~\cite{Keilmann2004nfm, Keilmann2009, Atkin2012n}

The demodulated signal is related to the polarizabilities $\chi^\nu(\omega, \ztip)$ we have been discussing in previous sections by
\begin{equation}
	s_{n}^\nu(\omega) = \mathrm{const}\,\times \chi_{n}^\nu(\omega) F^\nu(\omega)\,,
\label{eqn:s_n_def}
\end{equation}
where $\chi_{n}^\nu(\omega)$ is the $n$th Fourier harmonic of $\chi^\nu$:
\begin{equation}
  	\chi_{n}^\nu(\omega) =  \int\limits_{0}^{\pi} \frac{d \varphi}{\pi}\, \chi^\nu\bigl(\omega, \ztip(\varphi)\bigr) \cos{n \varphi}\,.
	\label{eqn:chi_n_def}
\end{equation}
One more element of the experimental protocol is normalization.
What is typically reported is $s_{n}^\nu(\omega)$ normalized against a certain reference material, e.g., Si or Au:
\begin{equation}
\bar{s}_{n}^\nu(\omega) = s_{n}^\nu(\omega)
                        / s_{n}^{\nu,\,\text{ref}}(\omega) \,.
	\label{eqn:s_n_bar}
\end{equation}
The normalization eliminates a number of physically uninteresting or poorly known factors, such as the constant in Eq.~\eqref{eqn:s_n_def}
that are related to the optical setup of the experiment.
The FFFs may also be canceled if
both the studied and the reference objects in the experiment are positioned nearby, so that the data for the two are taken at points no farther apart than the diameter $c / \omega$ of the radian sphere. 

The last point we wish to draw attention to is that the absolute value of the minimum probe-sample distance $z_0$ [Eq.~\eqref{eqn:z_tip_def}] cannot be determined very accurately.
Therefore, experimentalists have to measure the so-called approach curve,
which is the s-SNOM response as a function of $z_0$ at a fixed frequency.
They then identify the point $z_0 = 0$ as a point where a qualitative change in behavior in $s_{2}$ or $s_{3}$ appears.
The logic behind this procedure is that once the probe makes the mechanical contact with the sample, its oscillations become reduced in amplitude, marking an unambiguous change.
A potential flaw of this argument is that sharp changes in $s_n$'s may be generated by a rapid variation of electromagnetic coupling between the probe sample at short separation even \textit{before} making mechanical contact.
We will discuss this issue in more detail in Sec.~\ref{sec:result}.

\section{Case of local reflectivity: aluminum oxide}
\label{sec:result}

\begin{figure*}[htb]
\begin{center}
\includegraphics[width=\imagewidth]{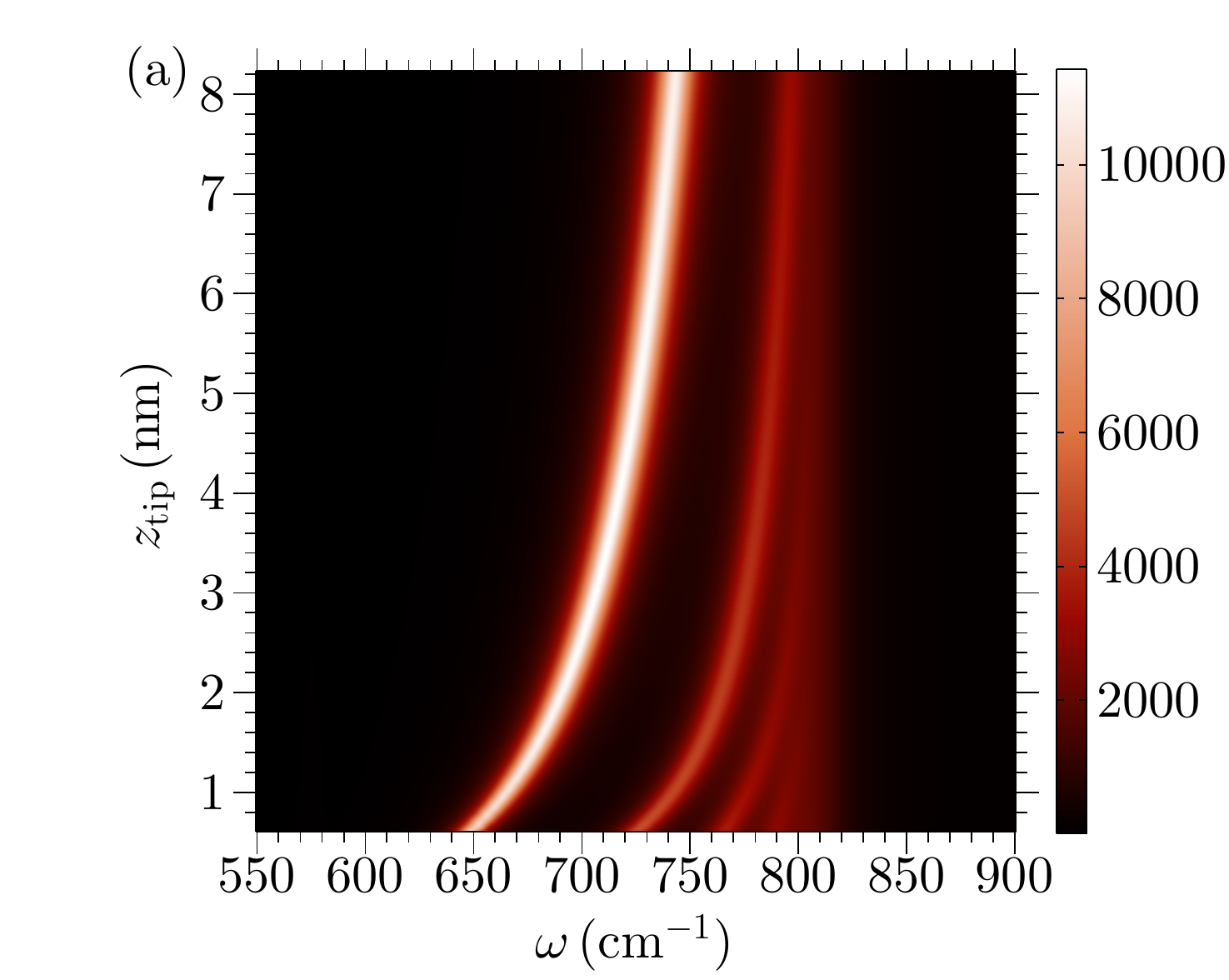}
\hspace{0.01 in}
\includegraphics[width=\imagewidth]{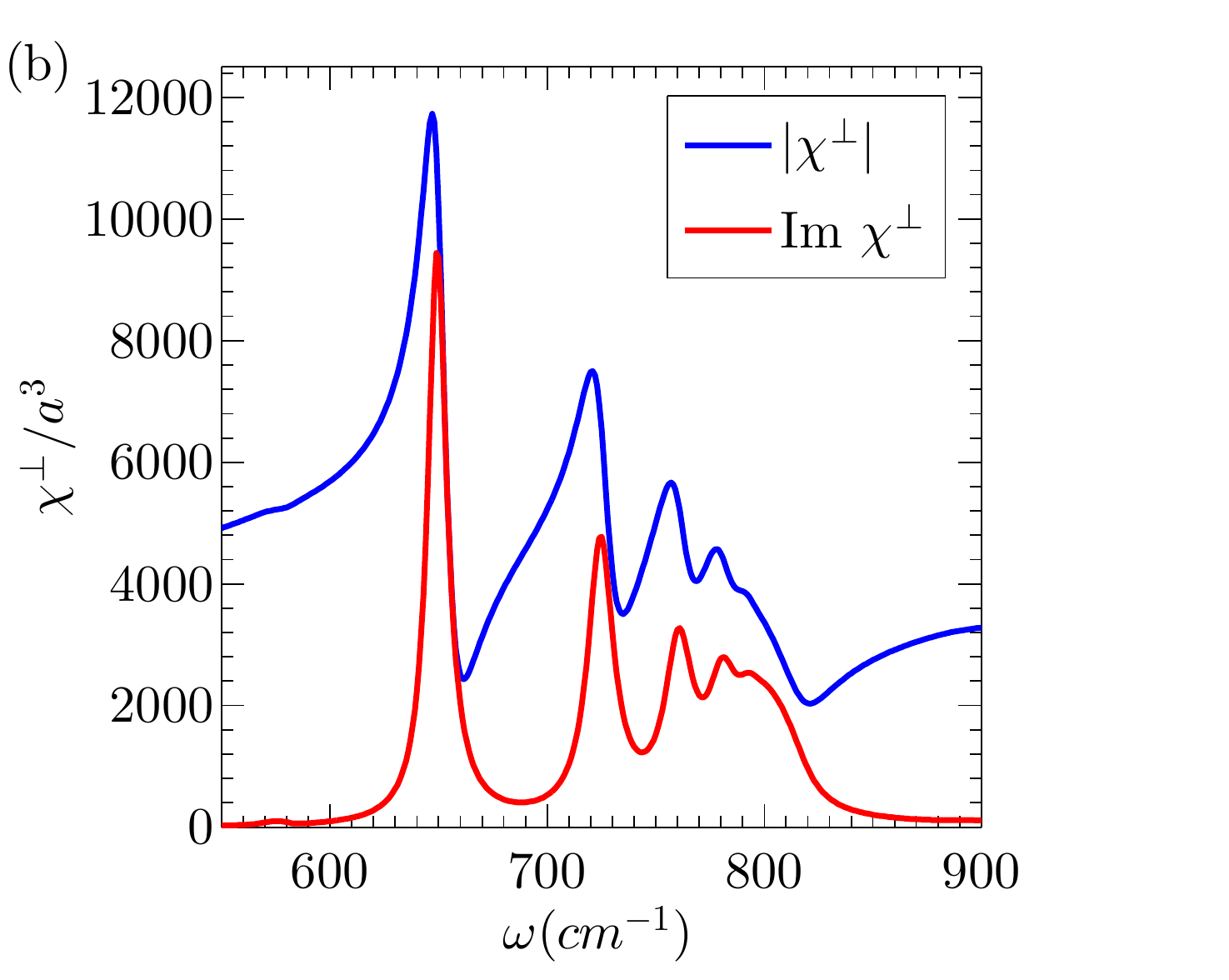}
\hspace{-0.25 in}
\includegraphics[width=\imagewidth]{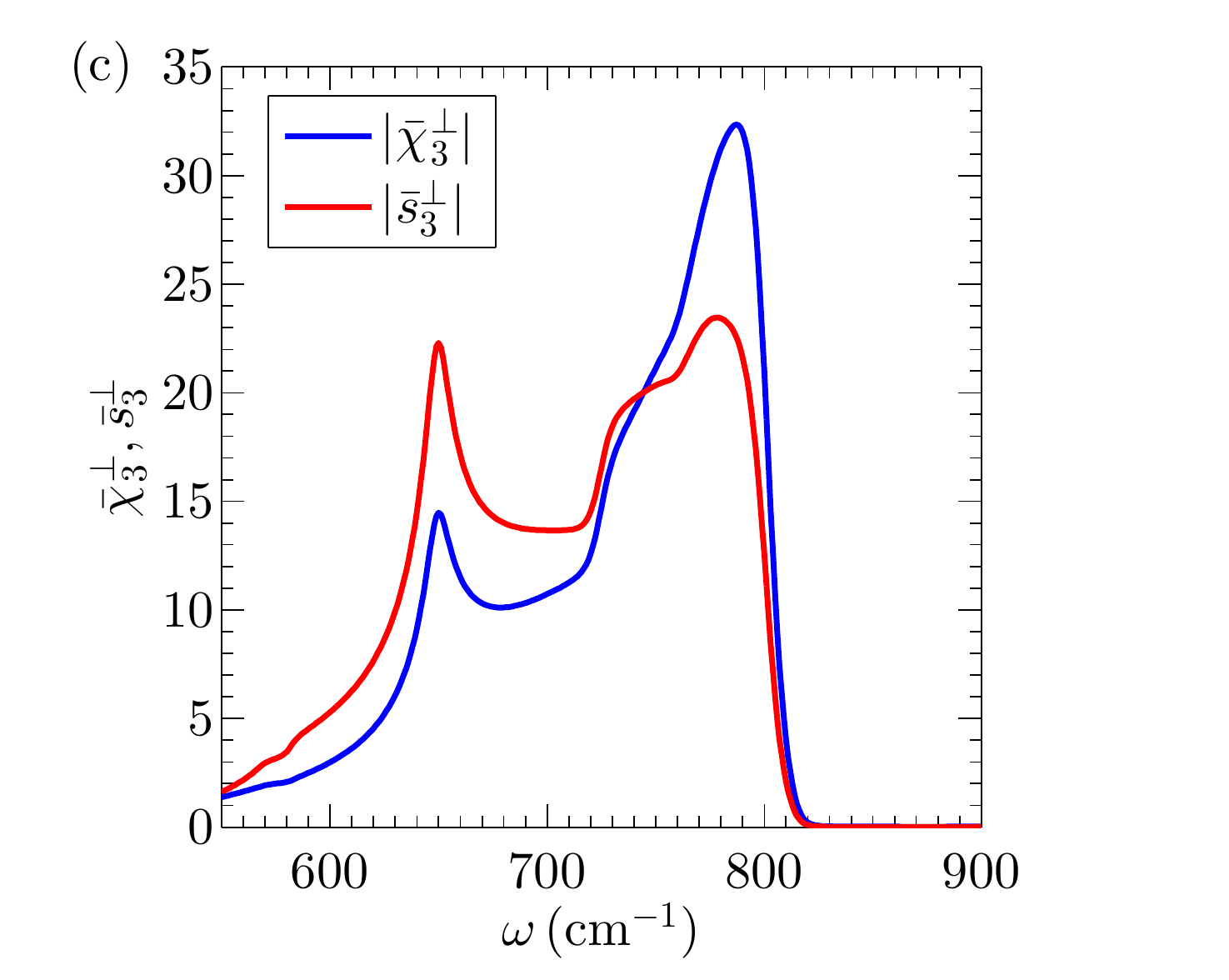}
\\
\includegraphics[width=\imagewidth]{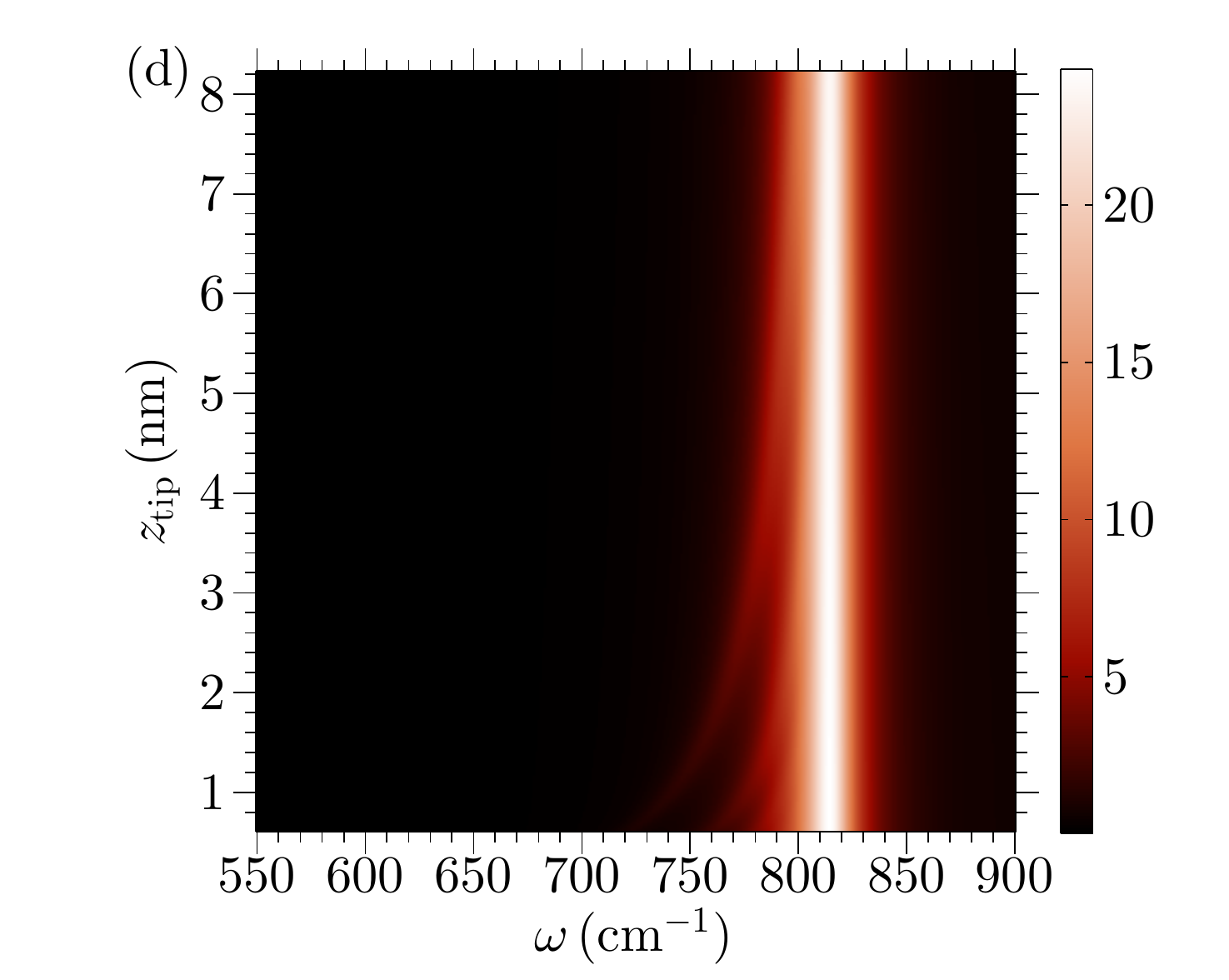}
\hspace{0.01 in}
\includegraphics[width=\imagewidth]{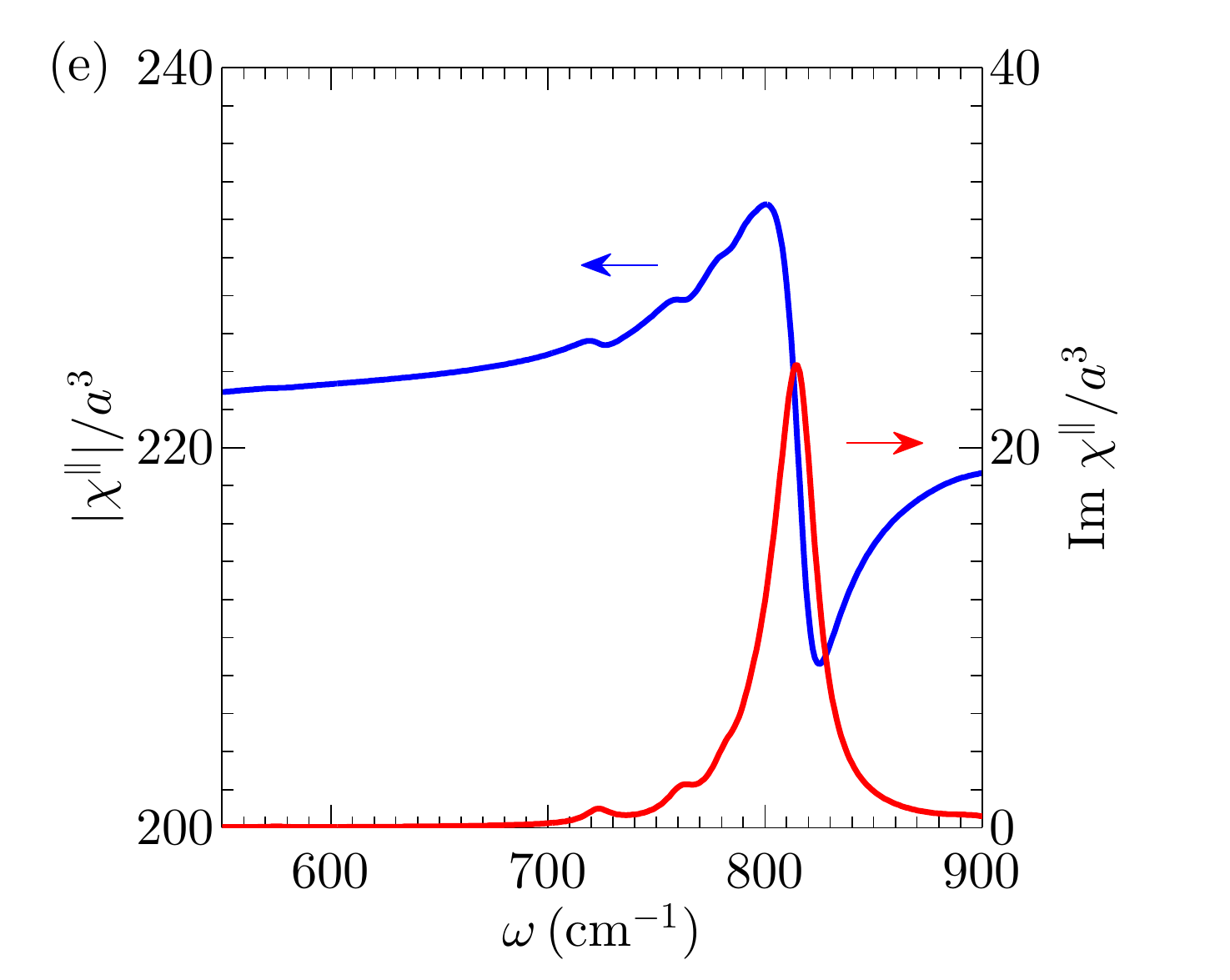}
\hspace{-0.25 in}
\includegraphics[width=\imagewidth]{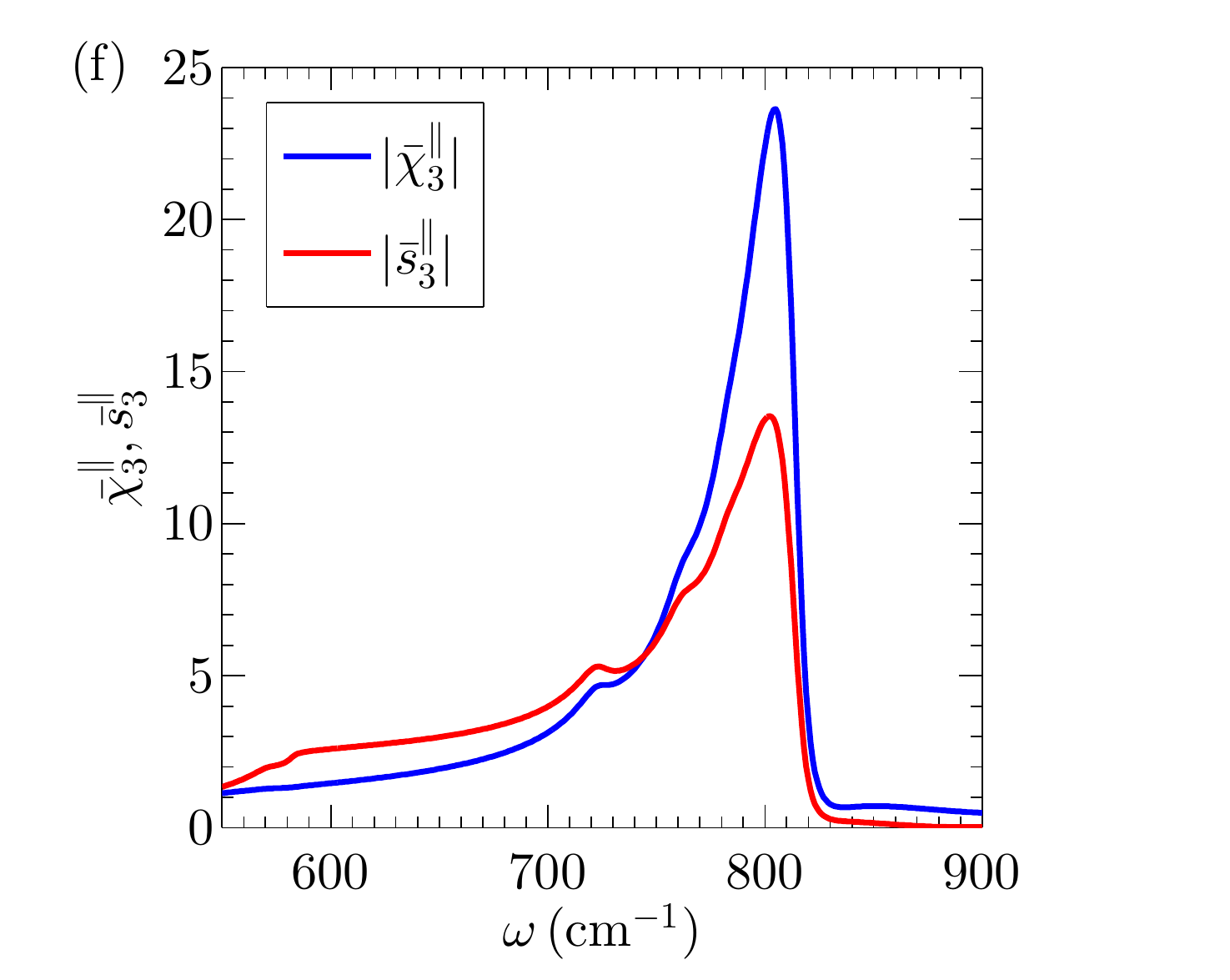}
\end{center}
\caption{(Color online) Response of a perfectly conducting spheroidal probe with $L = 25a$ and bulk Al$_2$O$_3$ sample.
(a) The false color plot of $\im \chi^\bot(\omega,z_\mathrm{tip})/a^3$. The bright curves correspond to the resonant modes, with $k = 0$ mode having the lowest frequency.
(b) The polarizability $\chi^{\bot}$ (absolute value and imaginary part) at $z_\mathrm{tip} = 0.02a = 0.6\,\mathrm{nm}$,
the smallest distance in panel (a).
(c) The absolute value of the demodulated polarizability $|\bar{\chi}_3|$ and scattering signal $|\bar{s}_3|$ for the tapping amplitude $\Delta z = 50\,\mathrm{nm}$ and $z_0 = 0.6\,\mathrm{nm}$.
The origin of the three peaks is discussed in the text.
(d)--(f) The counterparts of panels (a)--(c) for the parallel component, $\nu = \,\parallel$.
The plots again reveal multiple resonances.
However, the overall magnitude of
the polarizability is greatly reduced, $\chi^\parallel \sim 10^{-2} \chi^\bot$, and the resonances
are more strongly bunched near the surface phonon frequency $\omega\tsub{SP} = 818\,\mathrm{cm}^{-1}$.
}
\label{fig:Al2O3 s-SNOM Z}
\end{figure*}

In this and the following Sections we discuss the implications of our theory for near-field response of real materials.
We choose bulk $\alpha$-$\mathrm{Al}_2\mathrm{O}_3$, also known as sapphire or corundum, as our first example of highly resonant material with a momentum-independent reflectivity $\beta$ [Fig.~\ref{fig:Al2O3_beta}(a)].
Another material with these properties, silicon carbide, has been a subject of a recent s-SNOM study co-authored by two of the present authors.\cite{McLeod2014lrm}
Modeling results based on the BEM showing good agreement with the data were also reported in that work.
Realistic probe shapes and retardation effects have been taken into account in order to achieve that. The latter was necessary since the probe length $2 L \sim 20\,\mu\mathrm{m}$ in the experiments was in fact larger than the diameter of the radian sphere $c / \omega \sim 11\,\mu\mathrm{m}$.
Here we do not aim for a perfect agreement with a particular experiment but
instead
wish to illustrate how the general theory of multiple eigenmodes formulated in the preceding Sections can generate novel features in near-field observables.
We study mostly probes of an idealized spheroidal shape but
examine some other shapes as well.
We stay within the quasistatic approximation but
we will comment on retardation effects in Sec.~\ref{sec:shape}.

We use the following momentum-independent model for the reflection coefficient of the uniaxial $\mathrm{Al}_2\mathrm{O}_3$ crystal, 
\begin{equation}
\beta(\omega)=\dfrac{\epsilon_\mathrm{eff}-1}{\epsilon_\mathrm{eff}+1}\, ,
\quad
\epsilon_\mathrm{eff}(\omega)=\sqrt{\epsilon_o\epsilon_e}\, ,
\label{eqn:beta_Al2O3}
\end{equation}
where $\epsilon_\rho$ for $\rho = o$ (ordinary) and $e$ (extraordinary) axes
is given by 
\begin{equation}
\epsilon_\rho(\omega)=\epsilon_{\infty,\rho}
\prod_j
\dfrac{\omega_{j\mathrm{LO},\rho}^2-\omega^2-i\gamma_{j\mathrm{LO},\rho} \omega}{\omega_{j\mathrm{TO},\rho}^2-\omega^2-i\gamma_{j\mathrm{TO},\rho} \omega}\, .
\label{eqn:epsilon_Al2O3}
\end{equation}
The optical constants of Al$_2$O$_3$ reported in the literature~\cite{Harman1994oco, Zeidler2013oco} have slight variations,
presumably because of different crystal purity and processing. 
In our calculations we adopt the results of Ref.~\onlinecite{Zeidler2013oco} at room temperature, reproduced in Table.~\ref{tab:eps_al2o3}.
(For simplicity, the weak oscillator at $\omega_{\mathrm{TO},o} = 634\,\mathrm{cm}^{-1}$ is neglected.)
Due to smallness of the optical phonon linewidths $\gamma_\rho$ in this material, the near-field reflectivity of Al$_2$O$_3$ can be as high as $\beta \sim 10$.

\newcolumntype{3}{D{.}{.}{3}}
\newcolumntype{2}{D{.}{.}{2}}
\newcolumntype{0}{D{.}{.}{0}}
\begin{table}
\caption{Parameters of the optical constant of $\alpha$-Al$_2$O$_3$ used in our calculations, cf. Eq.~\eqref{eqn:epsilon_Al2O3}. The frequency unit is $1\,\mathrm{cm}^{-1}$. (From Ref.~\onlinecite{Zeidler2013oco} for temperature $T = 300\,\mathrm{K}$.)
}
\begin{tabular*}{3in}{@{\extracolsep{\fill} }c 2 c 0 2 0 2} 
\hline \hline\\[-0.1in] 
$\rho$ &
\multicolumn{1}{c}{$\epsilon_{\infty}$} & 
$j$ & 
\multicolumn{1}{c}{$\omega_{j\mathrm{LO}}$} &
\multicolumn{1}{c}{$\gamma_{j\mathrm{LO}}$} &
\multicolumn{1}{c}{$\omega_{j\mathrm{TO}}$} &
\multicolumn{1}{c}{$\gamma_{j\mathrm{TO}}$} \\ [0.5ex] 
\hline 
\\[-0.1in]
$o$ & 3.05 & 1 & 908 & 22.4 & 569 & 7.86\\
& & 2 & 482 & 2.96 & 439 & 3.23\\
& & 3 & 387 & 5.18 & 384 & 6.03\\
 [1ex] 
\hline \\
$e$ & 2.9 & 1 & 885 & 21.6 & 582 & 4.17\\
& & 2 & 481 & 3.21 & 482 & 3.42\\
& & 3 & 511 & 1.42 & 400 & 4.68\\
 [1ex] 
\hline \hline\\
\label{tab:eps_al2o3}
\end{tabular*}
\end{table}

We start by studying the behavior of the probe polarizabilities $\chi^\nu$ as a function of frequency $\omega$. 
In the mid-infrared range, the reflection coefficient $\beta$ of Al$_2$O$_3$ has a single peak centered at the surface-phonon frequency $\omega\tsub{SP}=818\,\mathrm{cm}^{-1}$, depicted in Fig.~\ref{fig:Al2O3_beta}(a).
As $\omega$ approaches $\omega\tsub{SP}$ from below,
$\re\beta(\omega)$ steeply rises.
Equation~\eqref{eqn:chi_eff_general_form} implies that
whenever $\re\beta$ is equal to a pole $\beta_k^{\nu}$,
$\im\chi^\nu$ has a local maximum as long as
the damping $\im\beta(\omega)$ is not too large.
The positions of three such underdamped resonances are indicated schematically in Fig.~\ref{fig:Al2O3_beta}(a).
Thus, a \textit{single} surface mode $\omega\tsub{SP}$ of Al$_2$O$_3$
may produce \textit{multiple} modes of the coupled probe-sample system.
These localized eigenmodes (resonances) have been discussed at length in the preceding Sections.
For example, they are depicted in Fig.~\ref{fig:modes}(b) for the case of a spheroidal probe.
Note that all the resonances are \textit{red-shifted} from the frequency $\omega\tsub{SP}$.
Since $\im \beta$ increases as $\omega$ approaches $\omega\tsub{SP}$,
higher-order resonances are progressively more broad.

The scenario above is described in terms of constant $\beta_k^{\nu}$.
However, the poles are functions of $\ztip$, and so the frequency of each resonance shifts with $\ztip$.
This is clearly seen in a false color plot of $\im\chi^\bot(\omega,\ztip)$ [Fig.~\ref{fig:Al2O3 s-SNOM Z}(a)], where each mode creates a bright curve.
All the curves are red-shifted from $\omega\tsub{SP}$ but converge to it at large $\ztip$.
The smallest $\ztip = 0.02 a$ in Fig.~\ref{fig:Al2O3 s-SNOM Z}(a) is limited by the accuracy of our numerical calculation.
Based on our analytical results we expect that at smaller $\ztip$
the resonance curves are shaped as parabolas that approach $\omega\tsub{TO} = 576\,\mathrm{cm}^{-1}$ where $\re\beta = 1$, cf.~Eqs.~\eqref{eqn:beta_k_short}, \eqref{eqn:beta_Al2O3}, and \eqref{eqn:epsilon_Al2O3}.
A horizontal line cut through Fig.~\ref{fig:Al2O3 s-SNOM Z}(a) taken at $\ztip = 0.6\,\mathrm{nm}$ is plotted in Fig.~\ref{fig:Al2O3 s-SNOM Z}(b) along with the absolute value of $\chi^\bot$.
The strongest peak in this plot corresponds to the $k = 0$ mode.
The multiple weaker peaks at higher frequencies are produced by
$k > 0$ modes.

\begin{figure*}[thb]
  \begin{center}
\includegraphics[width=\imagewidth]{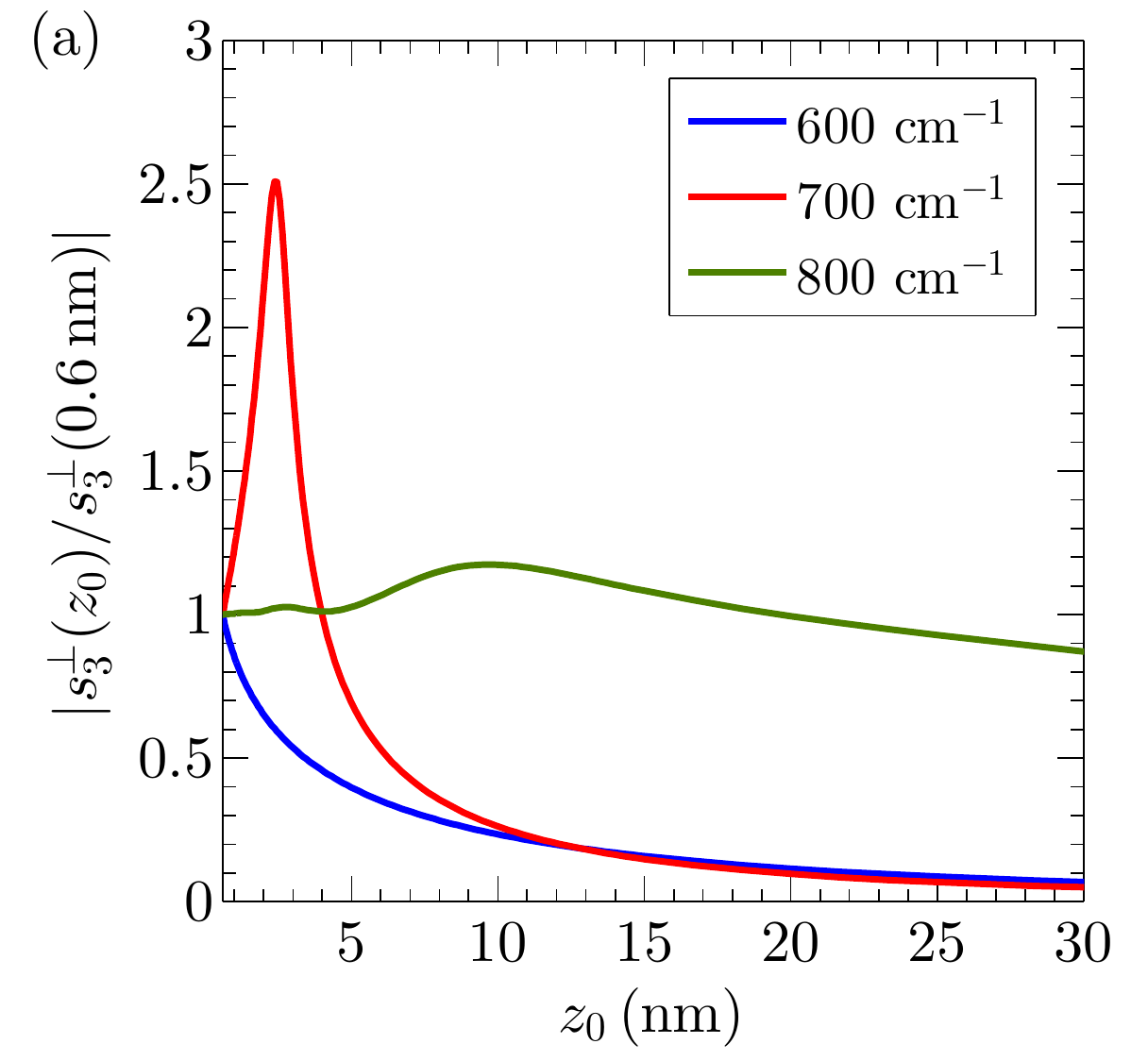}
\includegraphics[width=\imagewidth]{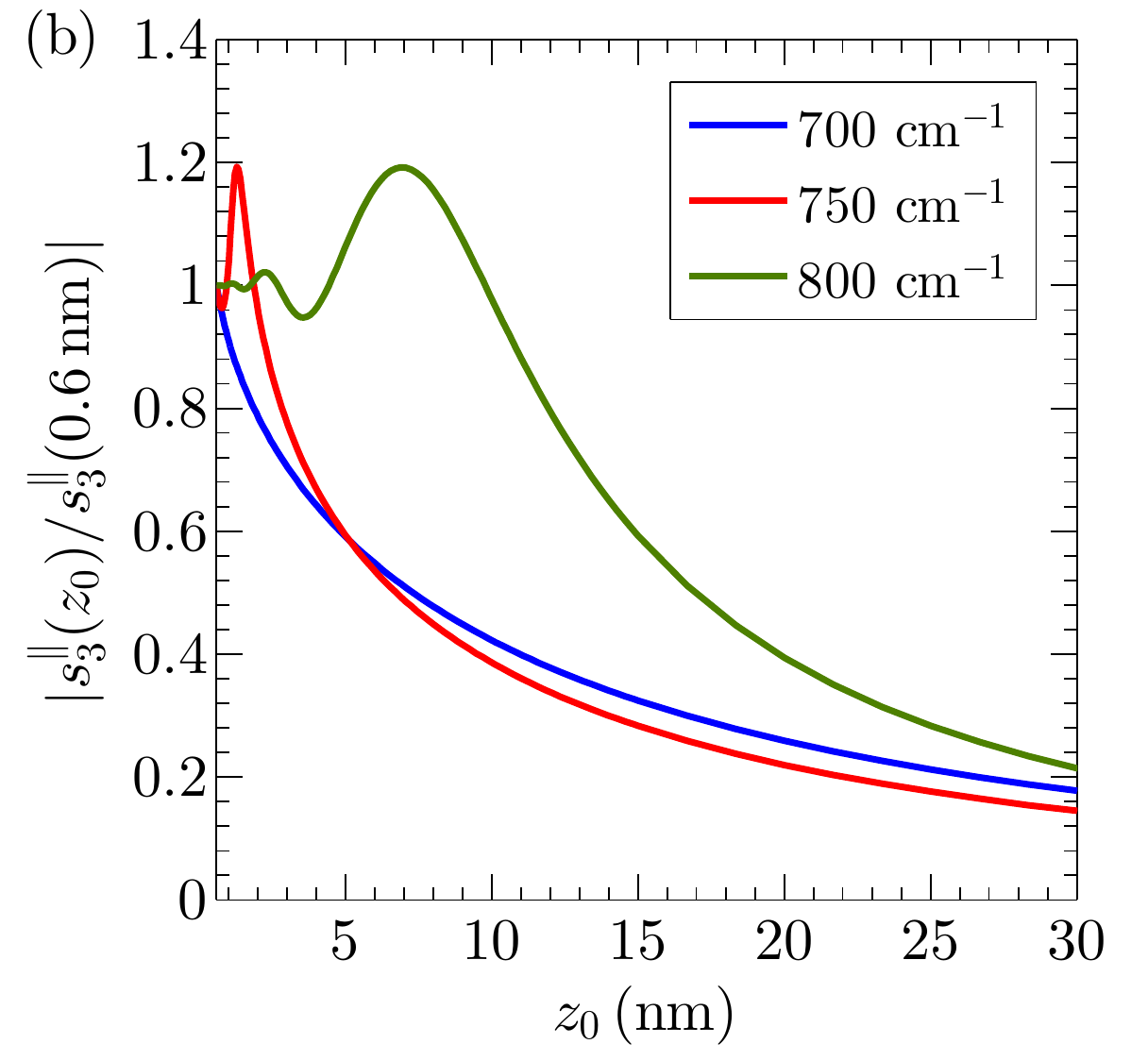}
  \end{center}
\caption{(Color online) Approach curves of $|s^\nu_3|$ for bulk Al$_2$O$_3$, normalized to the value at $z_0=0.6$ nm, for several characteristic frequencies.
The lowest frequencies in both (a) and (b) are such that no resonance curves are crossed during the probe tapping motion.
The approach curves are monotonic.
For the middle pair of frequencies one crossing (of the $k=0$ resonance) does occur.
At such crossing each approach curve has a peak.
The last pair corresponds to the frequencies where $|\bar{s}_3^\nu|$ is close to the maximum value in the spectral range studied.
The approach curves have several peaks because of multiple resonance crossings.
}
\label{fig:Al2O3 s-SNOM X}
\end{figure*}
\begin{figure}[htb]
  \begin{center}
    \includegraphics[width=\imagewidth]{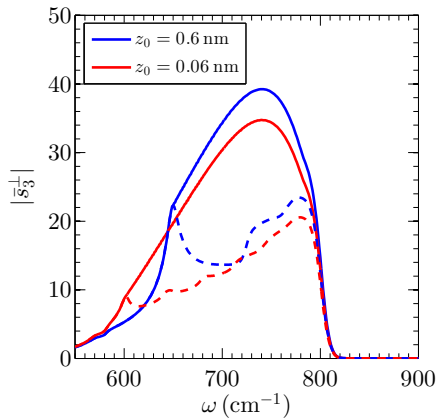}
  \end{center}
  \caption{(Color online) Comparison of spectra of the quantity $|\bar{s}_3^\bot|$ using two different experimental protocols, at two different minimum approach distance $z_0$. The value at each frequency is taken either from the maximum of the $|s^\bot_3|$ approach curve (solid) or from a fixed $z_0$ (dashed).}
  \label{fig:max_contact}
\end{figure}
\begin{figure*}[htb]
  \begin{center}
    \includegraphics[width=\imagewidth]{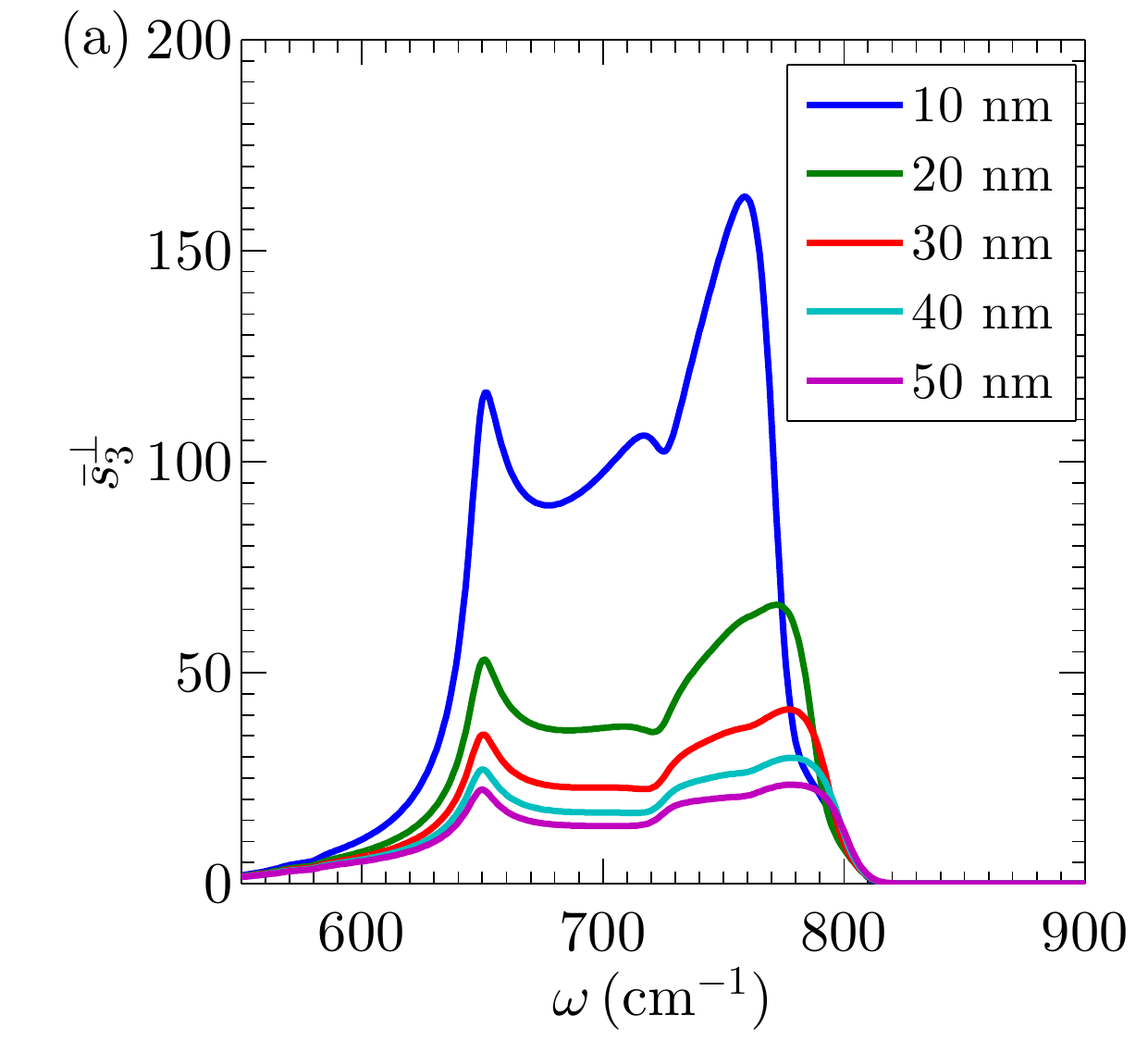}
    \includegraphics[width=\imagewidth]{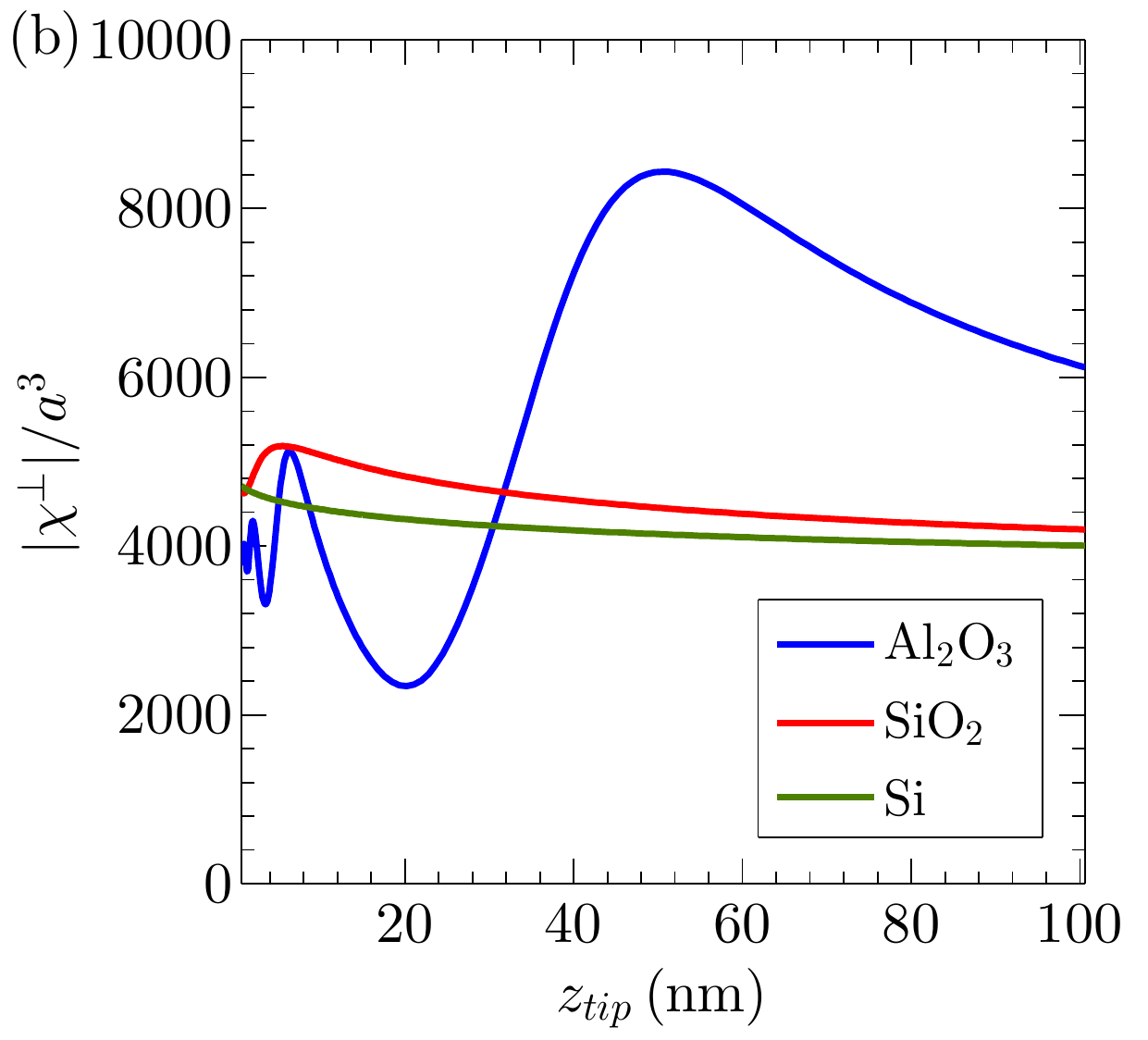}
  \end{center}
  \caption{(Color online) (a) Spectra of $|\bar{s}_3^\bot|$ for $z_0=0.6\,\mathrm{nm}$ and five different tapping amplitudes. The magnitude of $\bar{s}_3^\bot$ increases rapidly with decreasing $\Delta z$. (b) The $\chi^\bot$ approach curves for Al$_2$O$_3$, SiO$_2$ and Si, taken at frequencies corresponding to the largest peak in $|s_3^\bot|$ (790 cm$^{-1}$ for Al$_2$O$_3$, 1120 cm$^{-1}$ for SiO$_2$, and an arbitrary $\omega$ for the frequency-independent case of Si). For Al$_2$O$_3$ sample, $|\chi^\bot|$ shows multiple oscillations; for SiO$_2$ sample, it has a single maximum at small $\ztip$; for Si, the approach curve decays monotonically with $\ztip$. As $\Delta z$ decreases, the approach curves become increasingly different.
  }
  \label{fig:chi3_spec_dz}
\end{figure*}

Next we consider
the effects of demodulation on the s-SNOM signal,
which can be understood as follows.
As the probe oscillates, it spends most time at the minimum and maximum distances from the surface. One therefore expects peaks in $\chi^\nu_n$ at frequencies near those of $\chi^\nu(z_0, \omega)$ and $\chi^\nu(z_0 + 2 \Delta z, \omega)$.
This gives two frequencies per each resonant mode.
Actually, the number of observable peaks is smaller.
Indeed, from Figs.~\ref{fig:Al2O3 s-SNOM Z}(a) and \ref{fig:Al2O3 s-SNOM Z}(d)
one can see that all the resonance curves modes should merge together at $z = z_0 + 2 \Delta z$ for typical $\Delta z \sim 50\,\mathrm{nm}$.
Hence, all the modes should produce a single common peak in the demodulated signal from such $z$.
Furthermore, while the peaks of $\chi^\nu(z_0, \omega)$ are distinct, 
only a few strongest of them can survive the smearing effect of the demodulation. 
These expectations are supported by Fig.~\ref{fig:Al2O3 s-SNOM Z}(c), where we plot the normalized quantities $\bar{\chi}_{3}(\omega, z_{0}) \equiv \chi_3/ \chi_3\tsup{ref}$ and $\bar{s}_{3}$ for $\nu=\bot$, assuming tapping amplitude $\Delta z=50$ nm, $z_0=0.6$ nm, and Si as the reference material.
In Fig.~\ref{fig:Al2O3 s-SNOM Z}(c) we see only three peaks.
The peak at $650$ cm$^{-1}$ in $|\bar{s}_{3}|$
is produced by the dominant $k=0$ mode.
It has the same frequency as the $k=0$ peak in Fig.~\ref{fig:Al2O3 s-SNOM Z}(b).
The second peak near $725\,\mathrm{cm}^{-1}$ in $|\bar{s}_{3}|$ (which looks more like a shoulder in $\bar{\chi}_{3}$) is produced by the $k=1$ mode at the $\ztip = z_0$ point.
The remaining third peak at $787\,\mathrm{cm}^{-1}$ is produced collectively by all the modes.
A similar correspondence between the resonance curves of
the polarizability function and the peaks in the demodulated signal is found in the $\nu =\, \parallel$ component, cf.~Figs.~\ref{fig:Al2O3 s-SNOM Z}(d)--(f).
However, the lower $k = 1$ peak is now very weak and is considerably blurred by the demodulation, Fig.~\ref{fig:Al2O3 s-SNOM Z}(f).
Should we have considered a model with smaller dissipation, this and other high-order peaks would have been more clearly distinguishable in $|\bar{s}_3|$.
Note that although the normalized and demodulated signal strength is comparable for the two $\nu$ components, the polarizability for $\nu =\, \parallel$ is orders of magnitude smaller so its contribution can be safely ignored.

The discussion above pertain to horizontal cuts of $\chi^\nu(\ztip,\omega)$.
Taking a fixed-frequency (vertical) cut through Fig.~\ref{fig:Al2O3 s-SNOM Z}(a),
and performing the demodulation for a range of minimum distances $z_0$,
one obtains the $\nu =\, \bot$ approach curve for the scattering signal.
An intriguing result of this analysis is the possibility of a nonmonotonic dependence of the approach curve on $z_0$.
The nonmonotonicity is due to the crossing of the resonance curves of $\chi^\bot$ by the vertical line cut. Such crossings are found between $\omega_{\mathrm{TO}}$ where $\re\beta = 1$ and $\omega\tsub{SP}$ where $\re\beta$ reaches its maximum.
Near the low-frequency end of this interval, the $k=0$ mode should be again dominant. It is expected to produce a peak in the approach curve, which would follow the same trajectory as the $k=0$ curve in Fig.~\ref{fig:Al2O3 s-SNOM Z}(a), moving to larger $z_0$ as $\omega$ increases. 
Higher order modes should appear at frequencies closer to $\omega\tsub{SP}$ and produce weaker peaks at smaller $z_0$.
The amalgamation of these peaks give rise to the nonmonotonicity of the approach curve.

We show in Fig.~\ref{fig:Al2O3 s-SNOM X}(a) the $s_3$ approach curves for $\nu=\bot$ for three frequencies.
All the curves are normalized to their value at their left ends, $z_{0} = 0.6\,\mathrm{nm}$.
The approach curve for $\omega=600\,\mathrm{cm}^{-1}$ decays monotonically with increasing $z_0$ because the cut at such $\omega$ does not cross any of the resonances. 
In the approach curve for $700\,\mathrm{cm}^{-1}$, a strong peak is seen at around $2\,\mathrm{nm}$ due to the crossing of the $k=0$ resonance.
The last approach curve, for $800\,\mathrm{cm}^{-1}$ contains a series of oscillations at small $z_0$ and a broad hump at large $z_0$, due to the multiple resonance crossings.
The approach curves for $\nu =\,\parallel$ plotted in Fig.~\ref{fig:Al2O3 s-SNOM X}(b) exhibit the same general trends as those for $\nu =\, \bot$.

The striking multi-peak spectra and anomalous nonmonotonic approach curves we described above stem from the large $r\tsub{P}$ of Al$_2$O$_3$ and are not found in less resonant materials, see Sec.~\ref{sec:shape} and Ref.~\onlinecite{McLeod2014lrm}.
This rich structure is also quite sensitive to the choice of $z_0$.
If this parameter is too large, the peaks in the spectrum of the scattering signal merge together at $\omega = \omega\tsub{SP}$.
If $z_0$ is too small, the resonance curves become very flat at $\omega < \omega\tsub{SP}$,
so the corresponding peaks are smeared by demodulation and dwarfed by the $\omega\tsub{SP}$ peak.
Hence, there exists an optimal value of $z_0$ that allows one to resolve multiple peaks most clearly.
For our Al$_2$O$_3$ model this value is actually not too far from $z_0 = 0.6\,\mathrm{nm}$ used in Fig.~\ref{fig:Al2O3 s-SNOM Z}.
For example, the $s_3$ spectrum for a smaller $z_0=0.06\,\mathrm{nm}$ is shown in Fig.~\ref{fig:max_contact} (dashed lines), 
where the $k=0$ peak is much less pronounced while more higher order peaks become distinguishable and form small steps.
For even smaller $z_0$ the steps are further smoothed, eventually leaving only one peak near $\omega_\mathrm{SP}$.

In addition to the value of $z_0$, many other experimental parameters and procedures can significantly alter the resultant spectrum. 
For instance, the experimental determination of $z_0$ based solely on the s-SNOM approach curve can be inaccurate due to its possible nonmonotonicity, as discussed in the previous Section. 
It is generally incorrect to ascribe $z_0 = 0$ to the probe position at which the near-field signal has the highest amplitude.
Such a protocol effectively yields a frequency-dependent $z_0$.
The difference from the spectra taken for a truly constant $z_0$
can be drastic, as illustrated in Fig.~\ref{fig:max_contact}. 
Conversely, the strong sensitivity of the near-field signal to the value of probe-sample distance may perhaps be used for a more accurate measurement of $z_0$ (although this may require knowing the curvature radius $a$ and perhaps other details of the probe shape).

The tapping amplitude $\Delta z$ is another parameter that affects the spectrum.
When $\Delta z$ is small, the demodulation at $n$th order is roughly equivalent to taking the $n$th order derivative of $\chi^\nu(\ztip)$.
Therefore, a material with a sharply varying approach curve yields a stronger demodulated signal than the material with a smoothly varying one.
In our case the signal of Al$_2$O$_3$ is normalized against Si, whose polarizability decays monotonically with $\ztip$ [Fig.~\ref{fig:chi3_spec_dz}(b)].
As $\Delta z$ decreases, the polarizability of Al$_2$O$_3$ become increasingly oscillatory, while that of Si remains smooth.
This results in the increased contrast of the demodulated signal for the two materials for smaller $\Delta z$ [Fig.~\ref{fig:chi3_spec_dz}(a)].

\begin{figure*}[hbt]
  \begin{center}
\includegraphics[width=\imagewidth]{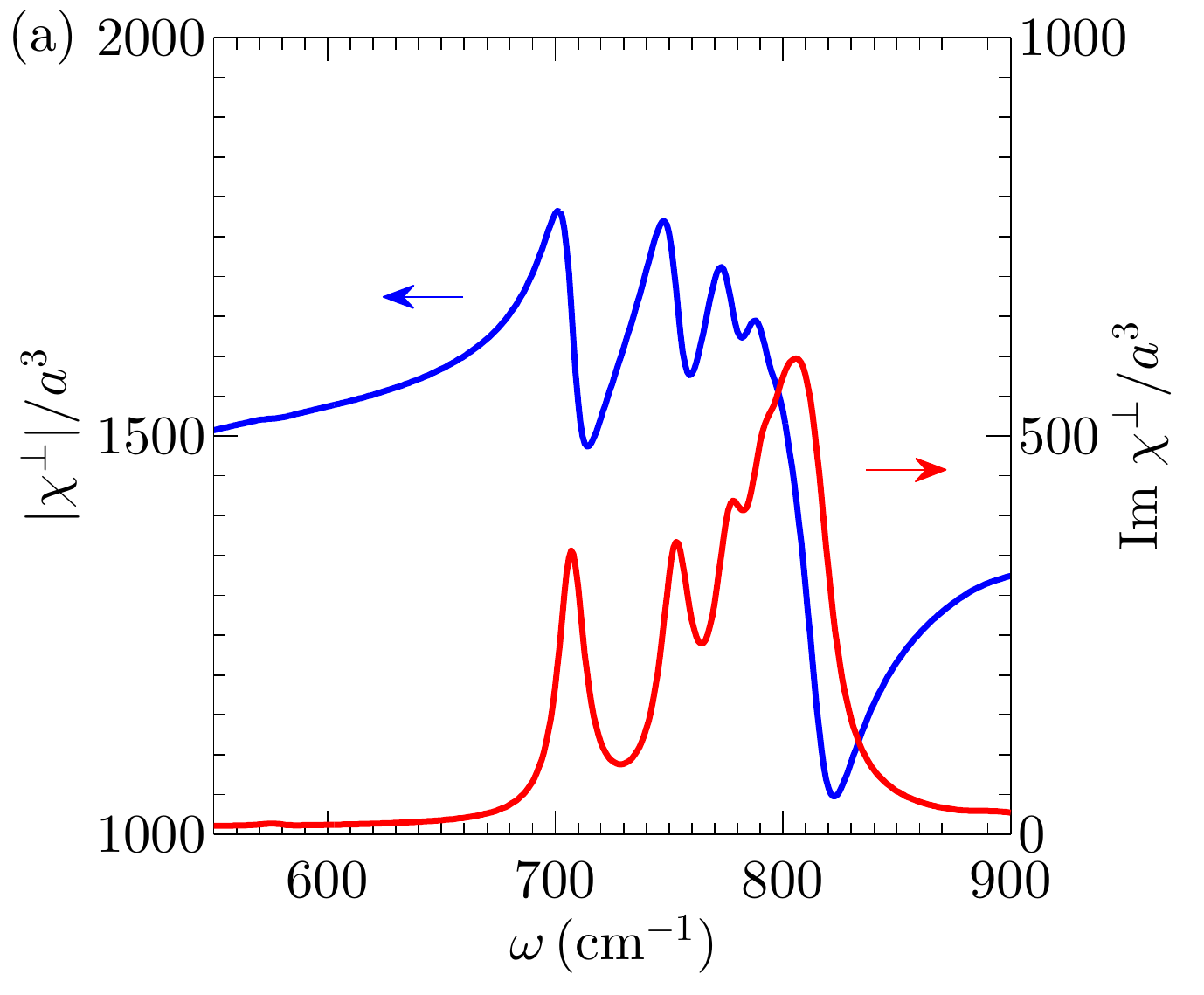}
\includegraphics[width=\imagewidth]{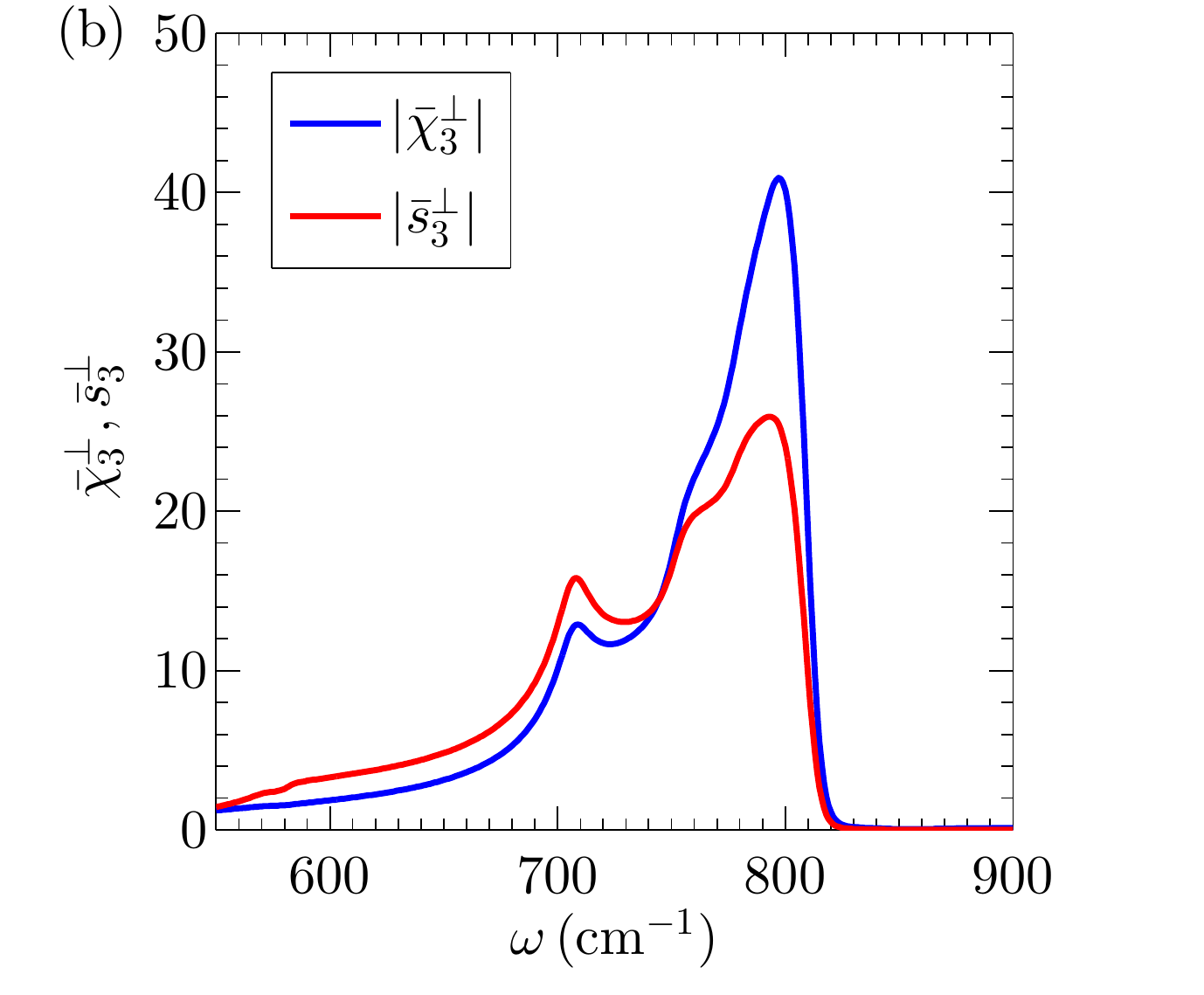}
  \end{center}
  \caption{(Color online)
  (a) $\chi^{\bot}$ and (b) $|\bar{\chi}^{\bot}_3|$ and $|\bar{s}^{\bot}_3|$ spectra of Al$_2$O$_3$ for the case of a Si probe. All other parameters are the same as in Fig.~\ref{fig:Al2O3 s-SNOM Z}(b) and (c). The spectra retain the same structure as for a metallic probe ($\epsilon_\mathrm{tip}=\infty$).}
  \label{fig:Al2O3-Si s-SNOM Z}
\end{figure*}

Other than these controllable parameters, the scattering signal is also dependent on the dielectric function of the probe itself.
The calculation in the preceding discussion is done for a perfectly conducting probe, $\epsilon_{\mathrm{tip}} = \infty$.
In practice, near-field probes often have a Si core and a layer of metallic coating whose thickness $\sim 20\,\mathrm{nm}$ can be smaller than the skin depth, i.e., the electric field penetration length of the metal.
In this case, it may be more appropriate to
set $\epsilon_{\mathrm{tip}} = \epsilon_{\mathrm{Si}} \approx 11.7$
in Eq.~\eqref{eqn:Lambda_def}.
Repeating the calculations, 
we find that while qualitative features in the signal are retained, there are major quantitative differences
(Fig.~\ref{fig:Al2O3-Si s-SNOM Z}).

The discussion above shows that the rich structure of the s-SNOM signal found for the case of Al$_2$O$_3$ sample is susceptible to many experimental parameters.
(Retardation effects, discussed later in Sec.~\ref{sec:shape},
introduce further significant dependence on the probe geometry.)
This presents a serious challenge to realistic modeling of s-SNOM experiments.
On the other hand, these strong dependences arise only for highly crystalline material with low dissipation.
For other, less resonant materials, the modeling can be quite robust,
as discussed in Sec.~\ref{sec:shape}.

\section{Nonlocal reflection function}
\label{sec:q-depend}

The example material of the previous Section is a bulk crystal with a local (momentum independent) reflectivity function.
However, in many other systems studied through s-SNOM, including thin films, graphene, and multi-layered systems reflection is inherently nonlocal.
Thus, it is imperative to study how the $q$-dependence of the reflectivity affects the probe-sample interaction.
As mentioned in Sec.~\ref{sec:weak-local}, a general description of such interaction is challenging because the series representation of the polarizability
\begin{equation}
\chi = \sum\limits_{k}\frac{\mathcal{R}_k}{\lambda_k}
\end{equation}
has generalized eigenvalues $\lambda_k$ and residues $\mathcal{R}_k$ that are now complicated functionals of $r_\mathrm{P}$, 
cf. Eqs.~\eqref{eqn:sphrd_char_eqns_matrix} and \eqref{eqn:sphrd_char_eqns_beta}.
Still, we can attempt to analyze these expressions using the simple perturbation theory developed in Sec.~\ref{sec:weak-local}, in which $\lambda_k$ are computed from the poles of the $q$-independent theory, with corrections obtained by integrating the weighting functions over the momentum. 
As shown below, this scheme produces qualitative agreement with the calculated s-SNOM response for graphene on bulk $\rm{Al}_2\rm{O}_3$.

\begin{figure*}[htb]
  \begin{center}
    \includegraphics[width=\imagewidth]{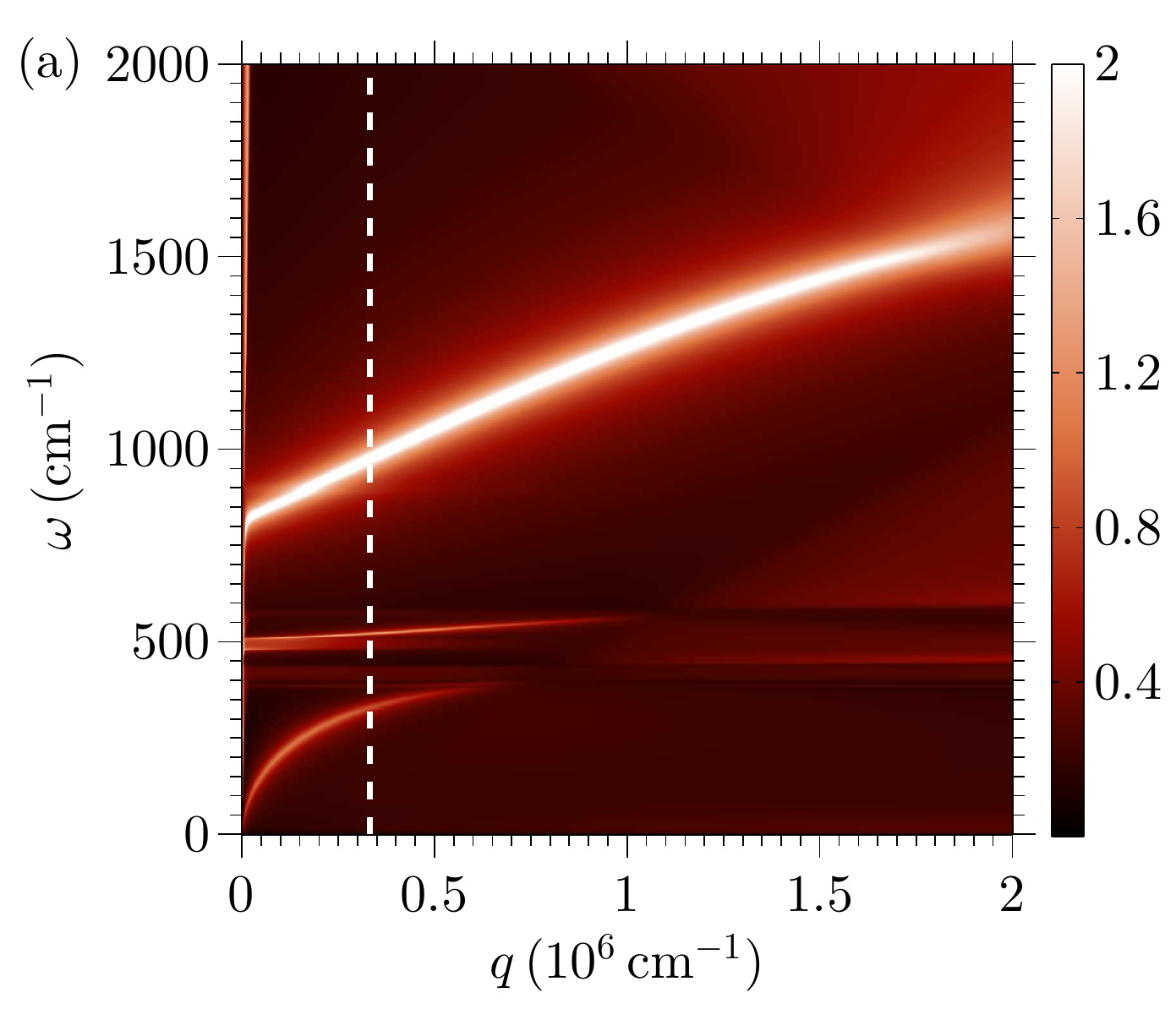}
    \includegraphics[width=\imagewidth]{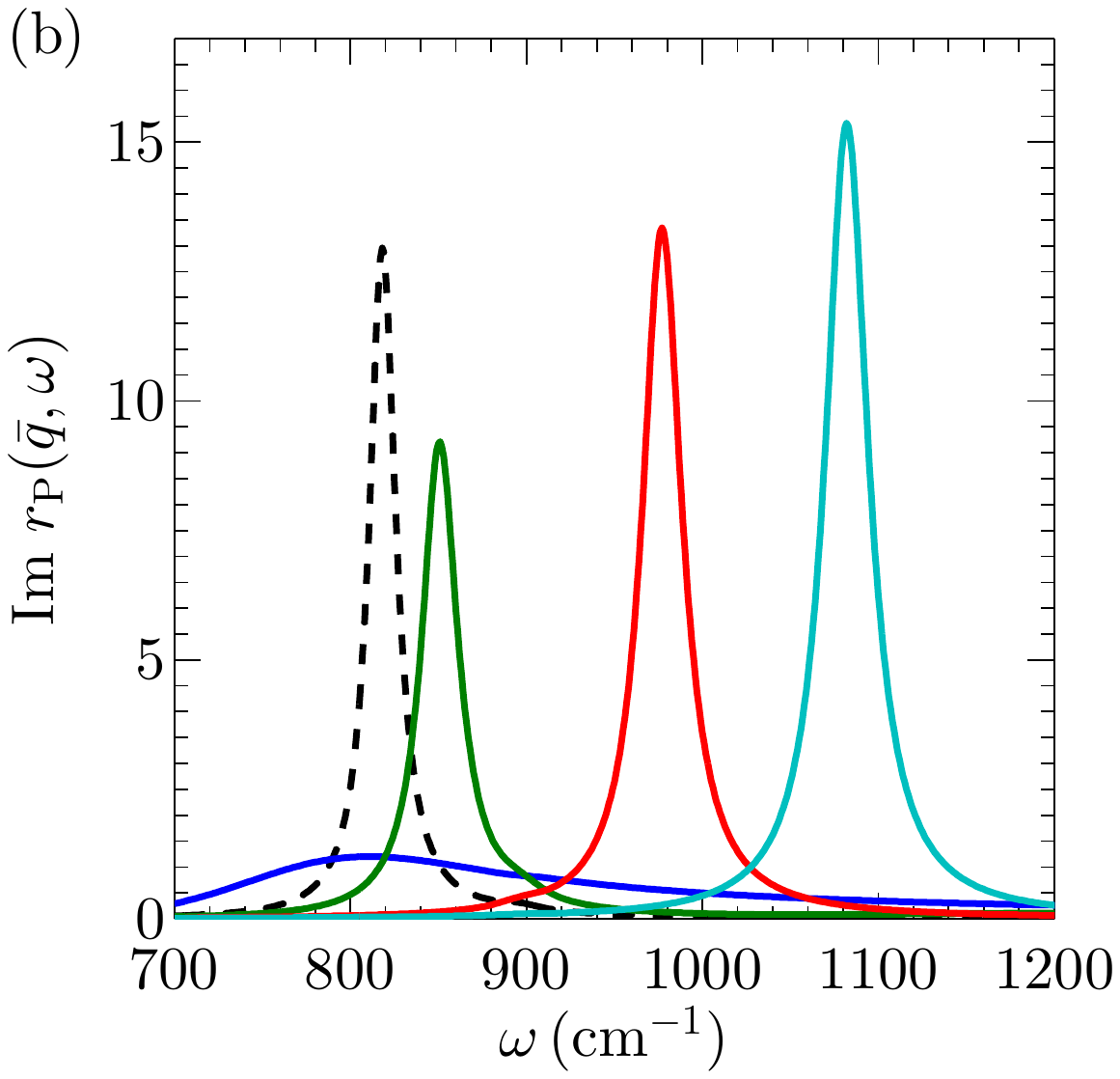}
    \\
    \includegraphics[width=\imagewidth]{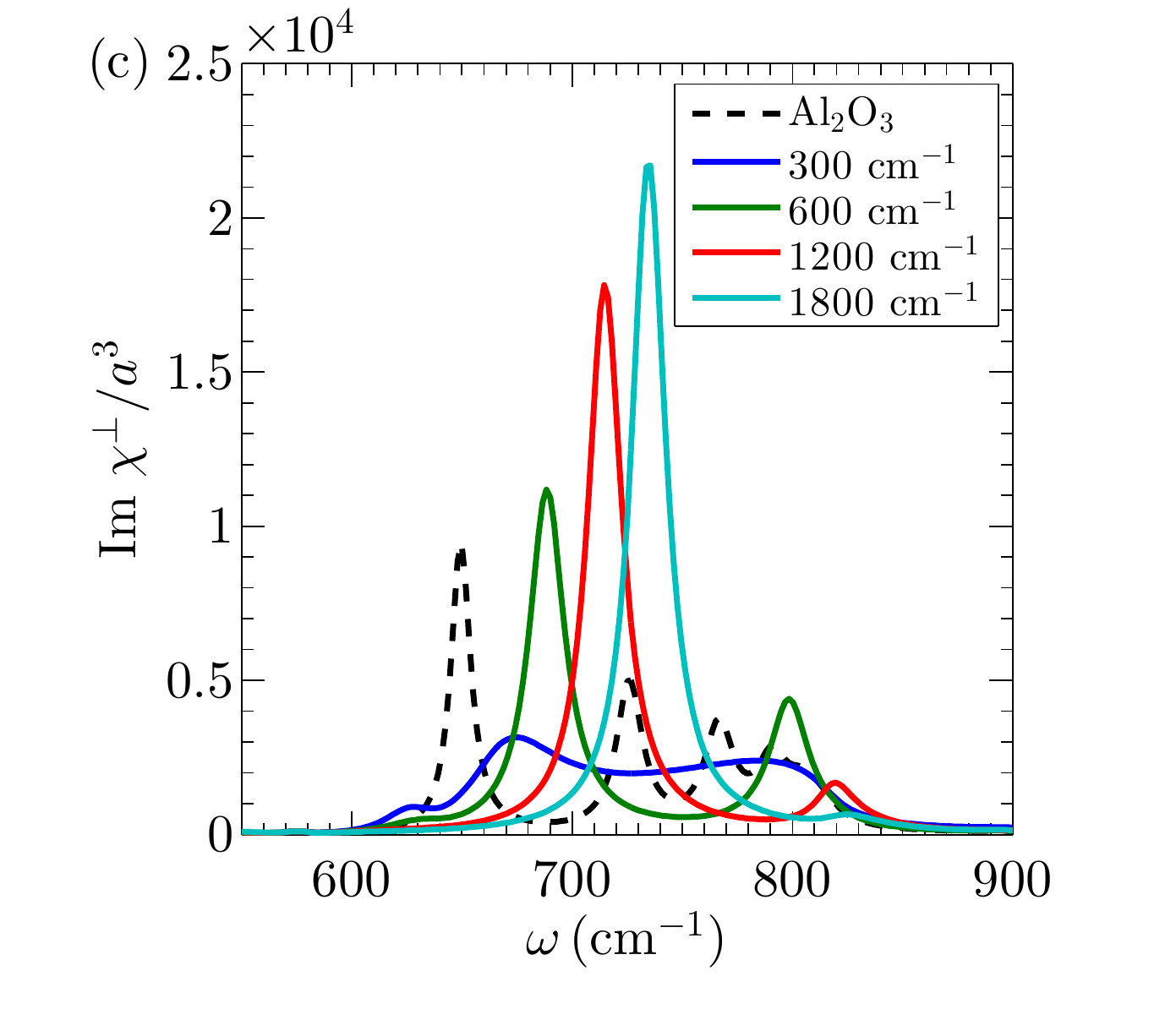}
    \includegraphics[width=\imagewidth]{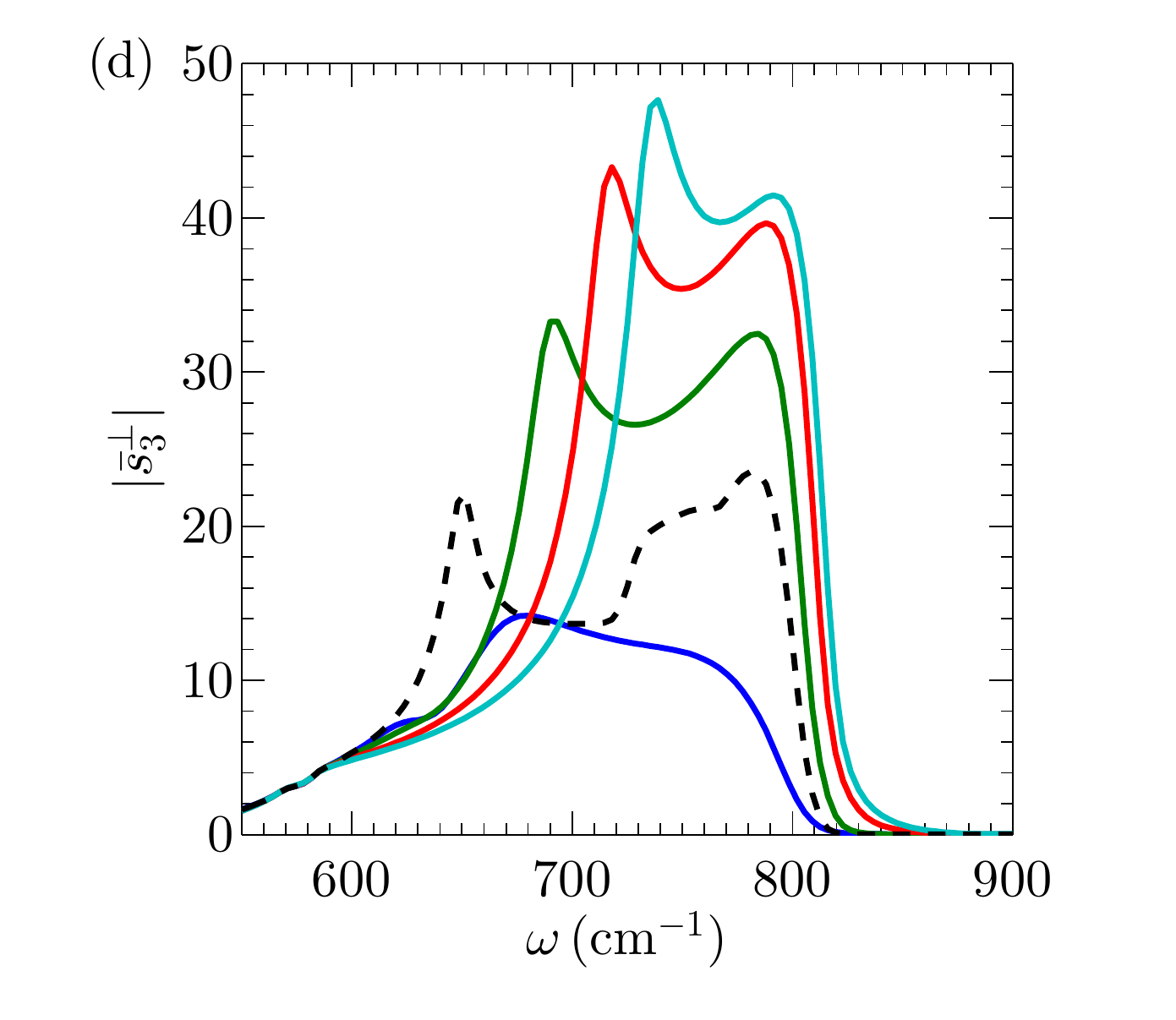}
  \end{center}
  \caption{(Color online) 
%
(a) Collective mode dispersion of graphene/Al$_2$O$_3$ system.
The mode repulsion between the graphene plasmon and the Al$_2$O$_3$ surface phonons are evident.
The false color stands for $\re \{[1 + r_\mathrm{P}(q,\omega)] q / k_0^z\}$,
which is a measure of power dissipation.~\cite{Ford1984eio}
This quantity is additionally raised to power $0.35$ to reduce the contrast.
The vertical dashed line marks $q = 1 / a$.
The faint curve just below $\omega=500\,\mathrm{cm}^{-1}$ is a weak surface phonon~\cite{Zeidler2013oco} that we do not discuss.
The chemical potential of graphene is $\mu = 1200\,\mathrm{cm}^{-1}$. 
(b) The solid curves are constant momentum $\bar{q} = 1/a$ line cuts through maps like (a) for several $\mu$.
The particular case of (a) is shown by the red curve (second solid curve from the right). 
The dashed curve is the same quantity computed for bulk Al$_2$O$_3$ without graphene.
(c) $\im \chi^\bot(\omega)$ and (d) $\bar{s}_3^\bot(\omega)$
computed using the $q$-dependent $ r_\mathrm{P}(q,\omega)$ at $\ztip=0.02a$ and $z_0=0.02a$, respectively.
Graphene chemical potentials $\mu$ for (b)--(d) are indicated in the legend of panel (c).
}
  \label{fig:GAl2O3_rP}
\end{figure*}

The Al$_2$O$_3$/graphene system has two collective modes (the upper and the lower one) that emerge from hybridization of the surface phonon of  Al$_2$O$_3$, originally at
$\omega_\mathrm{SP} \approx 750\,\mathrm{cm}^{-1}$
with the plasmon of graphene, $\omega(q) \propto \sqrt{\mu v_F q}$.
(Coupling of substrate phonons to graphene plasmons has been probed by s-SNOM experiments with graphene/SiO$_2$ systems.~\cite{Fei2011ino, Fei2012gto}
This and related work is reviewed in Ref.~\onlinecite{Basov2014cgs}.)
The modes share the optical weight and exhibit a level-repulsion that causes both to be dispersive.
Both features depend on the chemical potential $\mu$ of graphene.
Below we focus on the upper mixed mode and study its s-SNOM response for a range of $\mu$, and compare the results with the perturbation theory method.
To proceed, we need the formula for the reflectivity of the composite system.
This formula is well-known (see, e.g., Ref.~\onlinecite{Fei2011ino})
\begin{equation}
  r\tsub{P}(q, \omega) = \frac{\dfrac{\eps{1}}{\kz{1}} - \dfrac{\eps{0}}{\kz{0}}
  + \dfrac{4 \pi \sigma}{\omega} }
  {\dfrac{\eps{1}}{\kz{1}} + \dfrac{\eps{0}}{\kz{0}}
  + \dfrac{4 \pi \sigma}{\omega} } \,.
  \label{eqn:rP_q_omega_w_2d_sigma}
\end{equation}
Here $\eps{1} = \epsilon_\mathrm{eff}$ [Eq.~\eqref{eqn:beta_Al2O3}] is the permittivity of the lower half-space (Al$_2$O$_3$), $\eps{0} = 1$ is that of the upper half-space (vacuum), $\kz{j} = \sqrt{\eps{j} \frac{\omega^2}{c^2} - q^2}$ is the $z$-component of the wave vector in medium $j = 0, 1$,
and $\sigma = \sigma(q,\omega+i\tau^{-1})$ is the conductivity of graphene,
which we calculate within the random phase approximation~\cite{Wunsch2006dpo, Hwang2007dfs}
with a finite relaxation time $\tau^{-1}=25\,\mathrm{cm}^{-1}$.
For $q \gg \omega / c$, one finds $k^z_j\simeq iq$ and Eq.~\eqref{eqn:rP_q_omega_w_2d_sigma} reduces to
\begin{equation}
  r\tsub{P}(q, \omega) = \frac{\eps{1}  - 1
  +  4 \pi q \dfrac{i \sigma}{\omega} }
  {\eps{1} + 1 
  + 4 \pi q \dfrac{i \sigma}{\omega} } \,, 
  \label{eqn:rP_q_omega_w_2d_sigma_nf}
\end{equation}
which can be compared to Eq.~\eqref{eqn:beta_Al2O3}.
A convenient way to visualize the dispersion of the collective modes
is to plot the imaginary part of $r\tsub{P}(q, \omega)$,
which represents
the power dissipation in the system,~\cite{Ford1984eio} as a false-color map.
An example for $\mu=1200\,\mathrm{cm}^{-1}$ is shown in Fig.~\ref{fig:GAl2O3_rP}(a). 
In the low-$q$ regime ($\hbar v_F q \ll \hbar\omega \ll \mu$),\cite{Wunsch2006dpo,Hwang2007dfs} the lower bright curve is mainly the plasmon with dispersion $\omega \propto \sqrt{\mu v_F q}$, 
while the upper bright curve represents the dispersionless Al$_2$O$_3$ surface phonon.
(The additional bright curve around $\omega = 500\,\mathrm{cm}^{-1}$ is a weaker Al$_2$O$_3$ surface phonon, which we do not discuss.)
An increase in $\mu$ leads to a steeper dispersion of the plasmon,
which causes both hybrid modes to go up in frequency.
Decreasing $\mu$ has the opposite effect.
Additionally, if $\mu$ drops below $\hbar\omega_\mathrm{SP} /\,2 \approx 380\,\mathrm{cm}^{-1}$, the upper mode falls into the interband transition region of graphene, which results in strong damping of the surface phonon.
As we will see below,
this causes the $\mu=300\,\mathrm{cm}^{-1}$ curve
to look qualitatively different from the rest in Fig.~\ref{fig:GAl2O3_rP}(b).
Let us now discuss how the collective modes manifest themselves
in the s-SNOM response.

In the simplistic picture of the s-SNOM response,
the probe-sample interaction is dominated by a single momentum $\bar{q} = 1/a$.
If this assumption were accurate,
we could set $r_\mathrm{P}(\bar{q},\omega)$ as $\beta(\omega)$ and calculate the response using the set of poles and residues established previously.
We would then see peaks in the response generated by the upper hybrid mode.
However, this crude approximation leads to higher peak frequencies than the calculation using the full $r_\mathrm{P}(q,\omega)$,
as seen in Figs.~\ref{fig:GAl2O3_rP}(b), \ref{fig:GAl2O3_rP}(c), and \ref{fig:GAl2O3_rP}(d).
Indeed, we have shown in Sec.~\ref{sec:weak-local} that when the $q$-dependence in reflection is treated as a perturbation,
each mode has its own range of sensitive momenta due to the inherent length scales in its potential distribution.
The distributions change with an additional length scale --- the tip-sample distance $\ztip$, so that the momentum weighting functions are dependent on $\ztip$ as well, $G_k=G_k(q,\ztip)$.
For each mode, these functions provide a means to average over momentum and find an effective $q$-independent sample reflection $\beta_k^\mathrm{eff}(\omega)$, cf.~Eq.~\eqref{eqn:beta_eff_k}, so that we can again apply the established pole-residue decomposition.
Strictly speaking, the perturbative method cannot be applied here as the mixed mode may be strongly $q$-dependent.
Even so, we find a very reasonable agreement with the computed signal in the range of graphene chemical potentials $\mu=600$--$1800\,\mathrm{cm}^{-1}$ that we study.
We first consider peak frequencies in $\im \chi^\bot$, which can be predicted by invoking the resonance condition $\re \beta_k^\mathrm{eff}=\beta_k$.
For the lowest mode $k=0$ and $\ztip=0.02\,a$,
there is a systematic overestimate of the peak position by $20$--$30\, \mathrm{cm}^{-1}$ for $\mu=600$--$1800\,\mathrm{cm}^{-1}$.
The discrepancy is 
larger for higher $\mu$ at which the $q$-dependence of the upper hybrid mode is stronger.
This discrepancy is due in part to the well-known general tendency of the first-order perturbation theories to overestimate the lowest eigenvalues.
Next, for the $k=1$ mode, the resonance condition is satisfied only for $\mu=600\,\mathrm{cm}^{-1}$ at $\omega=797$ cm$^{-1}$ and $\mu$=1200 cm$^{-1}$ at $\omega=823$ cm$^{-1}$, which agree well with the smaller peaks in $\im \chi^\bot$. At these frequencies $\im \beta^\mathrm{eff}_1$ are larger than the $k=0$ case and the peaks have smaller magnitudes. For $\mu=1800$ cm$^{-1}$, the resonance condition is not met and the very small peak at $\omega=827$ cm$^{-1}$ in $\im \chi^\bot$ corresponds to where $\re \beta^\mathrm{eff}_1$ is largest and thus closest to $\beta_1$. Finally, for $k>1$, $\beta_k$ is larger than $\re \beta^\mathrm{eff}_k$ for all frequencies and no peaks in $\im \chi^\bot$ are found.
Seeing qualitative agreement in the polarizability, we proceed to analyzing the demodulated signal.

As inferred in Sec.~\ref{sec:result}, the demodulated signal is strongest near the peaks in $\chi^\bot(z_0,\omega)$ and $\chi^\bot(z_0+2\Delta z,\omega)$, where each peak is attributed to a resonant mode.
For the dominant $k=0$ mode,
we find a set of corresponding peaks in $s_3^\bot(z_0,\omega)$ at the same frequencies as those in $\chi^\bot(z_0,\omega)$, as shown in Fig.~\ref{fig:GAl2O3_rP}(c) and Fig.~\ref{fig:GAl2O3_rP}(d).
For the other set of peaks in the $s_3$ spectra we must consider how the situation is changed at $z_0+2\Delta z$.
At such distances $\ztip$ itself becomes the primary length scale and the sensitivity function $G_k$ is shifted toward smaller momentum, where the upper mode has a flatter dispersion and its frequency is close to $\omega_\mathrm{SP}$ of the bulk Al$_2$O$_3$ crystal.
Therefore, this set of peaks should all appear near $\omega_\mathrm{SP}$, which is indeed the case.
Repeating this procedure for the $k=1$ mode, we find that the peaks it contributes are inseparable from the set of higher frequency peaks produced by the $k=0$ mode as both  have frequencies very close to $\omega_\mathrm{SP}$.
Its contributions, however, alter the heights of these peaks.
For instance, the $k=1$ peak is strongest  in $\chi^\bot(z_0,\omega)$ for $\mu=600\,\mathrm{cm}^{-1}$ (among the four we used), so the high frequency peak in $s^\bot_3$ for this chemical potential has the largest relative magnitude with respect to the low frequency peak.
Thus we conclude the demodulated s-SNOM signal can be qualitatively explained by the perturbative method, albeit with inaccuracy in the lower peak frequency.
However, as we argued in Sec.~\ref{sec:result}, the lower frequency peak in the demodulated signal is mainly an artifact of the finite $z_0$ we are forced to use.
If $z_0$ were truly zero, only the peak near $\omega_\mathrm{SP}$ would survive.

\section{Model-dependent effects}
\label{sec:shape}
\begin{figure*}[tbh]
	\begin{center}
		\includegraphics[height=2.2 in]{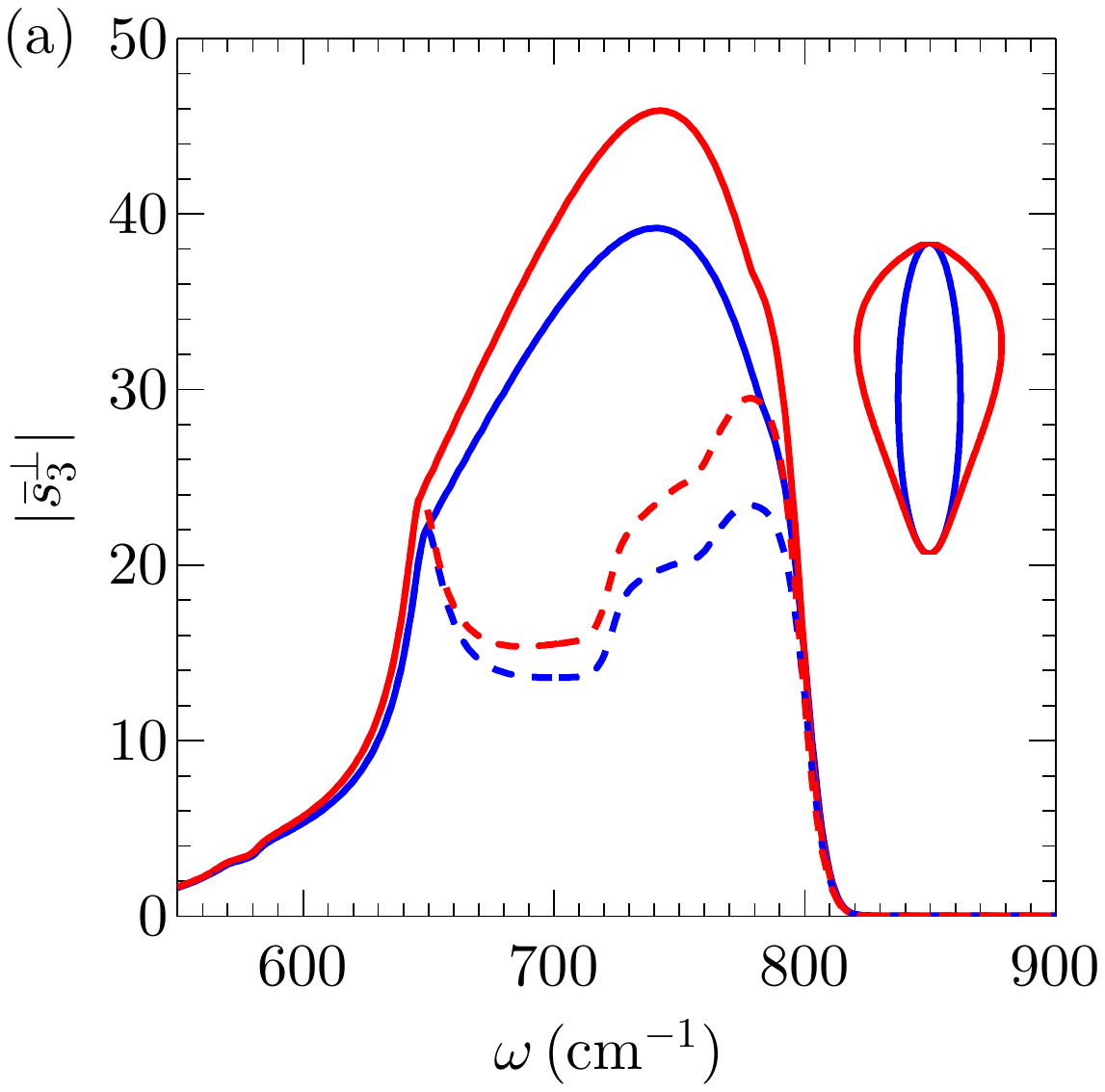}
		\includegraphics[height=2.2 in]{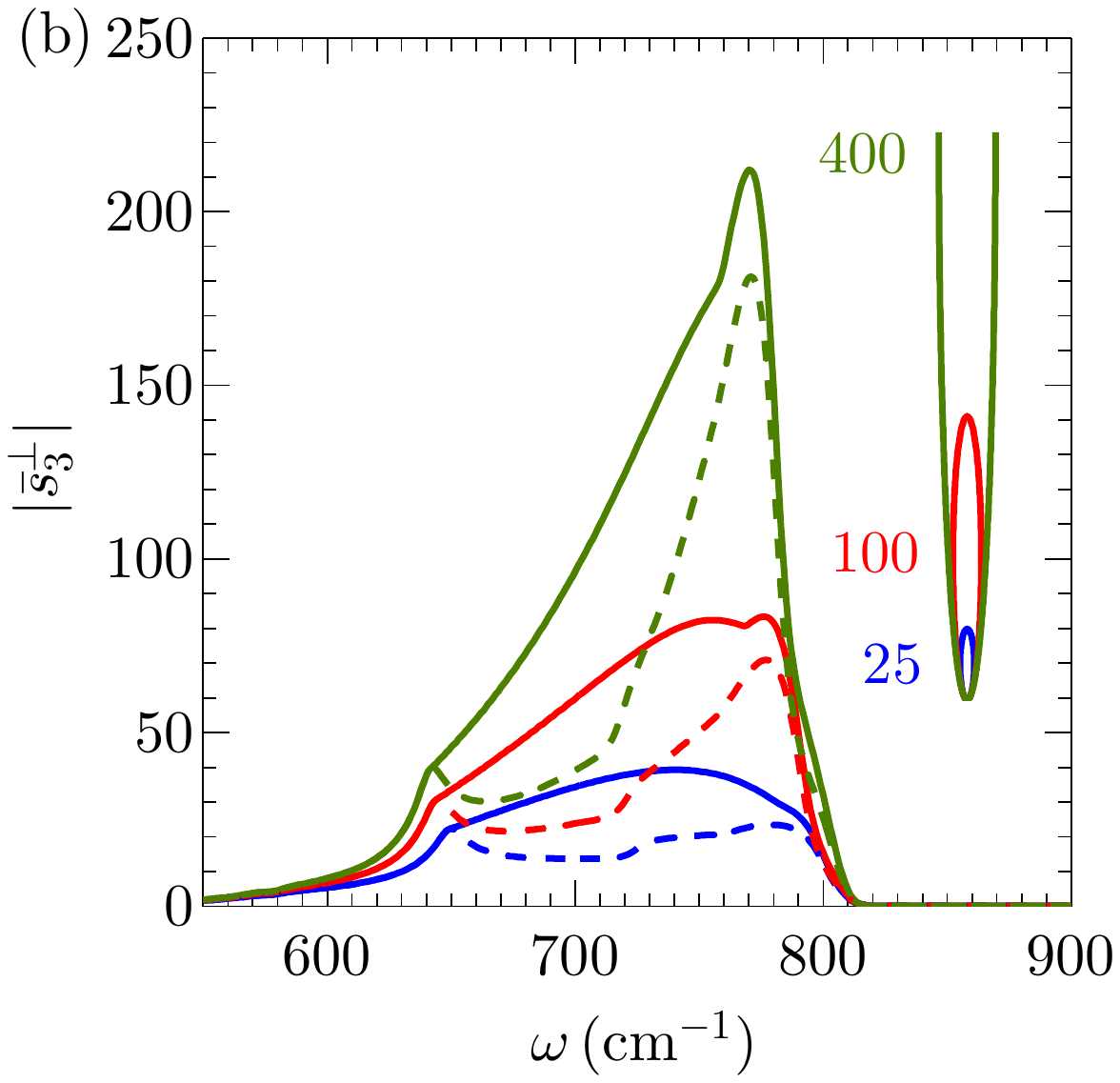}
	\end{center}
\caption{(Color online) (a) The s-SNOM signal $s_3^\bot$ computed for Al$_2$O$_3$
samples.
The inset shows the probe shapes used (spheroidal, with $L=25a$, and a pear-shaped). 
The two types of probes produce qualitatively similar but quantitatively different results.
(b) Spheroids of longer length have drastically increased signal strength.
The inset depicts the probe shape and the values of $L/a$ used. 
Note that this quasistatic calculation neglects radiative damping and antenna resonances, see Sec.~\ref{sec:shape}.
If included, such effects are expected to greatly reduce $s_3^\bot$. 
In all cases $\Delta z=50\,\mathrm{nm}$ and $a=30\,\mathrm{nm}$.
The value of $s_3$ is taken either from the maximum of the approach curves at each frequency (solid lines) or at the closest approach distance $z_0=0.6\,\mathrm{nm}$ (dashed lines). 
 }
\label{fig:shape_comp}
\end{figure*}

The spheroid model differs from real s-SNOM probes in two important ways: i) the real probe resembles an inverted pyramid, ii) at infrared wavelengths, the length $\sim 10\,\mu$m of the probe exceeds several times the diameter $c / \omega$ of the radian sphere.
In previous literature it was assumed that
these differences can all be neglected as the probe-sample interaction is focused around the apex of the probe [Fig.~\ref{fig:modes}(b)],
while contribution from the rest of the probe is canceled out during the process of demodulation and normalization.
Hence, the exact shape of the probe is unimportant and the only relevant physical quantity is the apex radius of curvature $a$.
Further, since the characteristic length scale $a$ is well within the radian sphere, a quasistatic description should suffice.
This simplistic argument is backed by previous agreement between the spheroid model and experiment.~\cite{Dai2014tsp,Fei2012gto,Zhang2012nfs}
However, we have shown that different probe shapes exhibit universal behavior only when $\ztip/a$ is of the order of a few percent (cf. Fig.~\ref{fig:pz_beta_resdu}(a)). This range is much smaller than typical tapping amplitudes, so the majority of the s-SNOM response lies outside the universality regime and should indeed be probe shape dependent.
Additionally, recent experiment and modeling have shown that a quasistatic formalism with \textit{ad hoc} probe shapes is insufficient for highly resonant materials such as on silicon carbide.~\cite{McLeod2014lrm}

In this Section we re-examine these issues by examining 
two materials, the highly resonant Al$_2$O$_3$ and the dissipative SiO$_2$, and study the probe shape dependence of their response as well as electrodynamic corrections.
We find that for dissipative materials shape dependence is weak and 
retardation effects are of less importance, so the spheroid model describes the s-SNOM experiment reasonably well.
This explains the success of our model in reproducing the response of various materials in experiment.
On the other hand, we find the response of resonant materials to be highly dependent on the probe shape and less well described within the quasistatic approximation.
For such materials a full electrodynamic treatment with the exact probe shape may be required. 
Common numerical methods suitable for 
electrodynamic treatment of light scattering by a spheroid near a surface include
$T$-matrix method~\cite{Wriedt1998lsp, Doicu1999ctm} and BEM.~\cite{McLeod2014lrm} 
For the case of a sphere near a surface, the calculation of necessary matrix elements can be done efficiently using recursion technique similar to what we use here.\cite{Bobbert1986lss}

We consider the probe shape dependence and the retardation effects separately.
To study the former,
we simulated the s-SNOM signal of Al$_2$O$_3$ samples
obtained with spheroidal probes of different length.
We also calculated (using BEM) the results for pear-shaped probes that may better mimic the inverted pyramids.
As shown in Fig.~\ref{fig:shape_comp}(a),
the signal for a pear-shaped probe is qualitatively similar to that for
the spheroid of the same length, but there are quantitative differences.
For spheroids, we find that the signal strongly increases and the peak frequencies steadily decrease as the length of the probe increases at a fixed apex radius,
as shown in Fig.~\ref{fig:shape_comp}(b).
These features can be explained by the scale invariance of the problem.
It implies that an increase in probe length is equivalent to a simultaneous decrease in tapping amplitude and the apex radius.
The decrease in radius produces changes in both the poles and residues.
The former explains the shift in peak frequencies.
The latter is mostly canceled out by normalization.
In turn, the decrease in tapping amplitude leads to a larger contrast between the sample and the reference as discussed in Sec.~\ref{sec:result}
[see Fig.~\ref{fig:chi3_spec_dz}(b)], so the signal strength is dramatically increased.

\begin{figure*}[tbh]
	\begin{center}
		\includegraphics[width=\imagewidth]{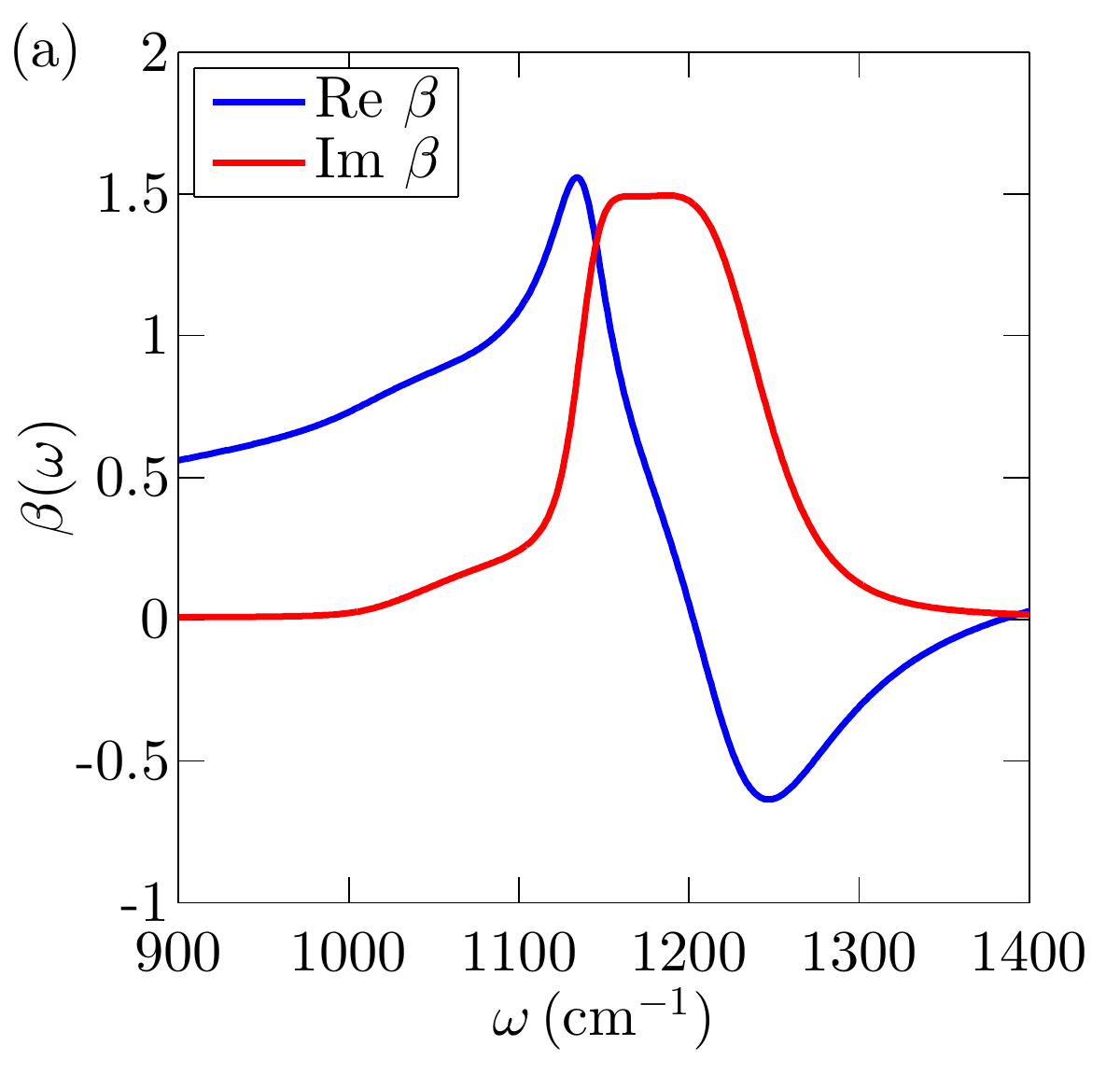}
		\includegraphics[width=\imagewidth]{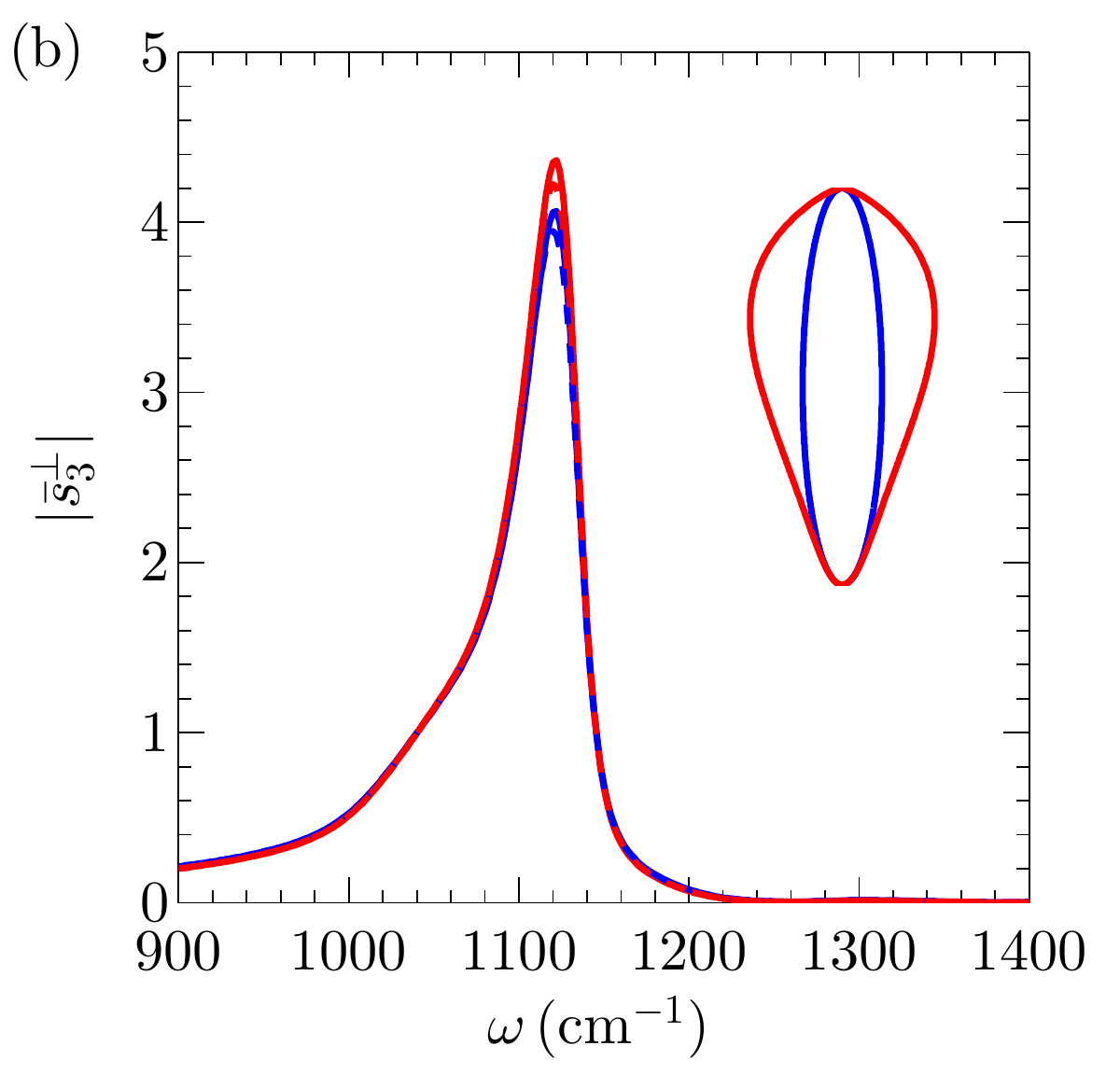}
		\includegraphics[width=\imagewidth]{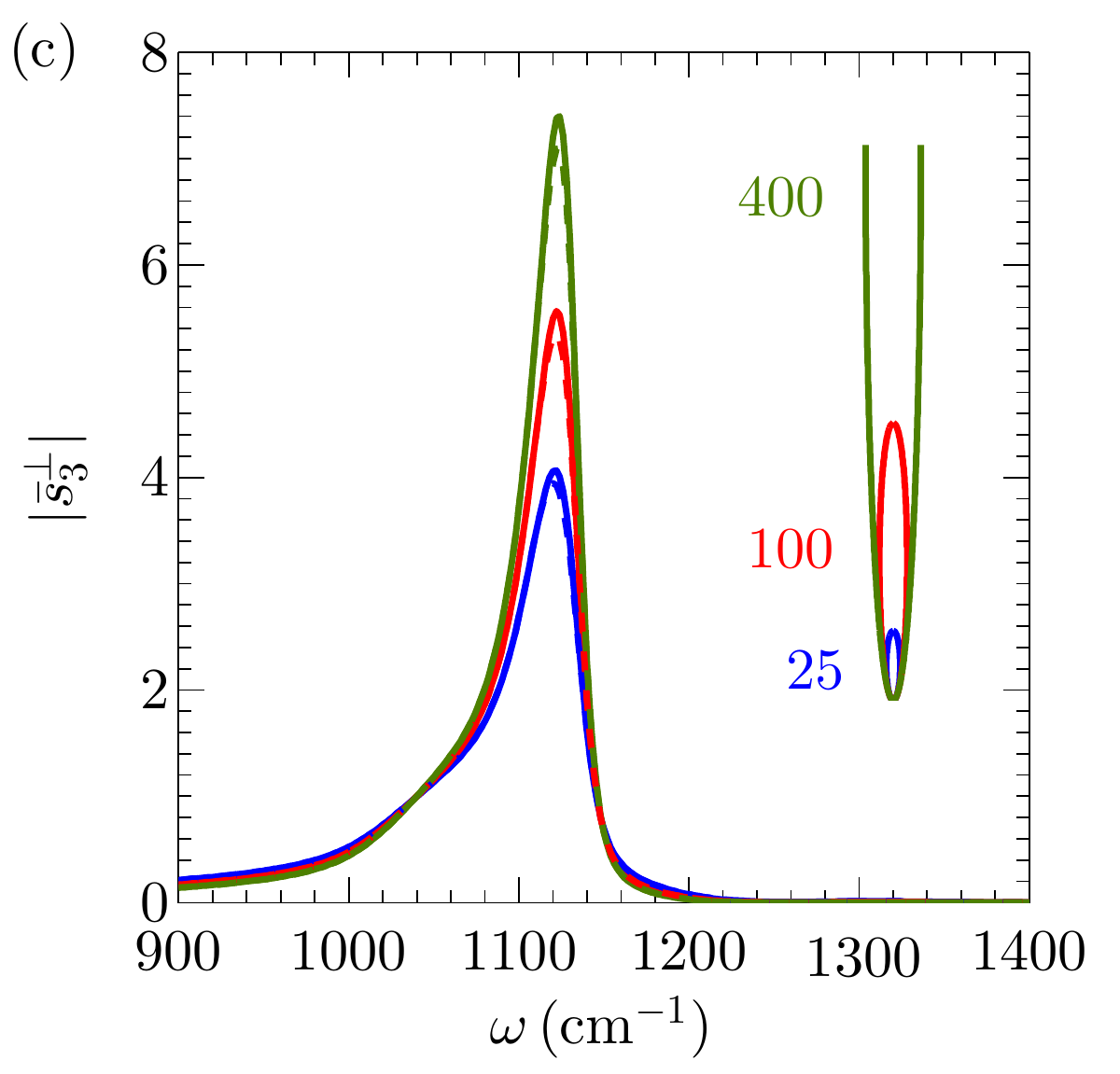}
	\end{center}
\caption{(Color online) (a) The reflection coefficient of SiO$_2$~\cite{Kucirkova1994iit} has a larger imaginary part than Al$_2$O$_3$ due to its inherent dissipation, leading to a 
weaker shape dependence in the s-SNOM signal. 
(b) The signal of the pear-shaped probe is very close to that produced by the spheroid.
(c) Increasing the probe length leads to a much smaller increase in the signal strength.
The overall shape of the spectrum is also preserved.
All geometric parameters are the same as in Fig.~\ref{fig:shape_comp}.}
\label{fig:SiO2}
\end{figure*}

The strong probe-shape dependence found above seem to suggest that theoretical modeling
of the s-SNOM experiments must always be done using the actual shape to be reliable.
In fact, such a sensitivity to the probe shape pertains only to the highly-resonant, i.e., large $\beta$ materials.
In Al$_2$O$_3$ this parameter reaches the maximum value of $|\beta| \approx 12$,
Fig.~\ref{fig:Al2O3_beta}(a).
For comparison, in Fig.~\ref{fig:SiO2}(b) and Fig.~\ref{fig:SiO2}(c), we show that the pear-shaped probe and the spheroid produced almost identical signals for amorphous SiO$_2$, a material with $|\beta| \leq 1.5$.
(For experimental studies of this material see, e.g.,
Refs.~\onlinecite{Amarie2011bia, Zhang2012nfs}.)
In this case, a factor of $16$ increase in the probe length leads to only a doubled signal strength, compared to a nearly tenfold increase for Al$_2$O$_3$ seen in Fig.~\ref{fig:shape_comp}(b).

The results above are obtained within the quasistatic approximation.
In reality, a probe half-length of $200a$ already exceeds the diameter $c / \omega$ of the radian sphere and one has to consider retardation effects.
Naively, contributions from such effects should be eliminated by demodulation, as they pertain to 
a length scale much larger than the tapping amplitude.
However, we show that one contribution --- the radiative damping --- survives demodulation.
The radiative damping has an effect similar to a finite $\mathrm{Im}\, \beta$, i.e., the dissipation in the sample.
Hence, for dissipative materials one can neglect radiative damping and still find reasonable agreement with experiment, while doing so for highly resonant materials may lead to qualitatively wrong results.
Let us illustrate these statements using the simplest model for the probe --- the point dipole.
The electrodynamic interaction between the dipole and the sample with the dielectric constant $\epsilon$ is given~\cite{Ford1984eio} by a modified version of Eq.~\eqref{eqn:g},
\begin{equation}
g^{\nu} = c^{\nu} \zfint 
\frac{i q^3}{k^z_0(q)}\,
\frac{\epsilon \kz{0}(q) -  \kz{1}(q)}{\epsilon \kz{0}(q) +  \kz{1}(q)}\,
 e^{-2q\ztip} d q \,,
\label{eqn:g_ed}
\end{equation}
where the second fraction in the integrand is the full form of 
the reflectivity $r_{\text{P}}(q, \omega)$. [It is obtained from Eq.~\eqref{eqn:rP_q_omega_w_2d_sigma} by setting $\sigma$ to zero.]

Suppose $\omega$ and $\ztip$ are fixed, then the above integral defines $g^{\nu}$ as a function of $\epsilon$, which is generally a complex number.
Alternatively, $g^{\nu}$ is a function of $\beta = (\epsilon - 1)/(\epsilon + 1)$.
The integration domain Eq.~\eqref{eqn:g_ed} includes momenta $q$ both inside and outside the light-cone.
The radiative damping effect arises from the integration over former, i.e., the momenta $q < \kz{0}$.
This part of the integral yields a negative imaginary contribution to $g^{\nu}$, which
shifts the pole of $\chi^{\nu}$ [Eq.~\eqref{eqn:chi_eff_pdp_q}] to the lower complex half-plane of $\beta$.
The real parts of the poles also change but this is less conceptually important, see below.
Consider now the remaining part of the integral, over momenta $q > \kz{0}$.
It is easy to see that if $\epsilon = -q^2 / (q^2 - \omega^2 / c^2)$, then
\begin{equation}
\epsilon \kz{0}(q) +  \kz{1}(q) = 0\,,
\label{eqn:surface_phonon_conditions}
\end{equation}
so that there is a pole on the integration path.
As a result, functions $g^{\nu}$ and $\chi^{\nu}$ have
branch cuts at $\epsilon \in (-\infty, -1]$ in the complex $\epsilon$ plane or equivalently at $\beta \in [1,\infty)$ in the complex $\beta$ plane.
These additional features are shown schematically in Fig.~\ref{fig:Al2O3_beta}(c).
The physical origin of both the poles and the branch cut is quite clear.
The discrete poles has been discussed at length in this article.
They correspond to the polariton modes localized near the tip,
Fig.~\ref{fig:modes}(b).
In turn, the branch cut corresponds to the continuum of \textit{delocalized} surface polaritons that exist without the probe. Indeed,
Eq.~\eqref{eqn:surface_phonon_conditions},
is the well-known equation for the spectrum of such excitations.~\cite{Ford1981eeo}

Of the two features, the branch cut is not expected to affect the signal as the small-momentum contribution is greatly diminished by demodulation.
Demodulation should also make less important the change in the real parts of the poles,
because these real parts vary greatly with $\ztip$ on account of the tapping motion of the probe.
However, the shift of the discrete poles away from the real axis is a qualitative change and its effects remain after demodulation. 
Our next objective is therefore to find this shift for the case of the spheroidal probe.

A free standing spheroid has an effective polarizability given by 
\begin{equation}
\chi_{0,\mathrm{eff}} = 
\frac{\chi_0}{1-i\frac23(\frac{\omega}{c})^3\chi_0}
\end{equation}
to the lowest order in $\omega/c$ when radiative correction is considered.~\cite{Wokaun1982rds,Moroz2009dfs}
Modifying $\Lambda$ accordingly [cf. Eq.~\eqref{eqn:R_k}], it is easily shown that this formula applies to our geometry as well.
Namely, the s-SNOM polarizability corrected for the
radiative damping is given by
\begin{equation}
\chi^\nu_{\mathrm{rad}} = \frac{\chi^{\nu}}{1 - i\frac23(\frac{\omega}{c})^3 \chi^{\nu}}\,,
\quad
\chi^{\nu}
  = \zfsum{k} \frac{R_{k}^{\nu}}{\beta_{k}^{\nu} - \beta} \,.
\label{eqn:chi_eff_ed}
\end{equation}
Viewed in the complex $\beta$ plane, this correction is equivalent to the shift of the poles $\beta_k$ into the lower half-plane by $-i(2 / 3)(\omega/c)^3 R_k$ (to the leading order in $\omega / c$).
Therefore, both the radiative damping and the intrinsic dissipation in the sample play a similar role: they increase the distance from the poles to the curve traced by the surface reflectivity $\beta$ of the sample as $\omega$ varies [Fig.~\ref{fig:Al2O3_beta}(c)].
For a dissipative material, the curve begins far from the poles,
and so further increase in the distance produces little change.
Conversely, for highly resonant materials the $\beta(\omega)$ curve passes close to the real axis, and so radiative damping may obscure or eliminate the fine features of the signals, such as multiple resonant peaks discussed in Sec.~\ref{sec:result}.
It is worth noting however that while it may be important for s-SNOM in infrared or visible domains, the radiative damping should be rather weak in the (experimentally more challenging) terahertz range, where typical s-SNOM probes would fit well inside the radian sphere.

Finally, a class of retardation effects we have not addressed here are antenna resonances arising when the length of the probe exceeds several times the diameter of the radian sphere.
They give rise to additional peaks in the s-SNOM signal as a function of $\omega$.
For most materials such resonances are removed once the s-SNOM signal is normalized to a reference sample; however, for strongly resonant materials such as SiC and presumably also Al$_2$O$_3$ we studied here, the cancellation is not complete.~\cite{McLeod2014lrm}

\section{Discussion and Conclusion}
\label{sec:Conclusions}

Further progress in the s-SNOM and related areas of near-field microscopy requires a quantitatively reliable procedure for determining the fundamental response function $r_{\text{P}}(q, \omega)$ from the amplitude and phase of the s-SNOM scattering data, from which one can proceed to the next step of inferring the optical constants of the studied sample.
Typically, materials with a higher absolute value of $r_{\text{P}}(q, \omega)$ produce a higher amplitude s-SNOM signal.
However, the peaks in the s-SNOM signal  are often red-shifted with respect to those in $|r_{\text{P}}(q, \omega)|$ or  $\text{Im}\, r_{\text{P}}(q, \omega)$.

Given additional information about the system, these inverse problems can be tackled by fitting the experimental data to the solution of the direct problem with a trial form of $r_{\text{P}}(q, \omega)$ as the input.~\cite{McLeod2014lrm}
Unfortunately, the direct problem is also difficult to solve.
The three-dimensional nature of this problem and the presence of widely different length scales make realistic simulations~\cite{Porto2003rse, Esteban2006soo, Esteban2009fso} of s-SNOM experiments very computationally intensive.
This led to popularity of simple \textit{ad hoc} approximations known as the point-dipole~\cite{Hillenbrand2000coc, Taubner2004nrt, Aizpurua2008sei} and the finite-dipole model,~\cite{Ocelic2007qnf, Amarie2011bia, Amarie2011ebi, Hauer2012qam} in which the actual charge distribution induced on the probe is approximated by a point-like image dipole or a combination thereof with additional point charges. 

The point-dipole model~\cite{Keilmann2004nfm} postulates
that Eqs.~\eqref{eqn:beta_0_dip_perp}--\eqref{eqn:g} that are rigorous in the asymptotic long-distance limit $\ztip \gg L$
remain qualitatively correct at much shorter $\ztip$ if the input physical parameters are suitably renormalized.
Thus, the bare polarizabilities $\chi_0^{\nu}$ become the adjustable parameters of the model. 
It is customary to assume that the in-plane polarizability $\chi_0^{\parallel}$ is negligible compared the out-of-plane one, which is taken to be
\begin{equation}
\chi_0^{\bot} = a^3,
\label{eqn:chi_0_pdp}
\end{equation}
where $a$ is of the order of the curvature radius of the tip.
Another adjustable parameter~\cite{Renger2005rls, Fei2011ino} $b \lesssim 1$ specifies the position of the effective dipole inside the probe:
\begin{equation}
\zp = b a + \ztip\,.
\label{eqn:z_tip_pdp}
\end{equation}
Clearly, the point-dipole model accounts only for the sharp tip and ignores the body of the probe, 
as $\chi^\nu$ for the point-dipole in Eq.~\eqref{eqn:chi_eff_pdp_q} 
is much smaller than $\chi_0^\nu$ for a tip with $L \gg a$.
If the point-dipole model were literally correct,
the radiating dipole of the probe in typical s-SNOM experiments would be so small that no measurable signal would be observed.

The finite-dipole model improves upon the point-dipole one by including the missing antenna-like enhancement approximately.
It assumes that the electric field of a spheroidal probe of length $2L$ is equivalent to that of several point charges of total zero charge that are positioned inside the spheroid near both of its ends.
For small $\ztip / L$, this model~\cite{Cvitkovic2007amf,
Amarie2011ebi} yields the following functional form of the probe polarizability:
\begin{equation}
	\chi\tsup{fdp} =
  \text{const} + \frac{R_{0}\tsup{fdp}}{\beta_{0}\tsup{fdp} - \beta}
	\eqc
	\beta_{0}\tsup{fdp} \approx 1.4 + O\! \left(\frac{\ztip^3}{L^3}\right) \,,
	\label{eqn:chi_eff_fdp}
\end{equation}
where $R_{0}\tsup{fdp} \propto a L^2$.
The finite-dipole model was shown to give a good qualitative agreement with s-SNOM data obtained for quartz, amorphous SiO$_2$, and SiC samples once
parameters $R_{0}\tsup{fdp}$ and $\beta_{0}\tsup{fdp}$
are suitably adjusted.~\cite{Amarie2011bia}
Thus, the best fit to the data was achieved choosing the length $2L = 600\,\mathrm{nm}$ of the probe, which is about one third of the diameter $c/\omega  \approx 1700\,\mathrm{nm}$ of the radian sphere.
Interestingly, this is approximately the value of $2L$ in the quasistatic calculation for which one obtains, in the case of SiO$_2$ sample, the same result for $s_3$
as one gets from the full electrodynamic calculation for a probe of a realistic (much longer) length.~\cite{McLeod2014lrm}

\begin{figure}[t]
\begin{center}
  \includegraphics[width=\imagewidth]{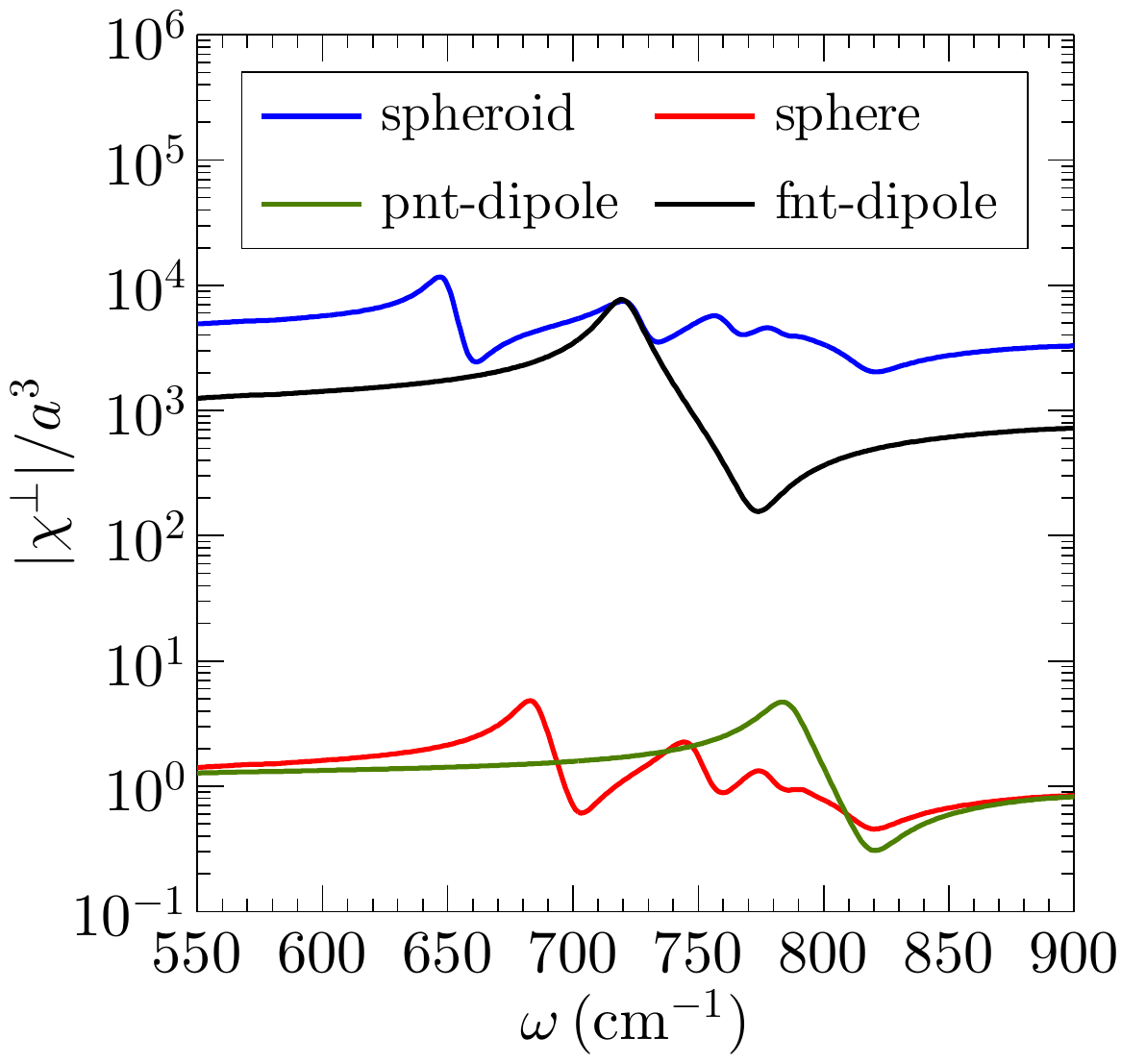}
\end{center}
\caption{(Color online) The spectrum of the probe polarizability $|\chi^{\bot}|$ for Al$_2$O$_3$ sample according to four different models. 
The point- and the finite-dipole models each predict a single peak in $|\chi^\bot|$.
The calculations for spherical and spheroidal probes reveal multiple peaks.
The sphere and the point-dipole models produce $\chi^\bot / a^3 \sim 1$.
The $L = 25 a$ spheroidal probe yields $\chi^\bot \sim 10^3$--$10^4$;
the finite-dipole of the same $L$ gives about an order of magnitude lower $\chi^\bot$.
These dramatic differences in both the form and the absolute magnitude of $\chi^\bot$ can however be significantly reduced in the usually reported $\bar{s}^\bot_3$, the normalized demodulated signal.
}
  \label{fig:Al2O3_models_compare}
\end{figure}

Agreement with the data notwithstanding,
from the theory point of view Eq.~\eqref{eqn:chi_eff_fdp} is unsatisfactory on at least three counts.
First, $R_{0}\tsup{fdp}$ does not follow the correct scaling $L^3 / \ln L$ as a function of $L$, thus underestimating the probe polarizability.
Second, the constant term in Eq.~\eqref{eqn:chi_eff_fdp} violates the general requirement that $\chi \to 0$ as $\beta \to \infty$, corresponding to the case when the applied field is screened completely by the induced charges in the sample.
Third, $\beta_{0}\tsup{fdp}$ goes to $\sim 1.4$ when $\ztip = 0$.
Instead, all smooth probe shapes must behave as a sphere at $\ztip \ll a$, and therefore yield $\beta_{0} = 1$ at $\ztip = 0$.
The fact that finite-dipole model violates these general requirements suggest its limited usability.
Figure~\ref{fig:Al2O3_models_compare} is an illustration of how widely different the predictions of the four discussed s-SNOM models can be for the case of Al$_2$O$_3$.
Additional examples of similarly large differences for SiO$_2$ and SiC samples can be found in previous works of the present authors and their collaborators.~\cite{Zhang2012nfs, McLeod2014lrm}
All these examples compel us to conclude that the prior success of the point- and finite-dipole models in fitting experimental data has to be due to insufficient range of the data, multitude of adjustable parameters, and also the demodulation and normalization procedures that mask the errors in both the functional form and the magnitude of the calculated signal.

Another way to explain the difference between the earlier \textit{ad hoc} models and our GSM is as follows.
For the case of a sample with a local reflectivity $\beta$, the exact scattering problem of a dielectric probe near a surface reduces to 
a generalized eigenproblem,~\cite{Agranovich1999gme} that has an infinite number of eigenmodes, as we discussed in Sec.~\ref{sec:Introduction}.
In contrast, both the point- and the finite-dipole models
attempt to approximate the \textit{infinite} number of eigenmodes by a \textit{single} one.

Since the real-space potential distribution of the eigenmodes [Fig.~\ref{fig:modes}(b)] depends on the shape and size of the probe and probe-sample distance but not on $\beta$,
we can describe interaction of the probe with an arbitrary sample efficiently
using the precalculated basis of such eigenmodes.
This allows one to use our GSM approach to model s-SNOM response for a wide range of materials.
However, calculations for realistic probe shapes are not always practical.
In search of a broadly applicable yet simple model, 
we have chosen the prolate spheroid to be our probe shape, as it captures the essential features of the actual probes --- 
a sharp apex and a strongly elongated shaft.
We quantified the eigenmodes of the probe-sample system in the form of poles and residues of the polarizability functions $\chi^\nu$ (Table~\ref{tab:table}), allowing an expedient, in fact,
instantaneous calculation of the s-SNOM reponse.
The point-dipole, finite-dipole and other \textit{ad hoc} models no longer have the advantage of computational speed
and should now be considered obsolete.

Recent work~\cite{McLeod2014lrm} has shown that
in the strong-coupling regime of the probe-sample interaction
a fully electrodynamic treatment using the BEM and realistic probe shape is necessary in order to reproduce the  measurements.
This regime is realized experimentally~\cite{Amarie2011bia, McLeod2014lrm} when using samples of SiC, a material for which $|\beta|$ can be as high as $15$.
The same considerations apply for Al$_2$O$_3$ for which $|\beta|$ can reach $12$, see Fig.~\ref{fig:Al2O3_beta}(a).
Our GSM theory gives analytical insight into near-field response of such materials.
We have shown that due to simultaneous excitation of 
multiple eigenmodes, novel features of the s-SNOM signal such as multi-peaked spectra and nonmonotonic approach curves can appear.
These features are however very sensitive to experimental parameters such as tapping amplitude, minimum approach distance, and even the data collection protocol.
Retardation effects, especially radiative damping can also qualitatively alter the signal and must be considered.
In order to observe the predicted anomalous approach curves and multi-peak spectra, it may be necessary to make efforts to minimize the radiative damping,
which requires working with shorter probes or at lower frequencies.
In contrast, in the weak- and moderate-coupling regimes, which are relevant for the vast majority of samples, the lowest-order eigenmode is dominant.
Hence, the approach curves should be monotonic in $\ztip$, while the spectra should be mostly insensitive to experimental details and retardation effects.
This is the regime where our spheroidal probe model can be used with the greatest confidence.

Our GSM theory also applies to a more complicated problem where the
sample reflectivity is nonlocal, i.e., momentum-dependent.
Here the salient advantages of our method are two-fold.
First, in the case of a weak nonlocality, our GSM provides a mapping of the nonlocal problem to a local one.
Thereby the sample-independent eigenmode decomposition is retained, providing an intuitive interpretation of the scattering signal.
Second, our numerical algorithm  (see Supplementary online materials) is much more efficient than the standard BEM because the number of necessary matrix element calculations scales linearly instead of quadratically with the matrix size.
It will be worthwhile to compare the actual computational speed of our algorithm with that of a recently developed and significantly more efficient BEM that utilizes pre-calculated matrix elements.~\cite{McLeod2014lrm}

We hope that the improved physical understanding of near-field probe-sample coupling enabled by the generalized spectral method advanced in this work as well as the numerical procedures we developed for its implementation can be of use for modeling and analysis of future s-SNOM and other near-field experiments.

The work at UCSD is supported by DOE grants DE-FG02-08ER46512 and DE-SC00122592 and by UCOP.
A.H.C.N. acknowledges the National Research Foundation, Prime Minister Office, Singapore, under its Medium Sized Centre Programme and CRP award ``Novel 2D materials with tailored properties: beyond graphene'' (R-144-000-295-281).
We thank A.~S.~McLeod and F.~Keilmann for valuable discussions and comments on the manuscript.

\appendix

\section{The electrostatic problem of a spheroidal probe}
\label{sec:spheroidal_model}

The electric field created by a spheroidal object is most conveniently described in the prolate spheroidal coordinates $(\xi, \eta, \phi)$ where the origin of the coordinate system is located at the center of the probe, as shown in Fig.~\ref{fig:pro_sph_coord}.
The relationships to the cylindrical polar coordinates $(\rho,\phi,z)$ are
\begin{equation}
	z = F \, \xi \eta
	\,,\quad
	\rho = F \sqrt{\xi^2 - 1} \sqrt{1 - \eta^2}\,.
	\label{eqn:psc_to_cpc}
\end{equation}
In the spheroidal coordinate each spatial position is specified by $\xi \in [1,\infty)$, $\eta \in [-1,1]$, and $\phi$ is the usual azimuthal angle.
Contours of constant $\xi$ are a series of concentric spheroids centered at the origin, with the major axis along the $z$ direction and common foci at $z=\pm F$.
For each such spheroid, $\xi$ is equal to the ratio of its major semi-axis and focal length.
We consider the case when the surface of the probe coincides with one of the spheroidal surfaces $\xi = \xi_0 = L / F$, where $L$ is the half-length or major semi-axis of the probe.
Related quantities such as the minor semi-radius $W$ of the probe  or the radius of curvature $a$ at the apex  are given by $W = \sqrt{L^2 - F^2}$ and  $a = W^2 / L$.

It is well known that Laplace's equation $\nabla^2 \Phi = 0$ has separable solutions in the prolate spheroidal coordinates.
In particular, we are interested in solutions outside a spheroidal probe
that decay at large $\xi$.
Their most general form is written in terms of a linear combination of spheroidal harmonics as follows:
\begin{equation}
	\Phi\tsub{sphd} \left( \xi, \eta, \phi \right)
	= \sum_{l=0}^{\infty} \sum_{m=-l}^{l}
		B_{l}^{m} \mathsf{P}_{l}^{m}(\eta) e^{i m \phi}
		\frac
		{ P_{l}^{m}(\xi_{<}) Q_{l}^{m}(\xi_{>}) }
		{ P_{l}^{m} (\xi_0) } \,,
  \label{eqn:Laplace_psc_sol}
\end{equation}
where $B_{l}^{m}$ are coefficients to be determined from boundary condition, $\mathsf{P}_{l}^{m}(\zeta)$ is the associated Legendre polynomial defined on the interval $[-1,1]$ and
$P_{l}^{m}(\zeta)$ and $Q_{l}^{m} (\zeta)$ are the associated Legendre function of the first kind and second kind (See, e.g. Ref.~\onlinecite{Gradshteyn_Ryzhik}) with
\begin{equation}
	\xi_{<} \equiv \min(\xi, \xi_0)
	\,,\quad
	\xi_{>} \equiv \max(\xi, \xi_0) \,.
\end{equation}
With the above definition of $\xi_{>,<}$, Eq.~\eqref{eqn:Laplace_psc_sol} covers 
both inside and outside the surface of the probe at $\xi_0$.


For the geometry considered in this paper [Fig.~\ref{fig:modes}(a)], the total potential $\Phi$ can be written as
\begin{equation}
	\Phi = \Phi_{0} + \Phi\tsub{plane} + \Phi\tsub{sphd}\,,
	\label{eqn:Phi_total}
\end{equation}
where
\begin{equation}
	\Phi_{0} (\bm{r}) = - \bm{E}_0 \bm{r}
	\label{eqn:Phi_0}
\end{equation}
is the potential of the external uniform field,
and $\Phi\tsub{plane}$ is the potential due to charges in the sample, which can be decomposed into evanescent plane wave as
\begin{equation}
	\Phi\tsub{plane} (\bm{r}) =
	\int B(\bm{q}) e^{-q z}
	e^{i \bm{q} \bm{\rho}} \frac{d^2 q}{4\pi^2} \,,
	\label{eqn:Phi_plane}
\end{equation}
where the position vector $\bm{r} = \bm{\rho} + z \bhat{z}$
is broken up into its cylindrical polar coordinate components.
We determine $B(\bm{q})$, as well as $B_l^m$ from boundary conditions.

To do that we quote two well-known mathematical results: the decompositions of evanescent plane waves in terms of spheroidal harmonics and vice versa.
The first reads
\begin{widetext}
\begin{equation}
	e^{i \bm{q} \bm{\rho} - q z} = \zfsum{l} \sysum[l]{m}
	\frac{2l+1}{2} (-)^{l}  i^m \frac{(l-m)!}{(l+m)!}
	\sqrt{\frac{2\pi}{q F }} I_{l+\frac12}(q F)
	P_{l}^{m}(\xi)\mathsf{P}_{l}^{m}(\eta)
	e^{im(\phi-\phi_q)} \,,
	\label{eqn:pln_wav_in_sph_har}
\end{equation}
with $\phi$ and $\phi_q$ being the azimuthal angles of $\bm{\rho}$ and $\bm{q}$, respectively.
The reverse is:
\begin{equation}
	Q_{l}^{m}(\xi) \mathsf{P}_{l}^{m}(\eta)
	= (-)^{l} i^{m} \frac{(l+m)!}{(l-m)!}
	\int \frac{\pi F}{q} \sqrt{\frac{2\pi}{q F }} I_{l+\frac{1}{2}} (qF)
	e^{i \bm{q} \bm{\rho} + q z} e^{- i m (\phi - \phi_q)}
	\frac{\dd^2q}{4\pi^2}\,,
	\label{eqn:sph_har_in_pln_wav}
\end{equation}
\end{widetext}
where $I_{\nu}(z)$ is the modified Bessel function of the first kind.~\cite{Morse_book}
These two relations follow easily from addition theorems of general Legendre functions such as those in Ref.~\onlinecite{Gradshteyn_Ryzhik} and~\onlinecite{Hobson1929tat}.

Near the sample surface $z = -\zp$, the boundary condition is
\begin{equation}
	\tilde{\Phi} = B(\bm{q}) e^{-q \zp} + \tilde{\Phi}\tsub{sphd} (\bm{q}, \zp)
	\propto e^{q \zp} - r_\mathrm{P}(q) e^{-q \zp}\,,
	\label{eqn:bc_plane_orig}
\end{equation}
where we use the notation
\begin{equation}
	\tilde{f}(\bm{q},z) = \int f(\bm{r}) e^{-i\bm{q} \bm{\rho}} d^2\rho
\end{equation}
for a partial Fourier transformation.
Eq.~\eqref{eqn:bc_plane_orig} implies:
\begin{equation}
	B(\bm{q}) = - r_\mathrm{P} e^{-q\zp} \tilde{\Phi}\tsub{sphd}(\bm{q},\zp)\,.
	\label{eqn:bc_plane}
\end{equation}
The other boundary condition for a uniform spheroidal probe with dielectric constant $\epsilon\tsub{tip}$ is:
\begin{equation}
	\left. \frac{\partial \Phi}{\partial\xi} \right|_{\xi \to \xi_{0}^{+}}
	= \left. \epsilon\tsub{tip} \frac{\partial \Phi}{\partial\xi} \right|_{\xi \to \xi_{0}^{-}} \,.
	\label{eqn:bc_spheroid}
\end{equation}

Boundary conditions in Eqs.~\eqref{eqn:bc_plane} and~\eqref{eqn:bc_spheroid} with the decompositions in Eqs.~\eqref{eqn:pln_wav_in_sph_har} and ~\eqref{eqn:sph_har_in_pln_wav} allow one to compute the unknown coefficients $B_{l}^{m}$.
The result can be summarized by first defining an infinite matrix:
\begin{equation}
	{H}_{ll\prm} \equiv 2 \pi \zfint
	r_\mathrm{P}(q) I_{l\prm+\frac12} (q \, F) I_{l+\frac12} (q \, F)
	e^{-2 q \zp} \frac{dq}{q}
	\,,
	\label{eqn:matrix_H}
\end{equation}
whose elements are integrals of $r_\mathrm{P}(q)$.
Then for each integer $m$ a quantity related to $B_l^m$, a column vector 
defined by:
\begin{equation}
	{A}^{m}{}_{l} \equiv (-)^{l+m}
	\frac{(l+m)!}{(l-m)!} \frac{B_{l}^{m}}{F}
	\,,
	\label{eqn:column_A}
\end{equation}
is the solution to the linear system of equations
\begin{equation}
	\ofsum{l} \left( {\Lambda}^{m} - {H} \right)_{ll\prm}
	{A}^{m}{}_{l\prm} = {b}^{m}{}_{l} \,.
	\label{eqn:char_eqns}
\end{equation}
The diagonal matrix elements $\Lambda^{m}{}_{ll\prm} = \Lambda^{m}_{l} \delta_{l l\prm}$ are defined by
Eq.~\eqref{eqn:Lambda_def},
%
and the numbers on the right-hand side of the equations are given by
\begin{equation}
	{b}^{m}{}_{l}
	= \frac{4}{3} \frac{(1+m)!}{(1-m)!}C^m \delta_{l1}
	\,,
	\label{eqn:column_b}
\end{equation}
where $C^0=-E_z$ and $C^{\pm 1}=(E_x\mp iE_y)/2$.
Thus, the form of matrix $\bm{\Lambda}$ is determined completely by the geometry of the probe (in terms of $\xi_0$) and its dielectric constant $\epsilon_\mathrm{tip}$, while $\mathbf{H}$ describes the interaction between the sample reflection function $r_\mathrm{P}$ and the momentum selectivity of the modes.
The column vector $\mathbf{b}$ describes the uniform external field.

The coefficients $B_l^m$ can be obtained directly from Eq.~\eqref{eqn:column_A} after one solves for ${A}^{m}{}_{l}$ from Eq.~\eqref{eqn:char_eqns}.
But for the purpose of determining the induced probe dipole moment, only $|m|\le 1$ cases are important.
By examining the asymptotic behavior of the electrostatic potential $\Phi(\bm{r})$, one obtains the total dipole moment of the spheroid probe.
Its Cartesian components are related to the components of $\left( {A}^{m} \right)_l$ by
\begin{equation}
	p_{{\rm sp},0} = p_{{\rm sp},z}
	= - \frac{F^3}{3} A^{0}{}_1
	\,,\quad
	p_{{\rm sp},1} = \fhalf{p_{{\rm sp},x} - i p_{{\rm sp},y}}
	= \frac{F^3}{3} A^{1}{}_{1}
	\,.
	\label{eqn:p_sph}
\end{equation}

\section{The spherical probe limit}
\label{sec:spherical_limit}
The spheroidal probe model presented in Appendix~\ref{sec:spheroidal_model} is quite general and can be a good model for tips of any aspect ratio $L/a$.
Here we explore a particular limit of $F\to 0$ and $\xi\to\infty$ while keeping the product $F \xi \to a$ constant.
This corresponds to the problem of a spherical probe of radius $a$.
The derivation in Appendix~\ref{sec:spheroidal_model} simplifies to that in Sec.~4.1 of Ref.~\onlinecite{Ford1984eio}, and by using the following asymptotic forms of various special functions:
\begin{subequations}
	\begin{align}
		\sqrt{\frac{2\pi}{q F}} I_{l+\frac12} (qF)
		& \simeq \frac{2^{2l+1}l!}{(2l+1)!}
		\left( \frac{qF}{2} \right)^{l}
		\,, \\
		Q_{l}^{m} (\xi)
		& \simeq (-)^{m} \frac{2^{l} l! (l+m)!}{(2l+1)!} \xi^{-l-1}
		\,, \\
		P_{l}^{m} (\xi)
		& \simeq \frac{(2l)!}{2^{l} l! (l-m)!} \xi^{l}\,,
	\end{align}
\end{subequations}
one can show that the decompositions Eqs.~\eqref{eqn:pln_wav_in_sph_har} and~\eqref{eqn:sph_har_in_pln_wav} reduce to Eq.~(4.9) and~(4.10) of Ref.~\onlinecite{Ford1984eio}.
The characteristic equation Eq.~\eqref{eqn:char_eqns} for $B^{m}{}_{l}$ reduces to:
\begin{align}
	&\ofsum{l\prm}
	\left\{ \frac{\delta_{ll\prm}}{\alpha_{l} a^{2l+1}}
	- \frac{(l+l\prm)!}{(l+m)!(l'-m)!} \mathcal{F}_{l+l\prm} \right\}
	\bar{B}^{m}{}_{l\prm}\notag\\
	 &= \frac{\delta_{l1}}{(l-m)!}C^m\,,
	\label{eqn:char_eqns_sphere}
\end{align}
where
\begin{equation}
	\bar{B}^{m}{}_{l} = (-)^{l+m} (l+m)!
	\frac{2^{l} l!}{(2l+1)!}
	F^{l+1} B^{m}{}_{l}
	\label{eqn:B_sphere_def}
\end{equation}
is similarly related to the induced charge distribution of the probe,
\begin{equation}
	\alpha_{l} = \frac{l(\eps{\text{tip}}-1)}{l(\eps{\text{tip}}+1) + 1}
	\label{eqn:alpha_l_sphere}
\end{equation}
is the multipole polarizability of the probe, and
\begin{equation}
	\mathcal{F}_{l} = \frac{1}{l!} \zfint r_\mathrm{P}(q) q^{l} e^{- 2 q d_{0}} dq 
	\label{eqn:F_L_sphere}
\end{equation}
is the integral that characterizes the interaction between 
the spherical probe and the sample with $d_0=a+\ztip$.
Eq.~\eqref{eqn:char_eqns_sphere}, the characteristic equation for a spherical probe, is derived in Ref.~\onlinecite{Ford1984eio} as Eq.~(4.20).
The solution to Eq.~\eqref{eqn:char_eqns_sphere} has some of the same properties as the spheroid case: $\bar{B}^{m}{}_{l} = 0$ for all $l$ and $|m|>1$; $\bar{B}^{0}_{l}$ is related to the charge distribution due to the $z$ component of the electric field and $\bar{B}^{\pm 1}_{l}$ are related to the charge distribution due to the $x$-$y$ component of the electric field.

For the case of $q$ independent $r_\mathrm{P}(q,\omega) = \beta(\omega)$, in which the integrals $\mathcal{F}_{l}$ reduces to:
\begin{equation}
	\mathcal{F}_{l} = \frac{\beta}{(2d_{0})^{l+1}} \,,
	\label{eqn:F_L_beta}
\end{equation}
there is an exact solution to the spherical characteristic equation.
Let
\begin{equation}
  \alpha = \arccosh \frac{d_{0}}{a}
  = \arccosh \left( 1 + \frac{\ztip}{a} \right)
  \,,
	\label{eqn:alpha_def}
\end{equation}
be a dimensionless parameter that characterizes the sphere-to-sample distance relative to its size, and let
\begin{equation}
	\sigma_{k}(\beta; \alpha) = \zfsum{m} \frac{(2m+1)^{k}}{e^{(2m+1)\alpha} - \beta} \,.
	\label{eqn:sigma_k_def}
\end{equation}
Using the following quantities:
\begin{subequations}
\begin{align}
	p_0 &= \chi_0 E_z\,, \quad \chi_0 = a^3\,,
	\label{eqn:p_0_def} \\
	q_0 &= \frac{p_0}{a}\left(\cosh \alpha - \sinh \alpha \frac{\sigma_{1}}{\sigma_{0}}\right)\,,
	\label{eqn:q_0_def} \\
	p_{n} &= p_0 \beta^{n} \left( \frac{\sinh \alpha}{\sinh (n+1)\alpha} \right)^{3}\,,
	\label{eqn: p_n_def} \\
	q_{n} &=  \frac{\beta^{n} \sinh\alpha}{\sinh (n+1)\alpha}
	\left[ q_0 -\frac{p_0}{a} \frac{\sinh n\alpha}{\sinh (n+1)\alpha}
	\right]\,,
	\label{eqn: q_n_def}
\end{align}
\end{subequations}
it can be shown that for a metallic sphere
\begin{equation}
	\bar{B}^{0}_{l} = (-)^{l} \zfsum{n} q_{n} (d_{0}-d_{n})^{l}
						- p_{n} l(d_{0}-d_{n})^{l-1} \,,
	\label{eqn:sphere_b_l_0}
\end{equation}
where
\begin{equation}
	d_0 - d_n = a \frac{\sinh n\alpha}{\sinh (n+1)\alpha}\,.
	\label{eqn:sphere_image_b_n}
\end{equation}

\begin{figure}[tb]
	\begin{center}
		\includegraphics[width=0.9\linewidth]{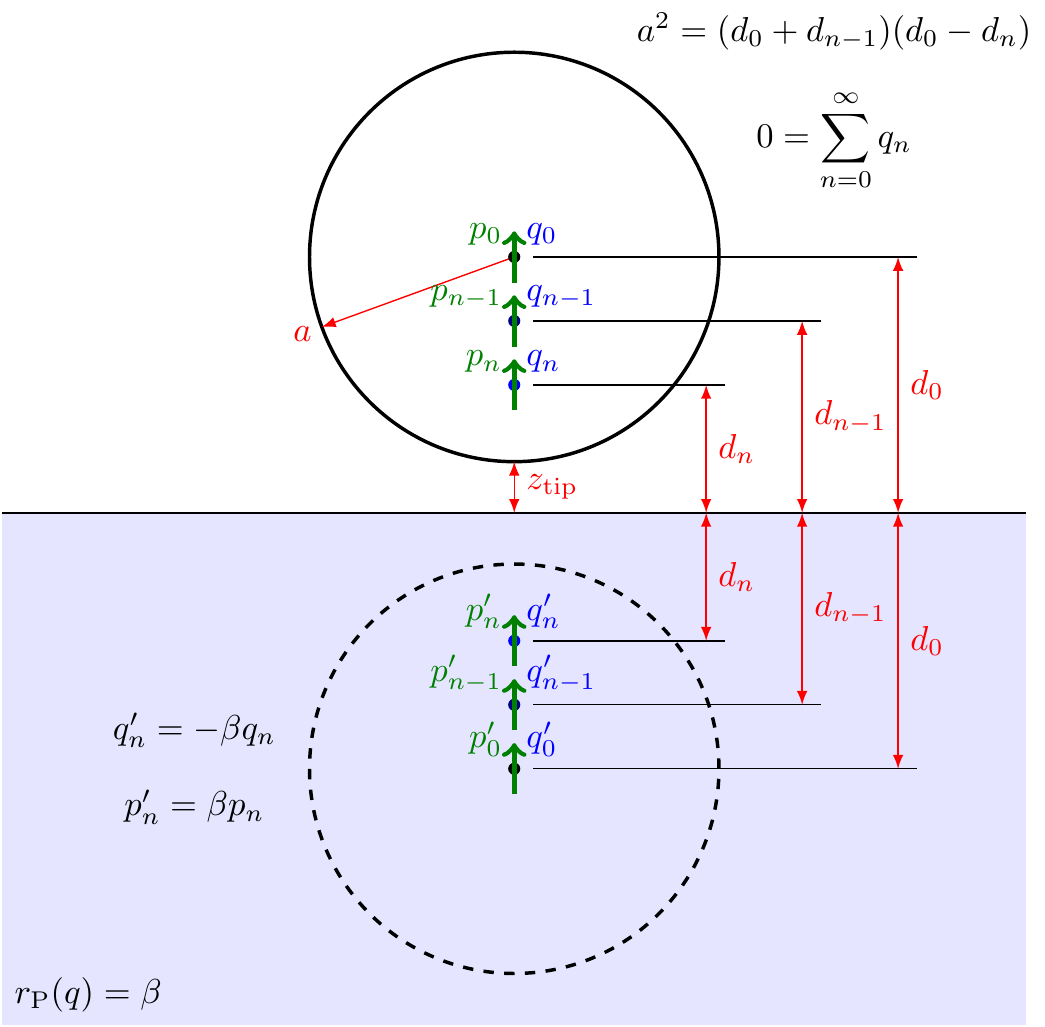}
	\end{center}
	\caption{The method-of-images solution of the problem of a metallic sphere above a dielectric half-space with the external field normal to the interface. 
	The method involves an infinite series of dipoles $p_{n}$ and point charges $q_{n}$ located inside the sphere at distances $d_{n}$ above the interface. The total charge inside the sphere is zero.}
	\label{fig:sphere_and_plane}
\end{figure}

The physical meaning of Eq.~\eqref{eqn:sphere_b_l_0} becomes clear when one treats the problem with method of images (Fig.~\ref{fig:sphere_and_plane}).
Suppose that an external electric field $E_z$ would have induced a bare dipole moment $p_0$ in the sphere.
This would induce an image dipole in the dielectric half-space, which would in turn induce an image dipole and an image charge in the sphere.
The position and strength of each successive image dipole and charge can be solved by recursion.
Setting the sample surface to $z=0$, the position of the center of the sphere is at $z=d_0$. At each position $z=d_n$ given by Eq.~\eqref{eqn:sphere_image_b_n} there is a point dipole $p_n$
and a point charge $q_n$.
Charge $q_0$ is determined by the neutrality condition $\sum_n q_n = 0$,
which yields Eq.~\eqref{eqn:q_0_def}.
Summing up all the contributions to the total dipole moment from both the dipoles \emph{and} the point charges inside the sphere, we get:
\begin{equation}
  \begin{aligned}
	\frac{\chi\tsup{sph,$\bot$}}{\chi_0}
  &\equiv
		\frac{p_{z}\tsup{total}}{p_0}
	= \frac{1}{p_0} \sum_{n=0}^{\infty}
		\left[ \vphantom{\sum} p_n + q_n (d_n - d_0) \right] \\
	&=	2 \sinh^3 \alpha
	\left( \sigma_2 - \frac{\sigma_1^{2}}{\sigma_0} \right) \,.
  \end{aligned}
	\label{eqn:sphere_P_total}
\end{equation}
with $\sigma_{k}$ given by Eq.~\eqref{eqn:sigma_k_def}.

The above analysis resulting in Eq.~\eqref{eqn:sphere_P_total} is for the case where the electric field is perpendicular to the sample.
For the case where the electric field is parallel to the sample, the analysis is simpler in that the positions of the image dipoles and their strength are the same, but no image point charges are present.
Therefore, in this polarization:
\begin{equation}
	\frac{\chi\tsup{sph,$\parallel$}}{\chi_0}
	\equiv \frac{p_{xy}\tsup{total}}{p_{0}}
	= \sinh^{3} \alpha \left( \sigma_{2} - \sigma_{0} \right) \,.
	\label{eqn:sphere_sigma_total}
\end{equation}

Both Eq.~\eqref{eqn:sphere_P_total} and~\eqref{eqn:sphere_sigma_total} conforms to our earlier assertion that $\chi$ has the form of Eq.~\eqref{eqn:chi_eff_general_form}:
\begin{equation}
  \chi\tsup{sph,$\parallel$}
  = \sum_{k=0}^{\infty}
  \frac{R_{k}\tsup{sph,$\parallel$}}
  {\beta_{k}\tsup{sph,$\parallel$} - \beta}
  \eqc
  \chi\tsup{sph,$\bot$}
  = \sum_{k=0}^{\infty}
  \frac{R_{k}\tsup{sph,$\bot$}}
  {\beta_{k}\tsup{sph,$\bot$} - \beta}\,.
\end{equation}
For horizontal electric fields, $\chi\tsup{sph,$\parallel$}$ is
singular whenever $\sigma_{2}$ or $\sigma_{0}$ is, so that
\begin{equation}
  \beta_{k}\tsup{sph,$\parallel$} = e^{(2k+3)\alpha} \,.
  \label{eqn:sphere_beta_m}
\end{equation}
The corresponding residues are:
\begin{equation}
	R_{k}\tsup{sph,$\parallel$} = 4(k+1)(k+2) \chi_0 \sinh^{3} \alpha \,.
\end{equation}
For electric fields perpendicular to the sample, the parenthesis in Eq.~\eqref{eqn:sphere_P_total} vanishes at each $\beta_{k}\tsup{sph,$\parallel$}$, so they are not poles of $\chi\tsup{sph,$\bot$}$.
Instead, $\beta_{k}\tsup{sph,$\bot$}$
occur at the zeros of $\sigma_{0}$ which has no simple analytic form.
The poles of the two polarizations, however, interleave:
\begin{equation}
  \beta_{k}\tsup{sph,$\parallel$}
  < \beta_{k}\tsup{sph,$\bot$}
  \lesssim \beta_{k+1}\tsup{sph,$\parallel$} \,.
  \label{eqn:beta nesting prop}
\end{equation}
\bibliography{bibliography}
\end{document}